\documentclass[3p,times,twocolumn]{elsarticle}

\pdfoutput=1

\usepackage{ecrc-ac,amsmath,widetext}
\usepackage{mathrsfs,latexsym}
\usepackage{booktabs}
\newcommand{\bew}{\begin{widetext}\begin{equation}\begin{aligned}}
\newcommand{\eew}{\end{aligned}\end{equation}\end{widetext}}

\usepackage{epsfig}
\usepackage{pstool}

\volume{00}

\firstpage{1}

\journalname{Nuclear Physics B Proceedings Supplement}

\runauth{Rainer Sommer}

\jid{nuphbp}

\jnltitlelogo{Nuclear Physics B Proceedings Supplement}

\usepackage{amssymb,bbm}
\usepackage{amsbsy,color}

\usepackage{macros_hs,miscdefs}

\begin{document}

\begin{frontmatter}



\dochead{ \small{\flushright{SFB/CPP-14-101 \\ DESY 15-006 \\}}}


\title{
Non-perturbative Heavy Quark Effective Theory:
\\
Introduction and Status
}


\author[DE,HU]{Rainer Sommer}

\address[DE]{John von Neumann Institute for Computing (NIC), DESY, Platanenallee~6, 15738 Zeuthen, Germany}
\address[HU]{Institut f{\"u}r Physik, Humboldt Universit{\"a}t, 
Newtonstr. 15, 12489 Berlin, Germany}

\begin{abstract}
We give an introduction to Heavy Quark Effective Theory (HQET). Our emphasis is on
its formulation non-perturbative in the strong coupling,
including the non-perturbative determination of the 
parameters in the HQET Lagrangian. In a second part we 
review the present status of HQET on the lattice, largely based 
on work of the ALPHA collaboration in the last few years.
We finally discuss opportunities and challenges.

\end{abstract}


\end{frontmatter}



\setcounter{tocdepth}{2} 
\tableofcontents

\vspace*{6mm}

\hrule

\section{Introduction}
\label{intro}

Heavy Quark Effective Theory (HQET) is an effective theory 
for QCD, the theory of strong interactions, 
in the limit where quark masses are large and other scales,
such as momenta are kept fixed. Understanding this limit is of great 
interest per se. 
In addition a control of HQET is very useful to arrive at phenomenological predictions
for B-meson properties and qualitatively also for D-mesons.
In particular B-meson decays need to be understood better
in order to further constrain the flavor sector of the 
standard model of particle physics.

In this article we give an introduction to 
HQET with an emphasis 
on its full non-perturbative formulation. 
Mostly we remain with the general ideas and an overview
of the present status and results. 
For more details concerning the basics
as well as the phenomenology
we refer to the literature \cite{reviews:neubert,books:m,books:G,reviews:Nara,LH:rainer}
with the non-perturbative aspects and particularly the 
discretisation on a lattice covered in the last reference.
Here, on the other hand, we give a more complete discussion 
of the status and of non-perturbative computations and the 
challenges for the future.

HQET, as discussed here, is an effective field theory for the low energy physics of energy levels or transition matrix elements with a single heavy quark or anti-quark
in initial and/or final state. We will label these hadronic 
states by H and the quark by h. The latter has a mass $\mh$.~\footnote{
We use the symbol $\mh$ generically when the precise definition
of the renormalized mass does not matter.
}
Usually we think of H as a B-meson, but it can be 
e.g. a baryon with beauty quantum number of one. 
We always consider a rest-frame where all spatial momenta $\vecp_i$ are small. The effective theory then yields the expansion 
of observables of QCD 
in powers 
\bes
    \label{e:exppar}
    (|\vecp_i| / \mh)^n\,, \;  (\Lambda / \mh)^n\,, \;
    (m_j / \mh)^n
\ees
where h is a heavy quark, while 
the masses $m_j$ of the  other quarks
are considered small.~\footnote{
At this stage we are interested mostly in the  
theory and less the phenomenology, where one
may ask whether the charm quark is to be treated 
as a light quark or a heavy one. Top quarks on the
other hand are not considered at all. They are heavy 
enough to safely be considered as decoupled.
}
\subsection{Mass scaling and phenomenology}
The HQET expansion is of a {\bf theoretical interest} because it 
describes the asymptotics of QCD as $\mh\to\infty$. 
For example we obtain statements such as
\bes
\label{e:expansion}
   &\cmqcd(\{\vecp_i\}, \{m_j\},\mh) = 
   \cm^\stat(\{\vecp_i\}, \{m_j\},\mh)  
   \nonumber \\ 
  &\phantom{\cmqcd(\{\vecp_i\}, \{m_j\},\mh) =}
  \times (1+\rmO(\mhinv))
\ees
with $\rmO(\mhinv)$ summarizing the terms of
\eq{e:exppar}. The intrinsic scale 
$\Lambda$ may be taken to be any low energy QCD scale.

The important content of \eq{e:expansion} is
that it gives the large mass scaling of observables 
$\cmqcd$ with $\mh$, in the form of the static (lowest order) of HQET prediction, usually
\be
    \cm^\stat \sim (\mh)^s  \label{e:massscal1}
\ee 
with the (not necessarily integer) power $s$ determined
by counting dimensions and adding anomalous ones in the static effective theory.
Understanding this scaling is clearly a very relevant part of 
understanding QCD.

\begin{table*}[t!]
\begin{tabular}{llllll}
\toprule
source of errors 
& 
\parbox[t]{0.15\textwidth}{cases}
&
asymptotics
&
\parbox[t]{0.2\textwidth}{\% effects}
&
References 
\\
\midrule
\parbox[t]{0.23\textwidth}{finite volume effect   
due to particle exchange around the periodic space} 
&
&
\parbox[t]{0.19\textwidth}{$\rmO(\exp(-m_\mathrm{gap} L))$}
&
\parbox[t]{0.2\textwidth}{$\mpi L \approx 4$}
&
\parbox[t]{0.09\textwidth}{\cite{Colangelo:2005gd,Colangelo:2003hf,Colangelo:2002hy}}
\\[7ex]
\parbox[t]{0.23\textwidth}{discretisation errors\\$(\rmO(a)$-improved)} 
&
\parbox[t]{0.15\textwidth}{lattice QCD \\ with b}
&
\parbox[t]{0.19\textwidth}{$\rmO((a E_i)^2)$\,, 
                           $E_i \sim \mpi \ldots \mB$}
&
\parbox[t]{0.2\textwidth}{extrapolation with \\
$a\leq0.025\,\fm$}
&
\parbox[t]{0.09\textwidth}{\cite{Symanzik:1983dc,Symanzik:1983gh,Luscher:1984xn}}
\\[4ex]
\parbox[t]{0.34\textwidth}{} 
&
\parbox[t]{0.15\textwidth}{lattice HQET \\ without c}
&
\parbox[t]{0.19\textwidth}{$\rmO((a E_i)^2)$\,, 
                           $E_i \sim \mpi \ldots |\vecp_i|, \Lambda$}
&
\parbox[t]{0.2\textwidth}{extrapolation with \\
$a\leq0.1\,\fm$}
&
\parbox[t]{0.09\textwidth}{\cite{Symanzik:1983dc,Symanzik:1983gh,Luscher:1984xn}}
\\
\bottomrule
\end{tabular}
\caption{\label{t:syst}Effects in lattice QCD computations due to 
infrared and ultraviolet cutoffs. The lowest particle mass is denoted by
$m_\mathrm{gap}$; in QCD with light quarks this is $m_\mathrm{gap}=\mpi$. 
In the column titled ``\% effects'' we list the condition needed
to have systematic errors around the 1\% level.}

\end{table*}

A second important motivation for studying (and computing in)
HQET is that the b-quark mass, say in the $\msbar$ scheme 
at $4\,\GeV$ renormalization scale, is of order $4 \GeV$.
It is an order of magnitude larger than the intrinsic 
QCD scale of around $\Lambda \approx 400\,\MeV$.
Indeed, if one wants the $\rmO(\mhinv)$ to give a first 
estimate of the numerical size of the corrections 
(without additional pre-factors) this value for $\Lambda$
is appropriate in \eq{e:exppar} as we will 
see in \sect{s:res}.
Consequently, as long as we keep momenta small,
static predictions are expected to be good at
the 10\% level and one has an accuracy at the 1\% level
when $\mhinv$ corrections are included. Thus HQET is 
a very interesting {\bf phenomenological tool}. 

\subsection{Heavy quarks in lattice QCD}
\label{s:difficulty}
One may still wonder why it is of interest in the 
context of lattice QCD. The reason is simply that 
a numerical lattice QCD computation necessarily is
done with an infrared cutoff $1/L$ through the linear extent
$L$ of the simulated $T\times L^3$ world on top
of the ultraviolet cutoff $1/a$ introduced by the 
lattice spacing $a$. The accessible physical energies 
$E_i$ have to be removed from these scales,
\be
   1/L \ll E_i \ll 1/a  \,,
\ee
otherwise properties of the associated states are distorted. 

In \tab{t:syst} we list the most relevant
effects that are at the origin of these bounds as well as the
errors which result from violating them.
Since the finite volume effects are exponential in $L$, 
the bound of $\mpi L = 4$ is  rather sharp. However it 
depends on the pion mass $\mpi$ which one has in the simulation. 
Reasonable values of $\mpi=300\,\MeV \ldots 150\,\MeV$
lead to $L\geq 2.5\,\fm \ldots 5\,\fm$. 

In contrast discretisation errors only disappear like $a^2$. 
Further they vary a lot depending on the quantity
and discretised action. They simply have to be studied by
changing $a$ and the difficult question is where the asymptotic
$a^2$ behavior sets in. From then on a factor 2 variation in
$a^2$ (or better more) is acceptable. We have included our
rough estimate where $a^2$ scaling sets in. 
Together with the required $L\geq 2.5\,\fm \ldots 5\,\fm$,
this shows that
$L/a$ has to be prohibitively large when the b-quark is 
simulated as a relativistic quark\footnote{We should mention 
that not everybody in the field agrees with this statement. 
There are lattice QCD computations with quark masses 
very close to the physical b-quark mass and $a\mh \lesssim 1$. Discretisation errors are fitted with polynomials in the lattice spacing and these representations of the data are used
to extrapolate the results to the continuum and the physical mass. As an example
we cite \cite{McNeile:2012qf}.}, while for HQET
lattices of size $L/a = 32\ldots 64$ seem sufficient.
For this reason lattice HQET is a very attractive 
phenomenological tool.

\subsection{Defining effective field theories beyond 
perturbation theory } \label{s:eft}
We now describe the general concept and formulation 
of an effective field theory. The special features of
HQET will be mentioned in the following subsection. 
We consider processes in a fundamental theory (QCD or the 
standard model of particle physics -- the important feature is the 
renormalizability of the theory) at low energy. 
In particular we
first focus on
processes (scattering, decay) of particles with masses of this low energy or below it (in HQET also the large mass particles {\em are}
involved as will be discussed soon). In this
situation, 
vacuum fluctuations involving much heavier particles
are suppressed and a true creation of the heavier 
particles is energetically forbidden. 
One therefore expects to be
able to describe the physics of 
these low energy processes by an effective field theory
containing only the fields of the light particles~\cite{Weinberg:1978kz}. The leading order Lagrangian of the theory
is formed first from the free field theory Lagrangians 
and all the renormalizable interactions. 
Assuming the usual power counting, all local
composite fields with mass dimension smaller or equal to 
four are allowed.  Let us denote the Lagrangian by $\lag{LO}$
and the Euclidean action is 
$\act{LO} = \int \rmd^4 x \lag{LO}(x)$. 
Correlation functions are then defined 
by the standard path integral 
\be   
  \langle \Oop \rangle_\mathrm{LO} = \frac{1}{Z_\mathrm{LO}} 
  \int_\mathrm{fields} \rme^{-\act{LO}} \Oop \,.
\ee
with  $\langle 1 \rangle_\mathrm{LO} =1$ and  $\Oop$ some 
multilocal 
product of fields such as $\Oop=\Phi(x) \Phi(y)$.
In this way we start with a {\em renormalizable} theory. For a 
lattice formulation this means that the {\em continuum limit}
of the theory exists when a finite number of renormalized parameters are kept fixed. The continuum limit is then also
expected to be universal, i.e. independent of the 
specific discretisation.

Higher order terms in the expansion of physical 
amplitudes (or correlation functions) in $\mhinv$ are given
by including fields with higher mass dimension, which is 
compensated by the appropriate factor of the large mass
in the denominator,
\be 
  \label{e:NLOlag}
   \lag{NLO} = \sum_i \omega_i \lagfield_i \,, \;\;
   \omega_i = \frac{1}{\mh } \tilde\omega_i 
\ee 
where the parameters $\tilde\omega_i$ are dimensionless.
The fields contained in the 
(multi-local) $O$ 
are expanded in the same way as the action,
$
 \Oop_\mathrm{eff}  =
 \Oop_\mathrm{LO}+\Oop_\mathrm{NLO}+\ldots\,.
$
We now have to deal  with interactions in \eq{e:NLOlag} which are 
not renormalizable (by power counting). However, 
we are only interested in the
expansion $\Phi = \Phi_\mathrm{eff}^\mathrm{LO} + 
           \Phi_\mathrm{eff}^\mathrm{NLO} + \ldots $ of observables $\Phi$ in $\mh^{-1}$. It is therefore
sufficient to define the theory with the 
weight in the path integral expanded, 
$\rme^{-\act{}}\to\rme^{-\act{LO}}\{1 - \act{NLO}+\ldots\,\} \,$.
At NLO accuracy the expansion is then given
by 
\bes
  \Phi_\mathrm{eff}^\mathrm{LO} &=& 
  \langle \Oop^\mathrm{LO} \rangle_\mathrm{LO}
  \\ 
  \label{e:NLOexpct}
  \nspace\Phi_\mathrm{eff}^\mathrm{NLO} &=& 
  \langle \Oop^\mathrm{NLO} \rangle_\mathrm{LO}\,
  \\ &&
  - \left(\langle \Oop^\mathrm{LO}\act{NLO} \rangle_\mathrm{LO} 
  - \langle \Oop^\mathrm{LO}\rangle_\mathrm{LO} \,
            \langle\act{NLO} \rangle_\mathrm{LO}\right)
            \nonumber
\ees
and $            
  \act{NLO}= \int \rmd^4 x \, \lag{NLO}(x)$.
The term $\Phi_\mathrm{eff}^\mathrm{NLO}$ (but not
the individual terms on the r.h.s. of \eq{e:NLOexpct}) is renormalizable
with a finite number of counter terms which are
equivalent to renormalizing the parameters $\omega_i$
(including the LO ones). Also
parameters in the fields $\Oop$ are part of the 
list of $\omega_i$. The reader may worry about
divergences in the form of contact terms between $\lag{NLO}$
and $\Oop^\mathrm{LO}$ in $\langle \Oop^\mathrm{LO}\act{NLO} \rangle_\mathrm{LO}$. These, however, can all be absorbed
into the $\omega_i$, see \cite{impr:pap1}
and \cite{LH:rainer}. 

Renormalizability is particularly important for a 
non-perturbative evaluation of the path integral 
in a lattice formulation. The continuum limit of
an effective theory only 
exists when we treat the higher dimensional interactions 
as insertions in correlation functions in the form
of \eq{e:NLOexpct}. 

\subsection{HQET}
HQET reaches somewhat beyond the situation discussed above. The 
difference is that we are interested in processes which
do involve the heavy quark h at small
momenta (remember we choose the rest frame properly).
It is therefore not immediately clear, what are the 
degrees of freedom to be kept at low energy: what is the complete basis 
of low energy fields? The answer to this question was found in various 
ways. We here sketch one line of reasoning.

\def\psibar{\overline{\psi}}
\def\rme{{\rm e}}
\def\LD{\lag{}}
\def\Dop{{\cal D}}
\def\Obkin{{\bar{\mathcal{O}}_\mrm{kin}}}
\def\Obspin{{\bar{\mathcal{O}}_\mrm{spin}}}
\def\nab#1{{\nabla_{#1}}}
\def\lnabstar#1{\overleftarrow{\nabla}\kern-0.5pt\smash
             {\raise 4.5pt\hbox{$\ast$}}\kern-4.5pt_{#1}}
\def\nabstar#1{\nabla\kern-0.5pt\smash{\raise 4.5pt\hbox{$\ast$}}
               \kern-4.5pt_{#1}}
\def\vecD{{\bf D}}
\def\vecB{{\bf B}}
\def\vecsig{\boldsymbol{\sigma}}
\newcommand{\vecg}{\boldsymbol{\gamma}}
\def\Dg{D_k\gamma_k}
\def\hub{\psibar_{\mrm{h},u}}
\def\hu{\psi_{\mrm{h},u}}
\def\ahub{\psibar_{\bar{\mrm{h}},{u}}}
\def\ahu{\psi_{\bar{\mrm{h}},{u}}}

\def\rot{{\Sigma}}
\newcommand{\lagh}[1]{{\mathscr{L}}^{\rm {#1}}_\mathrm{h}}
\newcommand{\laghb}[1]{{\mathscr{L}}^{\rm {#1}}_\mathrm{\bar h}}

For smooth fields, the Dirac Lagrangian  
\bes
 \lag{} &=& \psibar(\mh+D_\mu\gamma_\mu)\psi 
\ees
can be split order by order in $\mhinv$
into decoupled upper and lower compenent quark field contributions,
corresponding to the particle and the anti-particle field:
\begin{eqnarray}
  \lag{} 
        &=& \lagh{stat} \;+\;   \lagh{(1)} 
        \\ 
        &+& \laghb{stat} 
       +  \laghb{(1)}
       +\rmO(\frac{1}{ \mh^2}) \\[1ex]
 \lagh{stat} &=&   \psibarheavy(\mh+D_0)\psiheavy\,, 
       \label{e:laghclass}
                 \\
\laghb{stat} &=&  \aheavyb(\mh-D_0)\aheavy\,,\quad 
               \\[1ex]
\lagh{(1)} &=& -\frac{1}{ 2\mh}(\Okin + \Ospin ) \,.
\end{eqnarray}
The expansion is correct up to terms of order $1/\mh^2$,
assuming $D_0\psi =\rmO(\mh)$, $D_k\psi =\rmO(1)=G_\mu$. 
Here $G_\mu$ is the gauge field, $D_\mu$ the covariant derivative
and we introduced the higher dimensional 
fields
\bes
   \Okin(x)&=&\psibarheavy(x)\,\vecD^2\, \psiheavy(x)\,, \; 
  \\    \Ospin(x)&=&\psibarheavy(x)\,\vecsig\cdot\vecB(x)\, \psiheavy(x)\,,
\ees
with
\bes
  & \sigma_k = \frac12\epsilon_{ijk}\sigma_{ij}\,,\;
  \label{e:sigmak}
  \quad 
  B_k = i\frac12\epsilon_{ijk} [D_i,D_j]\,.\; 
\end{eqnarray}
The decoupling of the fields $\psiheavy,\aheavy$ is achieved by a Fouldy Wouthuysen-Tani (FTW) transformation 
 (see \cite{books:IZ,Korner:1991kf}) of the form
\bes
  \chi = \exp(i \rot(m)) \,\psi \,,\quad 
\ees  
{with } 
\bes
 \rot(m) = \frac{-i}{2\mh } D_k\gamma_k 
          + \frac{-i}{4\mh^2 } \gamma_k\gamma_0 [D_k,D_0]\,,
\ees
followed by a projection onto decoupled 
components
\bes
   \psiheavy=P_+ \chi\,, \quad \aheavy=P_- \chi\,, \; 
\ees
{with } 
\bes
   P_{\pm}=\frac12(1\pm\gamma_0)\,,\quad P_+P_-=0. 
\ees
Analogous expressions for $\psibarheavy$ and $\laghb{(1)}$ are skipped here. 
Indeed, in the following we 
do not consider processes involving the anti-quark field,
$\aheavy$, and
therefore drop all terms containing it.

It is worth summarizing some issues that arise in this formal derivation.
\bi
\item
  Assuming $D_k = \rmO(1)=G_\mu$ means that this is a classical derivation:
  in the quantum field theory path integral we integrate over rough fields,
  i.e. there are arbitrarily large derivatives. The renormalization
  of the derived classical Lagrangian 
  could then in principle 
  result in additional terms of a different structure. 
\item
  The derivation is perturbative in $\mhinv$, order by order. This is 
  all we want for an EFT. In this way we expect to obtain the {\em asymptotic} expansion 
  in powers of $\mhinv$. 
\item 
  There are other ways to ``derive'' the form of the Lagrangian.
  One may integrate out the components $P_-\psi, \; \psibar P_-$ in a path integral 
  and then perform a formal expansion of the resulting non-local action
  for the remaining fields in terms of a series of 
  local operators \cite{hqet:cont6}. Or one may perform a 
  hopping parameter expansion of the Wilson-Dirac lattice propagator.  
  The leading term gives the propagator of the static action.
\ei

\subsection{Heavy quark symmetries} 
The lowest order Lagrangian $\lagh{stat}$ has new 
symmetries. At each space-time point one may perform 
SU(2) rotations in the two-dimensional space spanned
by the two (non-relativistic) Dirac-components. This invariance 
is the spin-symmetry, which for example exactly relates
the correlation function of the vector current 
$V_k^\mathrm{stat}=\psibar_{\up }\gamma_k\,\psiheavy$
to those of the time component of the axial current 
$A_0^\mathrm{stat}=\psibar_{\up }\gamma_5\gamma_0\,\psiheavy$.
Furthermore, a phase-transformation 
$\psiheavy(x) \to \rme^{i\alpha(\vecx)}\psiheavy(x)$ leaves the LO action 
invariant. It means that the number of h-quarks is {\em conserved 
locally} at each space point. 
It is common to remove the mass term $\mh\psibarheavy\psiheavy$
from the Lagrangian. In Minkowski space this can be done by 
a time-dependent phase transformation of the quark fields.
In the Euclidean it turns into an exponential factor. 
The physical interpretation is (in both cases) that one just
shifts the energies of the states with a single h-quark by exactly 
$\mh$. We find it simpler to keep $\mh\psibarheavy\psiheavy$
because a term of this form appears upon renormalization 
anyway. The formulation without the mass-term, however, exposes
another symmetry, namely heavy-quark flavor symmetry which 
is present when more than one heavy quark are present. 
It has approximate phenomenological consequences. We do
not need it here, mainly since in Nature there is no obvious
partner of the b-quark which is heavy enough 
and forms bound states.\footnote{Of course, one often 
considers the charm quark in an HQET expansion, but on the
quantitative level large corrections have to be expected
with a mass of around a GEV.}

\begin{figure}[t!]
  \centering
  \includegraphics[width=0.42\textwidth]{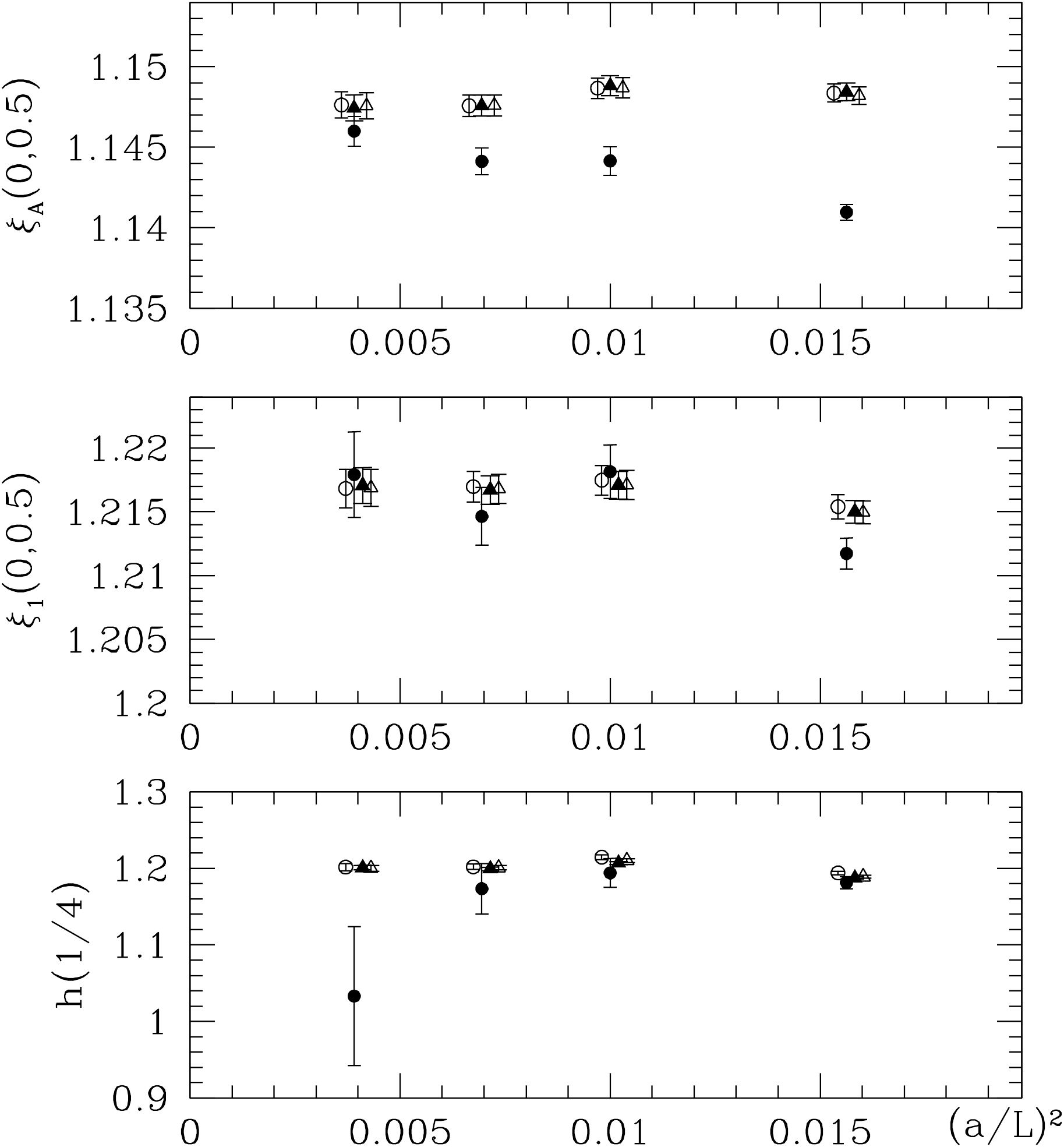}
  \caption{ 
The lattice spacing dependence of $\xi_{\rm A}(0,0.5)$, $\xi_1(0,0.5)$ and $h(1/4)$ in the quenched approximation\cite{DellaMorte:2005yc}. 
Some data points have been shifted in $a^2$ for visibility. 
Different symbols refer to different discretisation of the static  action with filled circles the original one by 
Eichten and Hill \cite{Eichten:1989zv}.  Graph from \cite{DellaMorte:2005yc}.}
\label{f:scal}
\end{figure}

\subsection{Theoretical status}
\label{s:status}
In \sect{s:eft} we have emphasized the importance of
the renormalizability of the 
lowest order of the effective theory. In HQET we 
depend on the renormalizability of the 
static theory. To our knowledge, 
this important property of the theory 
has not been proven to all orders in the coupling constant 
expansion -- in contrast to QCD. Simple power counting 
does not apply, since the static propagator does not fall off
in all directions in momentum space. An alternative strategy is
to prove the renormalizability after integrating out the 
static quark fields. The resulting non-local observables 
are then defined in QCD and are closely related to Wilson loops
whose renormalizability has been proven to all 
orders~\cite{
Dotsenko:1979wb,Gervais:1979fv,Arefeva:1980zd,Brandt:1981kf,Dorn:1986dt}.
Indeed, in \cite{Dorn:1986dt} also observables are considered which are 
very closely related to the non-local observables resulting from HQET. 
Furthermore, a significant number 
of loop computations have been performed, partially to
a high order \cite{Grozin:2004yc}. No problem with the assumed 
renormalizability has been found. Also the continuum limit
of the lattice theory has been studied in quite some 
detail (see e.g. \cite{DellaMorte:2005yc}). Its existence
is a non-perturbative ``proof'' of renormalizability. 
We use quotation marks, since it is a numerical proof only. 
Still, the quality of the numerical investigation is very good.
We demonstrate it by a graph from \cite{DellaMorte:2005yc}.
It shows three quantities in the static effective theory, 
which can be computed precisely. They are constructed from
correlation functions in a finite volume with Schr\"odinger functional boundary conditions \cite{Luscher:1992an,Sint:1993un,Sint:1995rb} i.e. Dirichlet boundary conditions in time, and periodic boundary conditions in 
space. For the quark fields, the spatial boundary conditions involve a phase $\theta$ via, 
\bes
    \psi(x+L\hat k) &=& \rme^{i \theta} \psi(x)\,,
    \\
    \psibar(x+L\hat k) &=& \rme^{-i \theta} \psibar(x)\,.   
\ees
Such boundary conditions are (precisely speaking at lowest
order in perturbation theory) equivalent to raising the lowest 
momentum of the finite volume modes from $\vecp =0$ to
$\vecp=\vecp_\theta \equiv \theta\times (1,1,1)$ for the quarks and 
$\vecp= - \vecp_\theta$ for the anti-quarks. 
One can then consider various types of correlation
functions, see also \sect{s:tests}.
Here we need $\fastat(\theta)$, where a
relativistic quark with $\vecp=\vecp_\theta$
and a static antiquark with $\vecp=-\vecp_\theta$
are created at time zero and are annihilated in the middle
of the $T\times L^3$ space-time through the static-light axial current $\astat$. This correlation function is depicted
on the left side of \fig{f:sfcorrel}.  Forming 
$\xi_\mathrm{A}=\fastat(\theta)/\fastat(\theta')$ gives
a first observable, which according to our naive
application of dimensional counting and symmetries 
needs no renormalization. It is also precisely computable. 
For $\theta=0,\theta'=1/2$ it is shown on the top of 
\fig{f:scal}. Letting the quark and the static anti-quark propagate from the boundary at $x_0=0$ to $x_0=T$, defines
$\Fone$ (rightmost graph in \fig{f:sfcorrel}) 
and $\xi_1$. Its continuum limit is shown for the
same kinematics in the middle graph of \fig{f:scal}. 
Finally, the ratio $h$ shown at the bottom of \fig{f:scal} is
constructed from the boundary-to-boundary 
correlation function of a static
quark-antiquark pair.
For all three cases, the lattice spacing dependence is shown for 
four different discretizations. One can see by eye that they 
tend to agree at $a=0$. We have good numerical evidence for 
the expected renormalizability of the theory. Much more 
such evidence was seen
in later works.

In summary, we have no doubt about the renormalizability of the 
theory.

\subsection{HQET parameters}

EFTs have a number of free 
parameters, the $\omega_i$ mentioned above. The number of parameters
usually grows rapidly as one increases the order of the expansion.
Concerning HQET, we provide an overview for the first two 
non-trivial orders in \tab{t:hqetparam}. Depending on the
application, i.e. the particular observable which is being
expanded, the number of parameters which actually contribute
may vary significantly. For example in the first row of the 
table there are three parameters for the Lagrangian at NLO. 
These generically contribute to any observable, but there are
exceptions. For example the splitting of two levels 
related by the spin symmetry of the theory 
(an exact symmetry at the static order) depends on a single parameter 
only, $\omegaspin$. And for the computation of the spectrum of
the theory, only the first row is needed anyway, since
the other rows arise from the expansion of specific local fields.

In general the growth of the numbers of parameters
with the order is problematic for phenomenological applications
of EFTs. They are usually fixed from experiments, limiting the 
predictivity enormously. For HQET the situation is actually worse
than e.g. for chiral perturbation theory (see 
\cite{LH:marten} and references therein for a recent review).
Even when the parameters $\omega_i$ are known, the observables cannot be determined by perturbation theory in the couplings
(not to be confused with the expansion in $\mhinv$): they are
non-perturbative in the QCD coupling irrespective of the 
order in $\mhinv$.
A lattice QCD formulation is needed. In its absence 
the predictions of the effective theory comprise of
approximate scaling (with powers of the quark mass) 
of certain results from charm to bottom and there are relations between matrix elements due to the enhanced symmetry of the static limit. 
Instead, with a lattice formulation, the theory becomes fully predictive.

Once one has the lattice formulation, it is also natural 
to solve the problem of the number of parameters by determining
them from lattice-accessible quantities instead of from experimental
ones. This is indeed possible. Starting from the general 
idea~\cite{Heitger:2003nj} a detailed strategy was developed
in a number
of works which we will discuss in the following. The basic 
procedure is just to require
\be
    \label{e:genmatch}
    \Phi_i^\hqet = \Phi_i^\qcd\,, \; i=1,\ldots,N_\omega \,,
\ee
for a number of observables equal to the 
number $N_\omega$ of parameters $\omega_i$ present at a certain 
order. This procedure is referred to as matching.  
The $\Phi_i$ do not need to be experimental observables.
It suffices that they are accessible to 
precise lattice simulations.
\begin{table*}[t!]
\begin{center}
\renewcommand{\arraystretch}{1.25}
\begin{tabular}{@{\extracolsep{0.2cm}}llllll}
\toprule
      \multicolumn{2}{c}{static (LO)} &  
      \multicolumn{2}{c}{$\rmO(\mhinv)$  (NLO)} & origin & application\\ 
      \midrule 
      Number & $\omega_i$ &  Number & $\omega_i$ \\ 
      \midrule \hline
      1 & $m_\bare$ & 2 & $\omegakin,\ \omegaspin$  & $\lag{HQET}$  & all
      \\[2mm] 
      1 & $\ln Z_{A_0}^\hqet$ & 
      2 & $\ceff{A}{0}{1},\ \ceff{A}{0}{2}$ & $A^\hqet_0$ & 
      $\mathrm{B} \to \ell \nu$: $\fB$,\, 
      $\mathrm{B_s} \to \ell \bar\ell$: $\fBs$
      \\[2mm]
      1  &  & 4 & & $A^\hqet_k$   
      \\[2mm]
      1  &  & 2 & & $V^\hqet_0$ & $\mathrm{B} \to \ell \nu\,\pi$: 
      form-factors $f_+,f_0$
      \\[2mm]
      1  &  & 4 & & $V^\hqet_k$ &  $\mathrm{B} \to \ell \nu\,\pi$: 
      form-factors $f_+,f_0$\\ 
\bottomrule
\end{tabular}
\caption{The number of free parameters in HQET at a given order, 
the specific fields where they appear (``origin'') as well
as some examples for applications where they contribute.
\label{t:hqetparam}
} 
\end{center}
\end{table*}

\section{Matching and mass scaling}
\label{match}
As simple as it is in principle, there are important issues to be
considered and to be understood when \eq{e:genmatch} is applied in practice.
\bin 
\item \label{it:hqet}
    Observables have to be chosen such that the accuracy of the 
    HQET expansion is not compromised: energies
    have to be sufficiently small. There is, however, 
    no point in going significantly below $\Lambda\approx 400\MeV$,
    since this scale is always present. 
\item \label{it:lat}
    The right hand side of \eq{e:genmatch} has to be computable in 
    lattice QCD, which means that $a \mh \ll 1$ has to be reachable.
\item \label{it:ambig}  
    When \eq{e:genmatch} is implemented non-perturbatively,
    it has to be imposed at a finite (the desired) quark mass.
    Since the left side is an approximation, 
    truncated at a given order,
    the effective theory
    parameters then do depend on the matching condition imposed. 
\item \label{it:prec} 
    The observables 
    $\Phi_i$  have to have a good statistical     
    precision in lattice computations, both in QCD and in HQET.
\ein

Clearly item \ref{it:lat}. 
is in conflict with \sect{s:difficulty}, where we explained that 
$a \mh \ll 1$ cannot be reached for large enough volumes where 
finite size effects are small. This is avoided by having a smaller
gap between infrared and ultraviolet cutoff, defining the observables
$\Phi_i$ in a small volume. Item \ref{it:hqet} suggests that 
the infrared momentum cutoff should be chosen around $\Lambda$,
namely $L = \rmO(1/\Lambda) = \rmO(\frac12\fm)$.

\subsection{One-loop perturbation theory}
Before coming to a general discussion, it is instructive to look at 
the simplest case of matching 
in perturbation theory. We consider the static effective theory. 
Its Lagrangian
\bes
 \lagh{stat} =   \psibarheavy(\mhbare+D_0)\psiheavy\,, 
\ees
contains a single parameter, $\mhbare$. In comparison to 
\eq{e:laghclass} we have added a label `bare' to indicate
the bare mass parameter. $\psiheavy$ etc are the bare fields
in the regularized path integral.
The explicit form of the heavy quark propagator, which we 
give later in \eq{e:stat_prop}, shows 
that $\mhbare$ drops out of all observables (at LO) except for the 
relation between the QCD quark mass and one energy level in
the static theory, say the mass of the B-meson. All energy
differences and all properly normalized\footnote{
A proper mass-independent non-relativistic normalization
has to be chosen. The standard one is $\langle  B(\vecp')| B(\vecp) \rangle 
= 2 (2\pi)^3 \delta(\vecp-\vecp')$. 
} matrix elements are independent of $\mhbare$. \label{footnotenorm}

Interesting, non-trivial, matching happens for composite fields.
We choose here the time-component of the axial current, 
\bes
  A_{0\mathrm{R}}(x) = \za A_0\,,\quad A_0=\psibar_\up(x) \gamma_5\gamma_0 \psi_\beauty(x)\,.
\ees
To distinguish it from the HQET field we label the 
heavy quark in QCD by $\beauty$.
The matrix element, 
\bes
    \label{e:fb}
    \langle 0| \opAR(\vecx=0)| B(\vecp=0) \rangle 
    =\mB^{1/2} \fB,
\ees
of the associated Hilbert space operator
defines the decay constant $\fB$, the only 
hadronic parameter determining the decay rate $B \to \ell \nu$. 
For matching, it is natural to consider more general matrix elements 
\bes
    \label{e:PhiA0}
    \cmqcd(L,\mbeauty) = \langle \Omega(L)| \opAR(\vecx=0)| B(L) \rangle,
\ees
to define a suitable quantity $\Phi_i$ ($i$ fixed) in \eq{e:genmatch}. 
In physical processes, $L$ is an
inverse momentum scale, but we will later use states 
in a finite periodic $L\times L \times L$ torus. 
The state $B(L)$ has the quantum numbers of a B-meson, while
$\Omega(L)$ has vacuum quantum numbers. 

In generic regularizations, e.g. in dimensional or the Wilson lattice one, the axial current is 
affected by a non-trivial renormalization 
$A_{\mu,\rm R} = \za A_\mu$. When the renormalization factor $\za$ 
is defined such that the current 
satisfies the chiral Ward identities \cite{curralgebra:MaMA,impr:pap4}, \eq{e:fb} gives correctly
the weak decay amplitude and thus $\fB$. The current is then also
scale independent.

For the moment 
 the relevant property of the states $|\Omega(L)\rangle$ and
$|B(L)\rangle$ is that $L$ is the only scale apart from
$\mbeauty$ and $\Lambda$.  Then, for sufficiently small
$L$, the relevant QCD coupling is small and 
there is a perturbative expansion
\bes
\cmqcd(L,\mbeauty) &=& \cm^{(0)}
            + \cmqcd^{(1)}(z) \, g^2 
            \\&& + \rmO(g^4) +\rmO(1/z)\,,\quad z=L\mbeauty\,. \nonumber
\ees
in terms of renormalized coupling and mass $g=\gbar,\mbeauty$.
We will specify their renormalization scheme and scale 
when it becomes relevant. 
For any finite $L$ and $\mbeauty$, the matrix elements are finite, 
but the large mass limit, $\mbeauty\to\infty$ with $L$ fixed
does not exist. It is logarithmically divergent \cite{Shifman:1987sm,Politzer:1988wp},
\bes
   \cmqcd^{(1)}(z) &\simas{\mbeauty\to\infty}& (- \gamma_0 \log(z) + \bqcd) \cm^{(0)} \,, \label{e:asym}
   \\ &&
   \gamma_0=-1/(4\pi^2)\,.  \label{e:gammazero}
\ees  
This behavior has to be reproduced by the effective theory. 
As a first step, the bare static-light current 
\bes
  \astat(x)=\psibar_\up(x) \gamma_5\gamma_0 \psi_\heavy(x)\,,
\ees
is just form-identical to the relativistic one. For the classical 
current this follows from the FTW transformation and beyond we just observe that there are no other dimension three (or lower) composite fields with the same quantum numbers. 

Unlike the relativistic
current, there are no chiral Ward identities which fix its 
renormalization. As a consequence the renormalized current
is scale dependent. For example in the lattice regularization we can 
renormalize it by lattice minimal subtraction,\footnote{
For comparison, the renormalization factor of the relativistic lattice current is 
\bes
  \za = 1 + \za^{(1)} g_0^2+ \ldots
\ees
with a pure number (no renormalization scale dependence) $\za^{(1)}$,
which can be chosen such that the chiral Ward identities hold.
}
\bes
  \astatlat(x;\mu)&=& \za^\mathrm{stat,lat}(\mu a, g_0)\, \astat(x)\,,
  \\
  \za^\mathrm{stat,lat} &=& 1 - \gamma_0 \log(a\mu)\, g_0^2 + \rmO(g_0^4)\,. 
  \label{e:zastatoneloop}
\ees
Here $a$ is the lattice spacing, $g_0$ is the bare 
coupling and $\mu$ is the renormalization scale.

The lowest order anomalous dimension $\gamma_0$ coincides with 
$\gamma_0$ from the mass-scaling  
defined above.
The matrix elements 
\bes \label{e:Mstat}
    \cmstat^\mathrm{lat}(L,\mu) = 
    \langle \Omega(L)| \za^\mathrm{stat,lat}\opAstat| B(L) \rangle,
\ees
corresponding to the above 
QCD ones have an expansion in the renormalized
coupling,
\bes
\nspace\cmstat^\mathrm{lat}(L,\mu) &=& \cm^{(0)}
            + \cmstat^{(1)}(\mu L) \, g^2 
            + \rmO(g^4)\,, 
            \\
\nspace\cmstat^{(1)}(\mu L) & = & (- \gamma_0 \log(\mu L) + \blat) \cm^{(0)} \,. 
\ees
For convenience we put the asymptotic QCD expression and the static one
together (up to $\rmO(g^4,1/z)$),
\bes
\cmqcd &=& (1 +(- \gamma_0 \log(\mbeauty L) + \bqcd)\,g^2)\cm^{(0)}\,,
\nonumber
\\
\cmstat^\mathrm{lat} &=& (1 +(- \gamma_0 \log(\mu  L) + \blat)\,g^2)\cm^{(0)} \,. 
\label{e:Mlat}
\ees
In this way one sees immediately that a finite renormalization 
of the static current 
\bes
\nspace  \astatmatch(x;\mbeauty)&=& \tilde C_\mathrm{match}(\mbeauty,\mu)\, \astatlat(x;\mu)\,,
  \\
  \tilde C_\mathrm{match}&=& 1 + c_1(\mu/\mbeauty)\, g^2 + \rmO(g^4)\,,
  \\
  c_1(\mu/\mbeauty) &=& \gamma_0 \, \log(\mu/\mbeauty)
      + (\bqcd -\blat) \,,
      \nonumber
\ees
brings QCD and the static effective theory into agreement,
\bes
  \label{e:currmatch}
  \cmqcd &=& \tilde C_\mathrm{match}\, \cmstat^\mathrm{lat} +\rmO(1/z)
  \\ &=& 
  \langle \Omega(L)| \opAmatch| B(L) \rangle_\mathrm{stat} +\rmO(1/z)\,.
  \nonumber
\ees
We emphasise the general structure and the important 
features of the example.
\bi 
\item The coefficient of the logarithm in \eq{e:asym}
and the anomalous dimension of the current in the static theory, \eq{e:zastatoneloop}, match. This matching cannot be enforced,
it is a property of the two theories.
And it is one of the conditions for the effective theory to
describe the asymptotics of QCD.
\item 
The relative one-loop coefficient \cite{BorrPitt,zastat:pap2},
\bes
   \bqcd -\blat &=& -0.137(1)\,, \label{e:Bastat}
\ees
is independent of the external states. They were chosen from
two rather different classes in the two cited references. 
Again this is a necessary condition for the effective theory
to describe QCD.
\item
Both $\cm^{(0)}$ and $\bqcd$ do depend on the external states. 
Let us label a matrix element for a different pair of states
by just a prime. In ratios of these matrix elements,
\bes
  {\cal R}_\mathrm{QCD} = \cmqcd'/\cmqcd \,,
\ees
the entire renormalization and matching of the 
current drops out, since it is multiplicative. One then has 
effective theory predictions
\bes
   {\cal R}_\mathrm{QCD} &=& {\cal R}_\mathrm{stat} +\rmO(1/z)
   \\
   {\cal R}_\mathrm{stat}&=& {\cal R}_\mathrm{stat}^{(0)}
   + {\cal R}_\mathrm{stat}^{(1)}\,g^2 +\rmO(g^4)\,,
   \\
   {\cal R}_\mathrm{stat}^{(0)}&=&  {(\cmstat')^{(0)} \,/\,\cmstat^{(0)}}\,,
   \\
   {\cal R}_\mathrm{stat}^{(1)}&=& 
     {(\cmstat')^{(1)} \over (\cmstat')^{(0)}} -
       {(\cmstat)^{(1)} \over (\cmstat)^{(0)}} \nonumber
       \\ &=&  \blat' - \blat =\bqcd'-\bqcd\,.
\ees
 
\item
The particular number of \eq{e:Bastat} depends on the 
renormalization scheme for the static current.
Here we chose minimal subtraction, a scheme which is not independent of
the regularization. Therefore, $\bqcd -\blat$ depends on the details of the regularization chosen in \cite{BorrPitt,zastat:pap2}.
It is valid for the $\rmO(a)$ improved Wilson lattice regularization. 
\item 
Of course the matrix elements of the matched static current in \eq{e:currmatch} do not depend on any details of the regularization. Their finite renormalization has been 
chosen to match QCD. This is unique. In \eq{e:currmatch}
we did not indicate the one-loop nature. Indeed, we expect this equation
to hold to all orders in the coupling and
also beyond, non-perturbatively.
\item
We have nowhere given the renormalization scale/scheme for
coupling and mass. At the one-loop order all expressions are 
independent of
it. The scheme only matters for the renormalization
factor of the current itself. 
In the following section we will also discuss a convenient choice of renormalization scales.
\ei

\subsection{Higher orders in the coupling and renormalization group invariants}
In this section we explain on the one hand what is known 
at higher orders in the coupling and on the other hand, how 
one passes to renormalization group invariants, which are independent
of schemes and scales. In particular they allow for a clean
factorization of observables into a non-perturbative matrix element 
and a multiplicative matching function, 
which has a perturbative expansion. 
This separation makes efficient use of the high order perturbative information accumulated over the years 
\cite{Shifman:1987sm,Politzer:1988wp,MS:4loop1,MS:4loop2,MS:4loop3,Czakon:2004bu,BroadhGrozin,ChetGrozin,Ji:1991pr,BroadhGrozin2,Gimenez:1992bf}.
We note, however, that we only know how to apply this strategy 
to the lowest order, static, effective theory. A hurried reader
may therefore skip this section and proceed to the following
one.

\subsubsection{RG functions and invariants}
It is well known, that fixed order perturbation theory,
where all observables are expressed in terms of a 
renormalized coupling and masses at one fixed renormalization scale
is not the best choice. In fact, if scales rather different 
from the renormalization scale are relevant in the observables, 
large logarithms multiplying powers of the coupling are
present and the accuracy of perturbation theory is not good.
This is a  reason to consider running coupling and mass,
i.e. coupling and mass as functions of the renormalization scale,
\bes
    \gbar(\mu)\,,\; \mbar_i(\mu)\,,
\ees
and the associated renormalization 
group equations (RGE).
Here $i$ runs over the different flavors and all masses are defined
in QCD, also the mass of the 
quark treated by HQET.  A 
consequent application of the RGEs is to use their solutions
and express all observables in terms of renormalization 
group invariants. 

Here we do not give an introduction to the RG
but just recommend Ref.~\cite{LH:peter}. We mainly 
describe the relevant formulae in order to apply them
to our matching problem.
We work in an unspecified massless
renormalization scheme, where the renormalization factors do not depend on the masses.
Consequently the renormalization group functions do not depend
on the masses.
Examples of massless schemes are (modified) minimal subtraction in 
dimensional regularization ($\ms, \msbar$) or lattice 
regularization (lat) or a Schr\"odinger functional scheme (SF)~\cite{Luscher:1992an,Capitani:1997xj}. The latter is independent
of the regularization.

Our renormalization group (RG) functions are defined through 
\bes
  &\mu {\partial \bar g \over \partial \mu} = \beta(\bar g) \enspace ,
     \quad 
  {\mu \over \mbar_i} {\partial \mbar_i \over \partial \mu} \;=\; \tau(\bar g) \enspace ,
     \label{e_RG_m}
  \\
  &{\mu \over  \cmstat^{s} } {\partial \cmstat^{s}  \over \partial\mu} = \gamma(\gbar) \,. \label{e_RG_Phione}
\ees
Apart from the running coupling and running quark mass we
here consider the matrix element $\cmstat^{s}$ of a (multiplicatively renormalizable) composite field renormalized at scale $\mu$. We label it with a superscript
$s$ for the scheme to remain consistent with previous notation. 
One may identify it with \eq{e:Mstat}, but other matrix elements
of general composite operators are possible as well.
The RG functions have asymptotic expansions
\bes
 \beta(\bar g) & \buildrel {\bar g}\rightarrow0\over\sim &
 -{\bar g}^3 \left\{ b_0 + {\bar g}^{2}  b_1 + \ldots \right\}
                      \enspace ,  \label{e_RGpert} \\ \nonumber
 &&b_0=\frac{1}{(4\pi)^2}\bigl(11-\frac{2}{3}\nf\bigr)
  \\ &&                 
   b_1=\frac{1}{(4\pi)^4}\bigl(102-\frac{38}{3}\nf\bigr) \enspace ,
 \\
 \tau(\bar g) & \buildrel {\bar g}\rightarrow0\over\sim &
 -{\bar g}^2 \left\{ d_0 + {\bar g}^{2}  d_1 + \ldots \right\}
                      \, , 
\\ &&  d_0={8}/{(4\pi)^2}
 \enspace ,  \label{e_RGpert_m}
 \\
  \gamma(\gbar)& \buildrel {\bar g}\rightarrow0\over\sim &
  - \gbar^2 \left\{ \gamma_0 + {\bar g}^{2}  \gamma_1 + \ldots \right\}\,. 
  \\ \nonumber
  \label{e_RGpert_Phi}
\ees

The integration constants of the solutions to the 
RGE define the RG invariants

\bew
 \Lambda &= \varphi_\mathrm{g}(\gbar)\, \mu \;=\; \mu \left(b_0\gbar^2\right)^{-b_1/(2b_0^2)} \rme^{-1/(2b_0\gbar^2)}
           \exp \left\{-\int_0^{\gbar} \rmd x
          \left[\frac{1}{ \beta(x)}+\frac{1}{b_0x^3}-\frac{b_1}{b_0^2x}
          \right]
          \right\} \enspace , \label{e:lammu} 
  \\
  M_i &= \varphi_\mathrm{m}(\gbar)\, \mbar_i \;=\; \mbar_i\,(2 b_0\gbar^2)^{-d_0/2b_0}
   \exp \left\{- \int_0^{\gbar} \rmd x \left[{\tau(x) \over \beta(x)}
     - {d_0 \over b_0 g} \right] \right\}  \enspace ,
  \\
  \cmRGI &= \varphi_\mathrm{stat}(\gbar)\, \cmstat^s \;=\; \cmstat^s \left[\,2b_0 \gbar^2\,\right]^{-\gamma_0/2b_0}
                   \exp\left\{-\int_0^{\gbar} \rmd x
                     \left[\,{ \gamma(x) \over\beta(x)}
                           -{\gamma_0 \over b_0 x}\,\right]
                     \right\} \, ,
\eew
\noindent
where $\gbar\equiv\gbar(\mu)$, $\mbar_i\equiv\mbar_i(\mu)$,  $\cmstat^s\equiv\cmstat^s(\mu)$. There are no corrections to eqs.~(\ref{e:lammu}) with 
exact RG functions.  Instead, when we only have access to their perturbative 
expansions such as \eq{e_RGpert} there are truncation
errors in the functions $\varphi_r(\gbar)$. For example, in order to
obtain a numerical
result for $\Lambda$ given a pair of numbers for $\gbar(\mu),\,\mu$, 
the integrand $[\,...\,]$ in \eq{e:lammu} may in principle
be expanded as $[\,...\,] = c_2 x + c_3 x^3 +\rmO(x^5)$. Knowing
$c_2$, the error made is (asymptotically) $\approx c_3\gbar^4/4$ and
the same is true when we insert the truncation of $\beta(x)$ and then
integrate. When we
discuss numbers, we will always expand the RG functions,
not the full integrands. 
 
Physical observables can be written as 
\bes
  {\cal P} = {\cal P}(\Lambda,\{M_i\}, \{p_i\})
\ees
making it manifest that they do not depend on a renormalization scale
$\mu$,
\bes
    \mu \frac{\rmd}{\rmd\mu} {\cal P}  = 0\,.
\ees
In the following section we express the lowest order HQET approximation
of matrix elements $\cmqcd$ in this way and define a RGI mass scaling function, which describes the mass-dependence beyond
the asymptotic form \eq{e:massscal1}. 

\subsubsection{Mass scaling}
As a first step we choose a suitable renormalization scale, 
\bes
  \mu = \mstar \,,
\ees
where  the solution of 
\bes
  \mstar = \mbar(\mstar)\,,  \label{e:mstar}
\ees
defines $\mstar$. We further use the shorthand
\bes
  \gstar=\gbar(\mstar)\,.
  \label{e:gstar}
\ees
The coupling $\gstar$ can be determined for any value of 
$M/\Lambda$ by combining the first two equations \eq{e:lammu}  to 
\bes
{\Lambda \over M} &=&  {\varphi_\mathrm{g}(g_*) \over \varphi_\mathrm{m}(g_*) }  \nonumber
\\ & = & \exp \left\{-\int^{\gstar(M/\Lambda)} \rmd x\;
          \frac{1-\tau(x)}{ \beta(x)}          \right\} \,. 
  \label{e:gstarM} 
\ees
The solution of this equation defines a function  $\gstar(M/\Lambda)$.

With the above choices the matching function simplifies to
\bes
  \tilde C_\mrm{match}(\mstar,\mstar) &\equiv&
  C_\mrm{match}(\gstar)
  \\ &=& 1 + c_1(1)\,\gstar^2+\ldots\,,
\ees
and we have
\bes 
   \cmqcd &=& C_\mrm{match}(\gstar) \times\cmstat(\mu) 
   +\rmO(\mhinv)\nonumber 
   \\ &=& \frac{C_\mrm{match}(\gstar)}{ \varphi_\mathrm{stat}(\gstar)}\,\cmRGI +\rmO(\mhinv)\,. 
\ees
The $\gstar$-dependence is equivalently to the mass dependence 
and defines another RG function $\gamma_\mrm{match}$,
\bes 
  \gamma_\mrm{match}(\gstar)  \equiv \left.{\mstar \over  \cmqcd} {\partial \cmqcd \over \partial \mstar}\right|_\Lambda  
  \label{e:gammamatch}\,,
\ees
with asymptotics
\bes
  \gamma_\mrm{match}(g) \simas{g\to0} -\gamma_0 g^2 - 
   \gamma_1^\mrm{match}g^4 + \ldots\,.
\ees
Because the leading order coefficient $\gamma_0$ is 
unchanged, ``match'' is just another renormalization scheme
for the axial current $\astat$. In this scheme, at scale $\mu=\mstar$, its matrix elements are equal to the QCD ones up to order $\mhinv$. We therefore refer to it as the matching scheme.

Finally, a complete transition to renormalization group invariants
is achieved by changing to the RGI mass scaling function
\bes
   \label{e:Mdep}
   \hspace*{-5mm} \rho_\mathrm{PS}(M/\Lambda) &\equiv& \left.{M \over  \cmstat} {\partial \cmstat \over \partial M}\right|_\Lambda = 
   \left.{M \over C_\mrm{PS} } {\partial C_\mrm{PS} \over \partial M}
   \right|_\Lambda
   \\
   &=& { \gamma_\mrm{match}(\gstar)\over 1-\tau(\gstar)}\,,
          \nonumber
\ees
and to
\bes
     \cmqcd &=& C_\mrm{PS}(M/\Lambda) \times\cmRGI 
      +\rmO(\mhinv)\,  \nonumber
\\
  C_\mrm{PS}(M/\Lambda) &=& 
  C_\mrm{match}(\gstar) \,/\, \varphi_\mathrm{stat}(\gstar)
  \nonumber \\ 
  &=& \exp \left\{\int^{\gstar} \rmd x\; 
        \frac{\gamma_\mrm{match}(x)}{\beta(x)}\right\} \,.
        \label{e:cpsrgi}
\ees
Everywhere it is understood that $\gstar=\gstar(M/\Lambda)$.
In the last equation, the integral at the lower bound is to be understood as 
in
\eq{e:lammu}.

At leading order in $\mhinv$ the conversion function
$\Cps$ contains the full (logarithmic) mass-dependence, while
the non-perturbative effective theory matrix elements, $\cmRGI$,
are mass independent numbers. 

An interesting application is the asymptotics of the decay constant
of a heavy-light pseudo-scalar (e.g. B):\footnote{Note the slow, logarithmic, 
decrease of the corrections in \eq{e:f_asymptotoics}. We will see below,
in the discussion of \fig{f:gammamatch}, that the perturbative
evaluation of $\Cps(\Mbeauty/\Lambda)$ is somewhat problematic.}
\bes
  \label{e:f_asymptotoics}
  F_\mrm{PS} &\simas{M\to\infty}& 
  { [\ln(M/\Lambda)]^{\gamma_0/2b_0} \over \sqrt{m_\mrm{PS}}} \cmRGI\ \\ &&
  \times[1+\rmO([\ln(M/\Lambda)]^{-1})]\,. \nonumber
\ees
In perturbation theory, the logarithmic corrections
are computed by solving \eq{e:gstarM} for $\gstar$ and
then integrating \eq{e:cpsrgi}.

\subsection{On the accuracy of perturbation theory}

\begin{figure*}[ht!]
  \centering
  \includegraphics[width=0.41\textwidth]{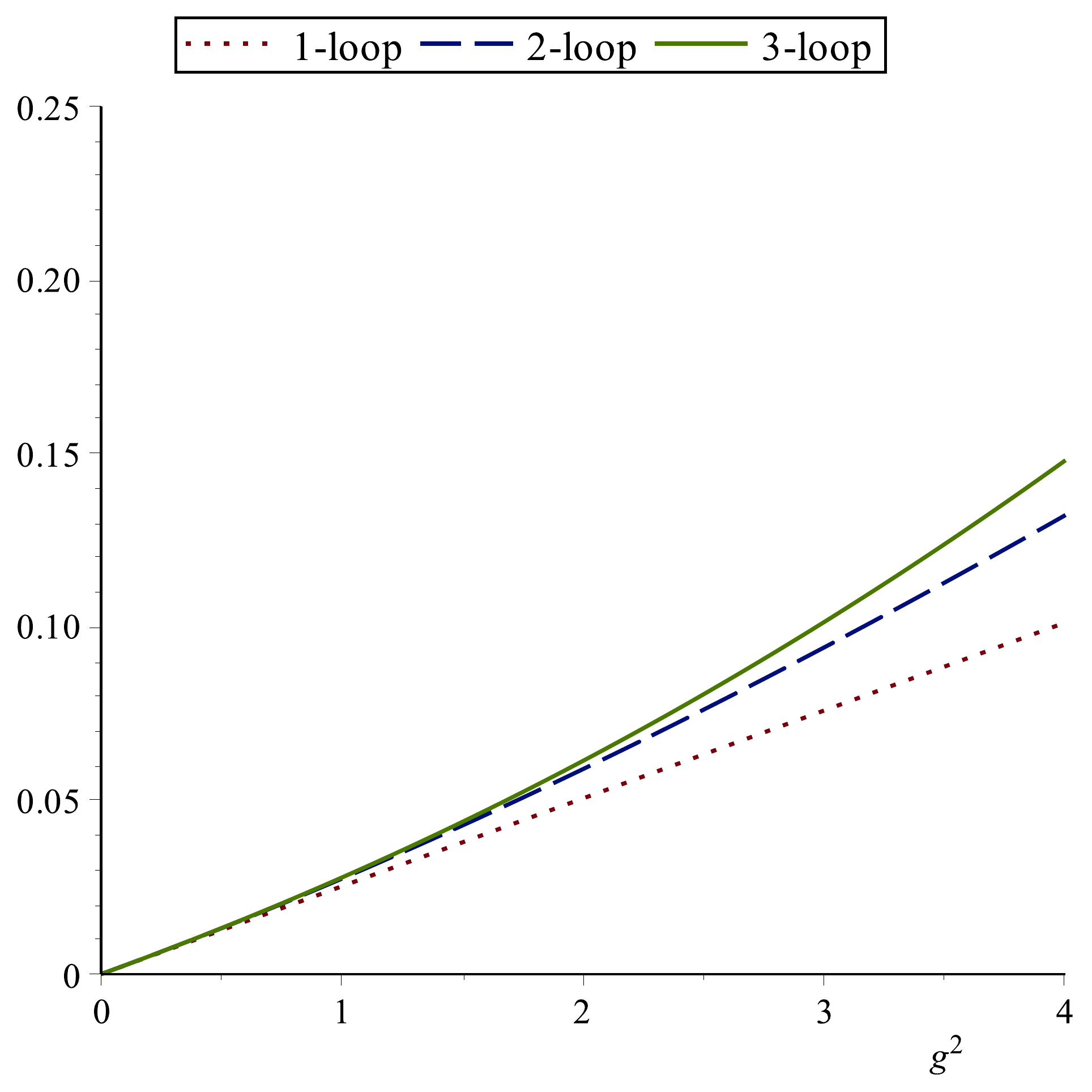}
  \hspace*{4mm}
  \includegraphics[width=0.41\textwidth]{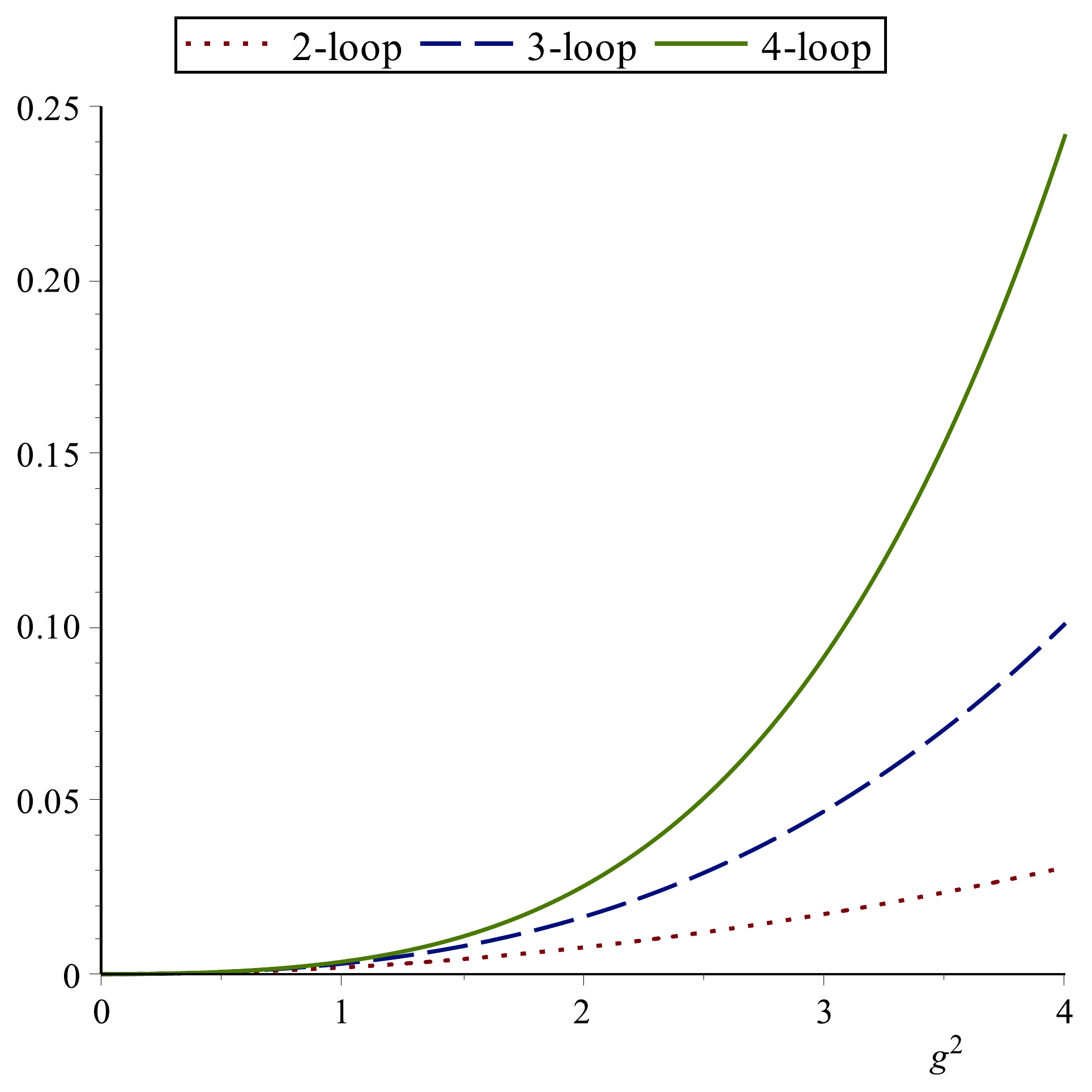}
    \caption{The function $\gamma_\mathrm{match}(g)$ as a function
    of $g^2=\gstar^2$ for $\nf=3$ flavors in the $\msbar$ scheme. 
    On the left we
    show $\gamma_\mathrm{match}=\gamma_\mathrm{match}^{A_0}$ for the 
    time component of the 
    axial current, on the right we show the difference 
    $\gamma_\mathrm{match}^{V_k}-\gamma_\mathrm{match}^{A_0}$. 
    Note that at  1-loop order the latter vanishes.
}
\label{f:gammamatch}
\end{figure*}

When one evaluates functions such as $\Cps$ in a given order
of perturbation theory, various quantities enter such as 
the beta-function, the quark mass anomalous dimension.
Apart from $\gamma_\mathrm{match}$, these all have a well behaved perturbative expansion in the $\msbar$ 
scheme, see appendix A.2.2 of \cite{LH:rainer} for a table
of the coefficients. Keeping this in mind, we just discuss $\gamma_\mathrm{match}$.
In the left graph in \fig{f:gammamatch} we plot different orders
of $\gamma_\mathrm{match}$. For the larger values $\gstar^2$ 
in the plot one may get worried about neglecting higher order terms. 
Note that $\gstar^2$ is around 2.5 for the b-quark and it
is out of the range of the graph for the charm quark.

However, a more serious reason for concern derives from the 
right hand side graph. There the difference of the anomalous dimensions for $V_k$ and $A_0$ is shown. For such differences perturbation theory 
is known to one loop higher \cite{Bekavac:2009zc} and the perturbative
coefficients do grow further. Asymptotic convergence seems
to be useful only for rather small couplings or masses far above
the b-quark mass. At the b-quark mass every known perturbative 
order contributes about an equal amount.
Since we do not understand the reason for this 
behavior, it raises concern about using perturbation theory for
the matching functions.

Let us emphasize, that the bad behavior is easily traced back to 
the function $C_\mathrm{match}$ and was noted in \cite{Bekavac:2009zc}.
We tried earlier \cite{LH:rainer} to rearrange the perturbative 
series in order to find a more stable perturbative prediction, 
but we did not succeed.

\subsection{Matching at NLO in $\mhinv$}

For quantitative phenomenological results one has to 
compute also $\mhinv$ corrections in HQET. Is it consistent to 
match perturbatively as we discussed in the previous sections? 
We saw that the uncertainty due to a truncation of the perturbative
matching expressions at $l$-loop order corresponds to
a relative {\em error}
\bes    
     {\Delta(\Cps) \over \Cps} &=& {\Delta(\cmqcd) \over \cmqcd} \;\propto \; [\gbar^2(\mstar)]^{\,l} \\
     &\sim&
        \left[{1 \over 2b_0\ln(\mstar/\Lambda_\mrm{QCD})}\right]^{\,l}
        \nonumber
\,.
\ees
As $\mstar$ is made large, this perturbative error decreases only
logarithmically. It becomes
dominant over the power correction which one wants to
include by pushing the HQET expansion to NLO,
\bes    
     {\Delta(\Cps) \over \Cps}  \ggas{\mstar \gg \Lambda}\; {\Lambda \over \mstar} \,.
    \label{e:deltacps}
\ees
With a perturbative matching function, one does not 
perform a consistent NLO expansion such that errors decrease as
$\mhinv^2$. 

A practically even more serious issue is 
that at NLO one has to deal with the mixing of 
operators with lower dimensional ones. For example $\Okin=\psibar_\heavy \vecD^2 \psi_\heavy$ mixes with $\psibar_\heavy D_0 \psi_\heavy$ and
$\psibar_\heavy \psi_\heavy$. In this situation
mixing coefficients are power divergent $\sim a^{-n}$.
In the example we have $n=1,2$. 
Subtracting power divergences in perturbation theory 
and then computing the matrix elements non-perturbatively 
always leaves a divergent remainder. The non-perturbative continuum 
limit of matrix elements of perturbatively subtracted 
operators does not exist.

We are lead to conclude that it is necessary to perform matching
and renormalization non-perturbatively. The only alternative is to 
supplement the theory by assumptions. Namely one may assume
that at the lattice spacings available in practice, power divergences
of the form $g_0^{2l} /(\mbeauty a)^{n}$ are small since $\mbeauty a > 1$.
This then has to be combined with 
the assumption that the b-quark is not large enough to be 
in the asymptotic region of \eq{e:deltacps}.

\subsection{Non-perturbative matching}
\label{s:strat} 
\subsubsection{Scope}
A true non-perturbative matching 
as in  \eq{e:genmatch} eliminates the problems
we just discussed: slow asymptotic convergence of PT, 
power divergent remainder terms in the subtraction of lower 
dimensional operators.  We do, however, still have 
to cope with the condition $\mbeauty a \ll 1 $ for the 
computation of $\Phi_i^\qcd$.
Before we explain the strategy to deal with this, 
we would like to emphasize a point about the magnitude 
of $\mhinv$ corrections which sometimes leads to confusions.

The size of these corrections does depend on the matching 
conditions used: after imposing a generic condition, \eq{e:genmatch}, the decay constant of the B-meson once computed 
at LO and once 
computed at NLO in $\mhinv$ differ by an amount $\rmO(\mhinv)$.
However, how much this difference is exactly,
depends on which set $\{\Phi_i\}$ was chosen
for matching the theories.
We are free to choose 
$\Phi_1 = \mB\,,\quad \Phi_2 = \langle 0| \opAR| B(\vecp=0 )\rangle$.
With this choice, the decay constant of the B-meson is exact at 
all orders in $\mhinv$. 

This rather trivial fact has to be remembered. The exact size
of the corrections is only defined, once one has fixed how the matching
is performed. What we said before implies that one has to define
non-perturbatively how the matching is performed. Only then
does the splitting of a HQET result into different orders 
acquire a precise meaning. 

\newcommand{\mBstar}{m_\mrm{B^*}}
\newcommand{\mbav}{m_{\rm B}^\mrm{av}}
\newcommand{\mbsplitt}{\Delta m_{\rm B}}

Let us illustrate the consequences on a frequently discussed
example, namely  the mass formulae
\bes
     {\mbav} &\equiv& {1\over 4} [\mB+3\mBstar] \\ &=& \mbeauty + {\bar \Lambda}
    + {1\over 2\mbeauty} \lambda_1 +\rmO(1/\mbeauty^2) \label{e:massformula}
  \\
     {\mbsplitt} &\equiv& \mBstar-\mB =  -{2\over \mbeauty} \lambda_2
     +\rmO(1/\mbeauty^2) 
\ees
with (ignoring renormalization)
\bes
   \lambda_1 = \langle B | \Okin | B \rangle \,,\quad 
   \lambda_2 = \frac13 \langle B | \Ospin | B \rangle \,.
\ees
The quantity ${\bar \Lambda}$ is referred to as ``static binding energy''
and $\lambda_1$ as the kinetic energy of the b-quark 
inside the B-meson. 
Also here, depending on how one formulates the matching conditions, 
one changes ${\bar \Lambda}$ by a term  
of order $\Lambda_\mrm{QCD}$. 
Similarly, the kinetic term $\lambda_1/ (2\mbeauty)$ has a non-perturbative 
matching scheme dependence of order 
$\Lambda_\mrm{QCD}^2/\mbeauty$ and thus 
$\lambda_1$ itself 
has a matching scheme dependence of order $\Lambda_\mrm{QCD}^2$. 
In fact we could set $\Phi_1=\mbav$. At static order 
this means $\mbeauty+\bar\Lambda=\mbav$ and at NLO it
is \eq{e:massformula} without the  $\rmO(1/\mbeauty^2)$ correction.
If both are to be valid, one has $\lambda_1=0$.

For this reason, our scope is not to compute (and therefore
first define) quantities such as $\bar \Lambda$ but to compute
physical observables such that they are correct up to 
corrections of order $\mhinv^2$. For this purpose we need to 
determine the bare parameters $\omega_i$. We do not need to define
renormalized ones. We ask the reader to keep in mind that since 
$\omega_i$ are bare parameters, they depend on the bare 
gauge coupling (equivalently the lattice spacing) and the 
heavy quark mass, where we may choose the RGI mass, $M$. 
Further dependences on the details of the discretisation are 
kept implicit.

\begin{figure*}[t!]
  \centering
  \includegraphics[width=0.76\textwidth]{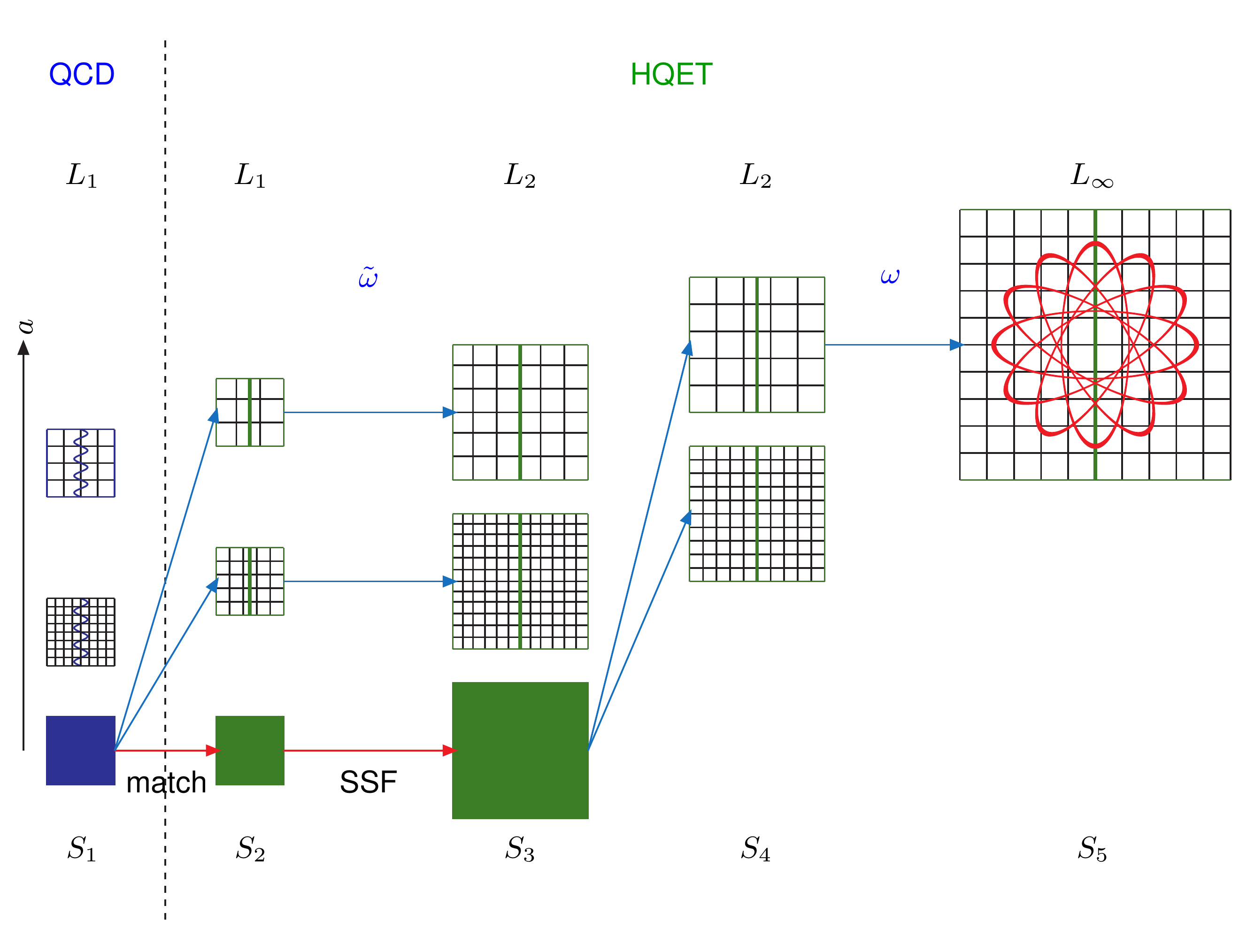}
    \caption{The ALPHA collaboration strategy for matching of QCD     
    and  HQET. The bottom row denotes the continuum limit 
    obtained by an extrapolation of the results in each column.
    Graph from \cite{Blossier:2012qu}.
}
\label{f:strat}
\end{figure*}

\subsubsection{Strategy}
\def\phistat{\eta}
\def\phimat{\varphi}

The general idea how to satisfy $\mbeauty a \ll 1 $ is to perform the 
matching step in a finite space-time volume of 
linear dimension $L=L_1$ \cite{Heitger:2003nj}, exactly in line with the
idea how to cover the scale hierarchy in the computation of running 
couplings~\cite{Luscher:1991wu}.
Choosing $L_1^{-1} \approx \Lambda$, the HQET expansion of correlation 
functions at distances $|x|=\rmO(L_1)$ is valid. 
$L_1\approx 0.5\,\fm$ is a 
reasonable choice. This allows for lattice spacings of $a = 0.02\,\fm$
and significantly smaller. One can then perform continuum extrapolations with several points satisfying $\mbeauty a < 1/2$ and determine
\bes
\Phi_i^{\rm QCD}(L_1,M,0) = \lim_{a\to0} \Phi_i^{\rm QCD}(L_1,M,a)\,,
\ees
as indicated on the left side of \fig{f:strat}.
One is now free to 
use the matching condition, \eq{e:genmatch}, to determine the $\omega_i$ at a given resolution
$L_1/a$. Explicitly, we write the HQET expansion in a matrix notation
\bes
\label{e:phihqet}
 \Phi^{\rm HQET} = \phistat + \phimat\,\omega\,, 
\ees
with a $N_\omega \times N_\omega$ matrix $\phimat$ and a 
homogeneous static part $\phistat_i, \; i=1,\ldots,N_\omega$. 
Both $\phistat$ and $\phimat$ can be computed by
numerical simulations of HQET. 
Solving \eq{e:phihqet} defines a first set of HQET parameters
\bes
\tilde\omega(M,a) &=&
\\&& \nspace\nspace  \nonumber
\phimat^{-1}(L_1,a) 
\Big(
\Phi^{\rm QCD}(L_1,M,0) -  \phistat(L_1,a)
\Big)\,.
\ees
For reasonable resolutions, say $L/a\geq 8$, where discretisation
errors may be assumed to be small, 
we have $a\leq L_1/8 \approx 0.06\,\fm$. Remembering that one should always have a 
range of lattice spacings in order to carry out a continuum extrapolation, 
we realize that such lattice spacings are still too small 
to perform the large volume HQET computations to determine
the physical energy levels and matrix elements. 

For this reason,
the full strategy includes a so called step scaling step, to reach
the same $\Phi^{\rm HQET}$ but for $L=L_2=2L_1$:
\bes
\label{eq:Phi_L2}
  &&\nspace\Phi^{\rm HQET}(L_2,M,0)
 \\
&&= \lim_{a\to0}
\Big[
 \phistat(L_2,a) + \phimat(L_2,a)\, \tilde\omega(M,a)
\Big]\,. \nonumber
\ees
This set of observables then serves to determine the 
desired parameters
\bes \label{e:omegafin}
&& \nspace \omega(M,a) =  \\ &&
\phimat^{-1}(L_2,a) 
\Big(
\Phi^{\rm HQET}(L_2,M,0) -  \phistat(L_2,a)
\Big)\;. \nonumber 
\ees
Here lattice spacings can be used which are suitable for large 
volume HQET computations. The overall strategy is depicted in 
\fig{f:strat}. It is explained in more detail in \cite{Blossier:2012qu}.

\subsubsection{The electroweak heavy-light currents}
For the phenomenological treatment of (semi-) leptonic B-decays in the 
standard model, one needs both the axial current and the vector current.
 Our discussion above has to be generalized
accordingly.  The finite renormalization (for matching) of
space and time-components is different, but in a minimal
subtraction scheme, all currents have the same anomalous dimension in the static effective theory because they are
related by spin symmetry and the chiral symmetry which emerges 
when the light quarks are massless. Details
can be found in \cite{LH:rainer}. 

Non-perturbative matching conditions for the full set of currents have been discussed in 
\cite{DellaMorte:2013ega}. They have partially been 
investigated in perturbation theory  
\cite{Hesse:2012hb,Korcyl:2013ara,Korcyl:2013ega}, see \cite{sfb:C1b} for 
more details.

\section{Discretized HQET}
\label{s:hqetlat}

\newcommand{\mhbarehat}{\hat\mhbare}
\newcommand{\Ptrans}{\mathcal{P}}
We need a lattice formulation in order to treat
HQET beyond perturbation theory in the QCD coupling. 
The static action discretized
on a hyper-cubic lattice is
\bes
   \label{e:llstat}
   \nspace \lagh{stat} &=& {1 \over 1+a\mhbare} \,\heavyb(x) [\nabstar{0}+\mhbare]
   \psiheavy(x)\,,
\ees
with $\nabstar{\mu}$ the gauge covariant backward derivative. 
Compared to the form written down first by Eichten and Hill \cite{Eichten:1989zv} we just added the mass term.
The static propagator in a gauge background is
\bes
G_{\rm h}(x,y) &=& \theta(x_0-y_0) \;
 \delta({\vecx-\vecy}) \; \nonumber
 \rme^{- \mhbarehat\,(x_0-y_0)}\; \\ && \times \;
\Ptrans(y,x;0)^{\dagger} \; P_+ \; ,
 \label{e:stat_prop}
\\ \mbox{with} && \mhbarehat=\frac1a \ln(1+a\mhbare)\,,
 \label{e:dmstathat}
\ees
where $\Ptrans(x,y;0)$ parallel transports fields in the fundamental 
representation from $y$ to $x$ along a time-like path and
the lattice  $\theta$-function is  
$ \theta(t)=0$ for $t<0$ and $ \theta(t)=1$ for $t\geq0$.
It is a simple exercise to obtain \eq{e:stat_prop} from the
defining equation 
$${1 \over 1+a\mhbare}(\nabstar{0}+\mhbare) G_{\rm h}(x,y) = \delta(x-y)\,,$$
where $\nabstar{0}$ is the covariant derivative with respect 
to the argument $x$.
The propagator \eq{e:stat_prop} is valid in any gauge field 
background and therefore shows that all 
correlation functions with a static quark have a mass-dependence
$C(x) = C(x)|_{\mhbare=0} \times \rme^{- \mhbarehat\,|x_0|}$. 
Equivalently we see 
that the mass $\mhbare$ in the Lagrangian
just yields an energy shift
\bes
  \label{e:energyshift}
      E^\mrm{QCD} &=&
      \left.E^\mrm{stat}\right|_{\mhbare=0}
      +\mhbarehat\,.
\ees

We now briefly discuss the most important features of the discretized static effective theory. 
The action is (non-perturbatively) $\rmO(a)$ improved 
without adding any counter-terms and therefore without
tuning any parameters. This is a consequence of the local
conservation of the b-quark number and the spin symmetry,
which are exactly preserved in the formulation 
\eq{e:llstat} on the lattice~\cite{Kurth:2000ki}. 
These symmetries do not allow for dimension five fields (containing
$\psiheavy,\psibarheavy$) in
Symanzik's effective Lagrangian describing the discretisation
errors. 
As a consequence
energies scale to the continuum with lattice spacing errors proportional to 
$a^2 \log(a\Lambda)^\eta$ at small lattice spacing\footnote{See \cite{Balog:2009np,Balog:2013vua} for a discussion of the logarithmic modifications.}.
Improvement terms of the form $a \delta {\cal O},\; [\delta {\cal O}]=4$  are, however, necessary for fields ${\cal O}$ such as
${\cal O}=\astat$. For the matrix element determining $\fb$,
there is a single such counter-term~\cite{Kurth:2000ki}. 
After including it with a properly chosen
coefficient, one again has $\rmO(a^2)$ scaling to the 
continuum limit. 

The above features are very attractive. 
There is, however,  a (numerical, Monte Carlo) issue with the static theory, 
\eq{e:llstat}. It has its origin in the fact that 
\bes \label{e:r1}
 \left.E^\mrm{stat}\right|_{\mhbare=0} &\sim & \left({{1}\over{a}} r^{(1)} + {\rm O}(a^0)\right) \; g_0^2+ \rmO(g_0^4) \,,
\ees
has a linear divergence (which is then cancelled by 
$\mhbare \sim - g_0^2  r^{(1)} / a$ in the physical energies).
The numerical value is
\bes
   r^{(1)} = 0.1685 \quad \mbox{for the action \eq{e:llstat}}.
\ees
The problem is not -- as originally was thought  -- 
to determine $\mhbare$ non-perturbatively. Rather
the statistical errors of, for example, 
two-point correlation fuctions of a static B-meson with
$\mhbare=0$ behave 
approximately as~\footnote{Remember that we can set 
$\mhbare=0$ due to \eq{e:energyshift}.} 
 \cite{Lepage:1991ui,LH:martin}
\bes
   \mbox{stat. error} \simas{x_0\to\infty} {\cal A_\mathrm{N}}\rme^{-\mpi x_0 / 2}
\ees
while (at least at small $a$) the correlation function
scales as
\bes
   \mbox{correlator} \simas{x_0\to\infty} 
   {\cal A_\mathrm{S}} \rme^{- E^\mrm{stat}x_0} 
   \sim {\cal A_\mathrm{S}} \rme^{- {r^{(1)}g_0^2 x_0/a + \ldots}} \,.
\ees
We then have
\bes
    \label{e:SN}
    \frac{\rm signal}{\rm noise} \sim \frac{\cal A_\mathrm{S}}{\cal A_\mathrm{N}} \rme^{- [E^\mrm{stat}-\mpi/2]\,x_0}\,,
\ees
where as before $E^\mrm{stat}$ for $\mhbare=0$ enters. 
While methods to use translation invariance with the help of
``stochastic sources'' \cite{Sommer:1994gg} can reduce
the pre-factor $ {\cal A_\mathrm{N}}$, the correlators inevitably
disappear in the noise at some time $x_0=\rmO(a/r^{(1)})$. We here set 
$g_0\approx 1$ as appropriate for the considered 
gauge action. Inserting further that small
lattice spacings means $a\leq 0.05\fm$, we see that it is very
difficult to have reasonable precision around and beyond one fm
distance. However, this is the time separation where one can be 
confident that the desired 
ground state dominates. This problem brought progress in
the static theory (and therefore in HQET) to a stop in the beginning
of the nineties. 

Our description of the issue already suggests the solution. 
The coefficient $r^{(1)}$ is regularization dependent. Acceptable
alternative discretizations can be found which have much smaller
values \cite{DellaMorte:2005yc,DellaMorte:2003mn}, for example
\bes
   r^{(1)} = 0.0352 \quad 
   \mbox{for ``HYP2'' action \cite{DellaMorte:2005yc}} \,.
\ees
Our previous estimate is now pushed to time separations
of around 4~fm. In practice it turns out that the
estimate is overly optimistic, since it neglects
the finite terms in $E^\mathrm{stat}$.
The effect of the size of $r^{(1)}$ is also seen in \fig{f:scal}
where the results with the Eichten Hill action, \eq{e:llstat},
have much
larger statistical errors and these do grow towards the continuum 
limit, where the time separation in the correlation functions
diverges when it is measured in lattice units.

It is interesting to compare 
the HQET situation (HYP2) to the one for Nucleon matrix elements.
The energy difference $E^\mathrm{stat} -\mpi/2$ has to be compared to
$m_\mathrm{nucleon} -3\mpi/2$. At a pion mass of $\mpi=350$MeV and
a lattice with a small lattice spacing, $a=0.05$~fm we compare about 900~MeV for the former to 600MeV
for the nucleon.  For HQET this gets (slowly) worse for even smaller
lattice spacings and for the nucleon it gets quickly 
worse towards smaller pion masses. In both cases it is an
unpleasant mass scale that governs the time-dependence
of the signal-to-noise ratio.


%
\begin{figure*}[t!]
\hspace{1.0cm}
\begin{minipage}{.3\linewidth}
\vspace{3mm}
\psfragfig[scale=.45]{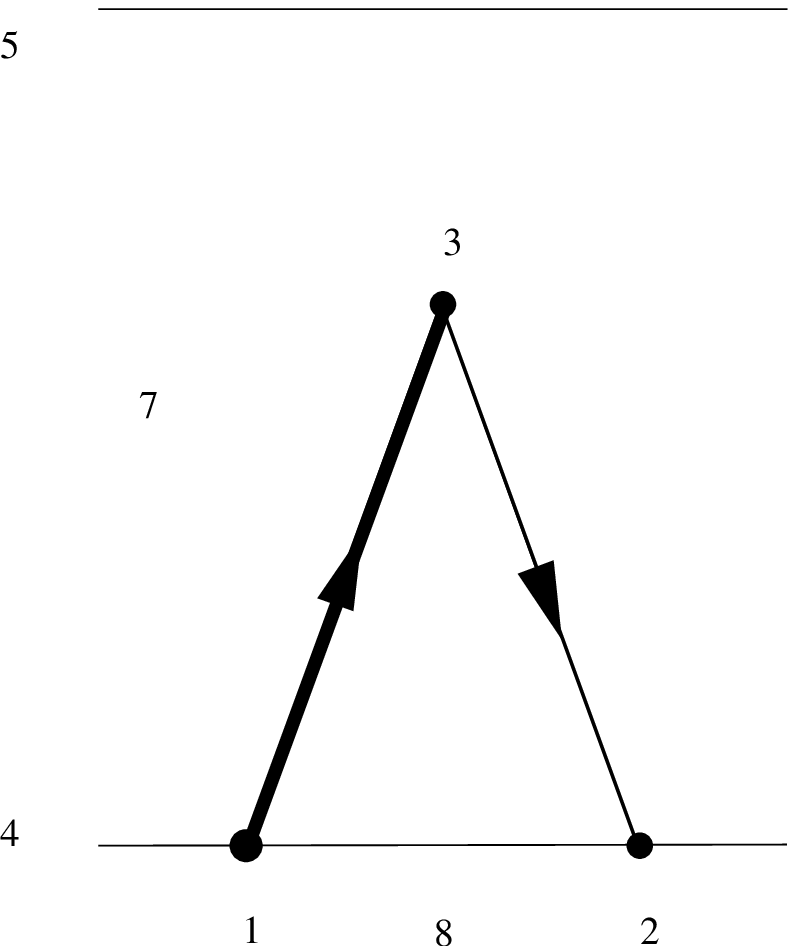}{
\psfrag{1}[c][c][1][0]{$\overline{\zeta}_{\rm b}$}
\psfrag{2}[c][c][1][0]{$\zeta_{\rm l}$}
\psfrag{3}[c][r][1][0]{${A}_0(x_0)$}
\psfrag{4}[r][t][1][0]{$x_0=0$}
\psfrag{5}[r][r][1][0]{$x_0=T$}
\psfrag{7}[c][c][1][0]{\large$\fa$}
\psfrag{8}[c][c][1][0]{$\gamma_5$}
}
\end{minipage}
\begin{minipage}{.3\linewidth}
\vspace{3mm}
\psfragfig[scale=.45]{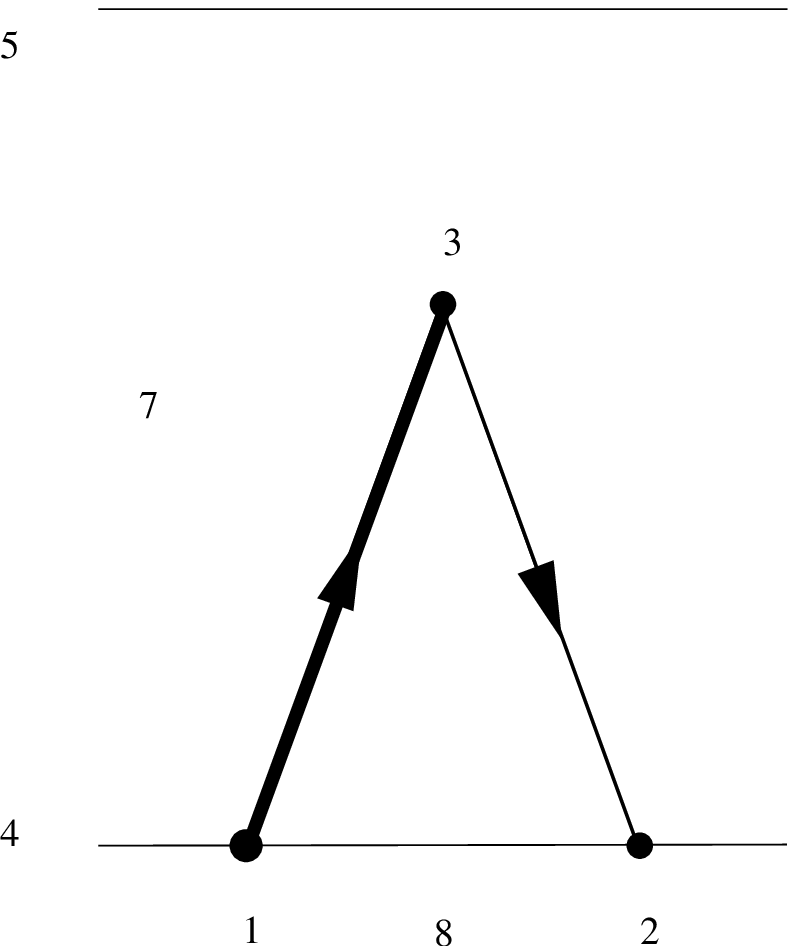}{
\psfrag{1}[c][c][1][0]{$\overline{\zeta}_{\rm b}$}
\psfrag{2}[c][c][1][0]{$\zeta_{\rm l}$}
\psfrag{3}[c][r][1][0]{${V}_k(x_0)$}
\psfrag{4}[r][t][1][0]{}
\psfrag{5}[r][r][1][0]{}
\psfrag{7}[c][c][1][0]{\large$\kv$}
\psfrag{8}[c][c][1][0]{$\gamma_k$}
}
\end{minipage}
\begin{minipage}{.3\linewidth}
\vspace{-1.15mm}
\psfragfig[scale=.45]{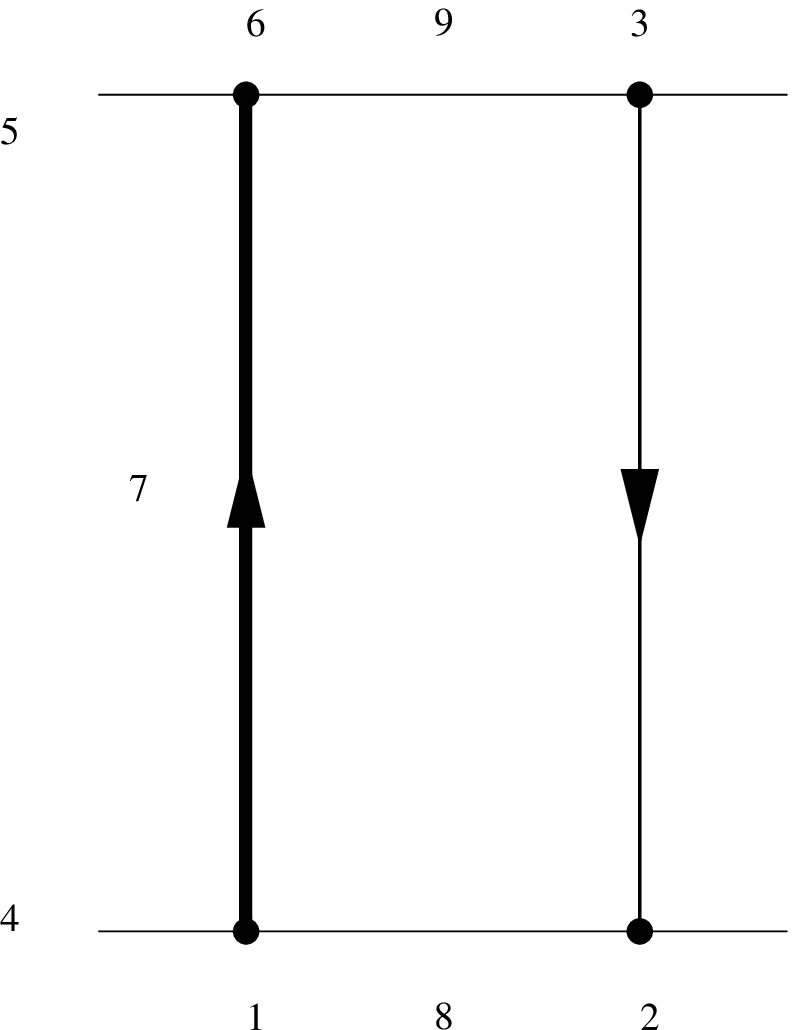}{
\psfrag{1}[c][c][1][0]{$\overline{\zeta}_{\rm b}$}
\psfrag{2}[c][c][1][0]{$\zeta_{\rm l}$}
\psfrag{3}[c][c][1][0]{$\overline{\zeta}^\prime_{\rm l}$}
\psfrag{6}[c][c][1][0]{$\zeta^\prime_{\rm b}$}
\psfrag{4}[r][t][1][0]{}
\psfrag{5}[r][r][1][0]{}
\psfrag{7}[c][c][1][0]{\large$F_1$}
\psfrag{8}[c][c][1][0]{$\gamma_5$}
\psfrag{9}[c][c][1][0]{$\gamma_5$}
}
\end{minipage}
\vspace{0.125cm}
\caption{
The Schr\"odinger functional correlation functions $f_{\rm A}$, 
$k_{\rm V}$ and $f_1$.
For $\kone$, in the rightmost diagram $\gamma_5$ is replaced by 
$\gamma_k$.
}\label{f:sfcorrel}
\end{figure*}
%

Given that correlation functions at 2-3 fm are not really 
accessible, one would like to have methods which work reliably at 
smaller distances. 
A technique that accelerates the  approach
of masses and matrix elements to the ground state quantities is
the use of correlation function matrices constructed from a few smeared interpolating fields~\cite{Teper:1987wt,Gusken:1989ad}.
These correlation functions can be analysed 
by the ``GEVP'' method \cite{Michael:1982gb,Luscher:1990ck,Blossier:2009kd} and then a systematic acceleration 
of the (asymptotic) convergence to the ground state is present.
Physically it means that one uses trial wave functions in a variational calculation. The method is easily modified
to be applicable also to matrix elements 
\cite{Blossier:2009kd,Bulava:2011yz} and to effective theories such as 
HQET. We note that the acceleration of the asymptotic convergence
is proven mathematically \cite{Blossier:2009kd}, but it is less
clear where asymptotia sets in in practice. The numerical results of \sect{s:res} are based on the GEVP technique.

\section{Verification of HQET} 
\label{s:tests}
In a first step in non-perturbative investigations of
HQET, it is of considerable interest to verify
the mass-scaling of QCD observables predicted by the lowest
order effective theory. This can only be done with the 
help of lattice gauge theory, where we can change the quark masses.
As for the matching strategy discussed above, a natural choice
is to focus on observables in a finite volume. For reasons to be explained
shortly, Schr\"odinger functional boundary conditions were chosen
with a $T\times L\times L\times L$ geometry, where periodic boundary conditions
(up to a phase $\theta$ for the quarks, see \sect{s:status})
are present in the $L\times L\times L$ space.
Without dynamical fermions the verification was carried out 
with $L\approx 0.2\,\fm\; T=L$ \cite{Heitger:2004gb} and
with $\nf=2$ dynamical fermions  with 
$L\approx 0.5\,\fm, \; T=L$  \cite{test:nf2}.

We here focus on $\nf=2$ observables constructed from the correlation functions
\bes
  &&\hspace*{-10mm} \fa(x_0,\theta) =  \nonumber
  \\
  && -{1 \over 2}\int{\rmd^3\vecy\, \rmd^3\vecz}\,
  \left\langle
  A_0(x)\,\zetabar_{\rm b}(\vecy)\gamma_5\zeta_{\rm l}(\vecz)
  \right\rangle  \,, \label{e_fa} \\
  &&\hspace*{-10mm} \kv(x_0,\theta) = \nonumber
  \\
  &&-{1 \over 6}\sum_{k}\int{\rmd^3\vecy\, \rmd^3\vecz}\,
  \left\langle
  V_k(x)\,\zetabar_{\rm b}(\vecy)\gamma_k\zeta_{\rm l}(\vecz)
  \right\rangle  \,, \label{e_kv}
\ees
as well as the boundary-to-boundary correlations
\bes
  \fone &=& -{1 \over 2L^6}\int{\rmd^3\vecy\, \rmd^3\vecz\,
                             \rmd^3\vecy'\,\rmd^3\vecz'}\,
  \nonumber \\ && \qquad \left\langle
  \zetabarprime_{\rm l}(\vecy')\gamma_5\zetaprime_{\rm b}(\vecz')\;
  \zetabar_{\rm b}(\vecy)\gamma_5\zeta_{\rm l}(\vecz)
  \right\rangle  \,, \label{e_fone} \\
  \kone &=& -{1 \over 6L^6}\sum_{k}\int{\rmd^3\vecy\, \rmd^3\vecz\,
                             \rmd^3\vecy'\,\rmd^3\vecz'}\,
  \nonumber \\ && \qquad \left\langle
  \zetabarprime_{\rm l}(\vecy')\gamma_k\zetaprime_{\rm b}(\vecz')\;
  \zetabar_{\rm b}(\vecy)\gamma_k\zeta_{\rm l}(\vecz)
  \right\rangle  \,. \label{e_kone}
\ees
They are illustrated in \fig{f:sfcorrel}. Fields $\zeta,\bar\zeta$ 
can be thought of as quark fields at the boundary $x_0=0$ while 
$\zeta',\bar\zeta'$ are located at $x_0=T$. The above correlation 
functions are gauge invariant despite the different locations
of  $\zeta$ and $\bar\zeta$. This is due to the fixed gauge fields at
the boundary; it is a very useful feature of 
the  Schr\"odinger functional which allows for the projection 
onto quark momenta
\bes 
  \vecp=\vecp_\theta \equiv (\theta,\theta,\theta)/L
\ees
in the above correlators (and $\vecp=-\vecp_\theta$ for the antiquarks).

As an example we discuss the function $\fa(x_0,\theta)$ in some detail.  
It describes the creation of a (finite-volume) $\vecp=0$ heavy-light \ps 
meson state, $|\varphi_{\rm B}(L)\rangle$, through quark and antiquark
boundary fields which are separately projected onto 
momenta $\vecp=\pm\vecp_\theta$. This state ``propagates'' an
interval $x_0$ in Euclidean time. From the upper boundary 
a state with vacuum quantum numbers propagates a time-distance
$T-x_0$.
The correlation function $\fa(T/2,\theta)$ can thus be written in terms of Hilbert space matrix elements,
\bes
  \fa(T/2,\theta) &=& {\cal Z}^{-1} \langle \Omega(L)| \opA |B(L)\rangle\,,
  \\
  |B(L)\rangle &=& \rme^{-T\ham /2 } |\varphi_{\rm B}(L)\rangle \,,
  \\
  |\Omega(L)\rangle &=&\rme^{-T\ham/2 } |\varphi_{0}(L)\rangle \,.  
\ees
where $\ham$ is the QCD Hamiltonian and 
\bes
  {\cal Z} =  \langle \Omega(L)| \Omega(L)\rangle\,.
\ees
$|\varphi_{0}(L)\rangle$ denotes the \txtSF intrinsic boundary
state. It has the quantum numbers of the vacuum. 
All states appearing in our analysis are 
eigenstates of total spatial momentum with eigenvalue zero. Their dependence
on $\theta$ has been suppressed.
The time evolution 
operator $\rme^{-T\ham/2 }$ suppresses high-energy states
exponentially. 
When expanded in terms of eigenstates of the Hamiltonian,
$| \Omega(L)\rangle$ and $|B(L)\rangle$ are thus dominated by 
contributions with energies of at most $\Delta E = \rmO( 1/L)$ above the 
ground state energy in the respective channel (recall that we take 
$T=\rmO(L)$). 

This explains why, at large time separation $x_0\gg \mhinv$, 
HQET is expected to describe the large-mass behavior of the correlation 
function, also in the somewhat unfamiliar framework of the 
Schr\"odinger functional.\footnote{ 
More generally, HQET applies to correlation functions at large 
Euclidean separations. 
}

Equations similar to the above hold for $\kv$; one only needs to replace 
\ps states by vector ones. 
Finally, the boundary-to-boundary correlator is represented as
\bea
 \fone = {\cal Z}^{-1} \langle B(L)| B(L)\rangle \,.
\eea
Since the boundary quark fields $\zeta,\zetabar,\ldots$ are 
multiplicatively renormalizable \cite{Sint:1995rb}, this holds also for 
the states $|\varphi_{0}(L)\rangle$ and $|\varphi_{\rm B}(L)\rangle$.

\begin{figure}[ht!]
  \centering
  \includegraphics[width=0.47\textwidth]{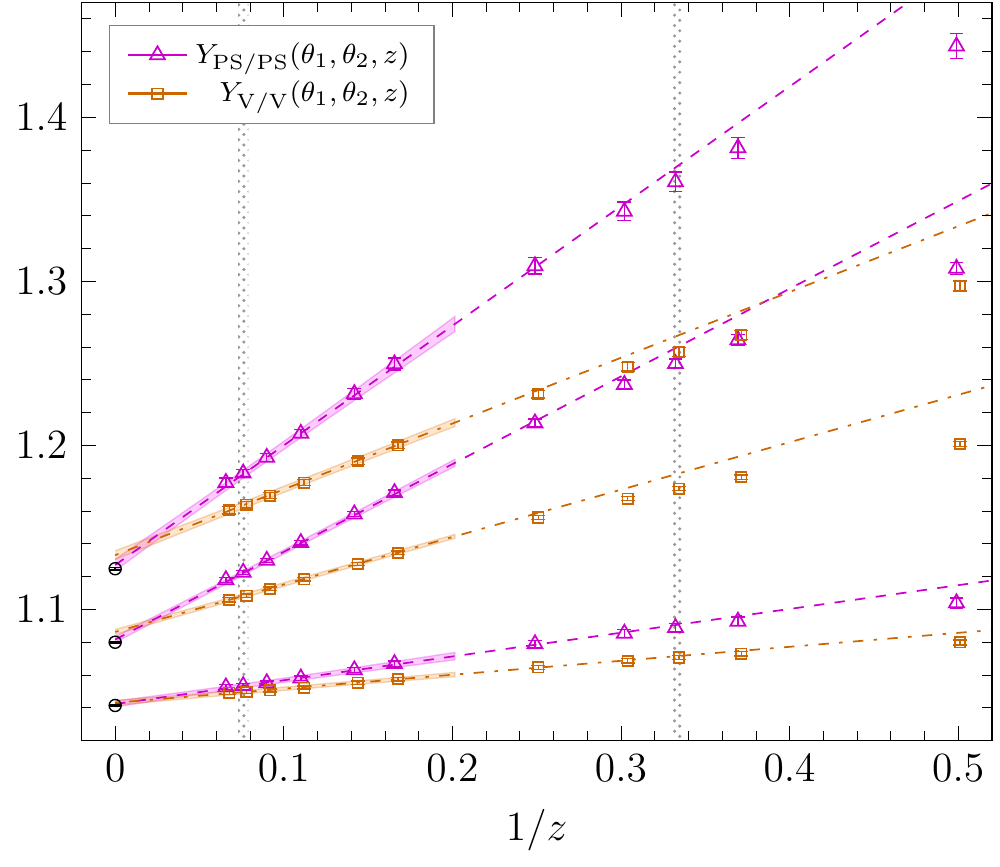}
    \caption{Extrapolations of $Y_{\rm PS/PS}(\theta_1,\theta_2)$ and $Y_{\rm V/V}(\theta_1,\theta_2)$
 to the $M\to\infty$ limit for three combinations of $(\theta_1,\theta_2)$.
 The dimensionless heavy quark mass $z=ML$ is used as a variable.
 Black circles  show the continuum  results of the corresponding quantity computed  in the static approximation. All data points were first extrapolated
           to the continuum limit~\cite{test:nf2}. Three different combinations of $\theta_1,\theta_2$ are shown, where $0\leq \theta_i\leq 1$. Graph provided by P.~Fritzsch, based on \cite{test:nf2}.
}
\label{f:RY}
\end{figure}

It now follows that the ratios
\bes
\Yr(\theta) &\equiv& \za\frac{\fa(T/2,\theta)}{\sqrt{\fone(\theta)}} \,,\quad \\
\Yv(\theta) &\equiv& - \zv\frac{\kv(T/2,\theta)}{\sqrt{\kone(\theta)}} \,,\quad  \\
\Yrr(\theta_1,\theta_2) &\equiv& \frac{\Yr(\theta_1)}{\Yr(\theta_2)}\,, \label{e_ratios}
\\
\Yvv(\theta_1,\theta_2) &\equiv& \frac{\Yv(\theta_1)}{\Yv(\theta_2)} \,,
\ees
are finite quantities. 
As is immediately clear from the foregoing discussion,
\bea
  \Yr(L,M) = { \langle \Omega(L)| \opA |B(L)\rangle 
                \over 
                ||\;| \Omega(L)\rangle\;|| \;\;
                ||\;| B(L)\rangle\;|| }
\eea
(\,or $\Yv(L,M)$\,) becomes proportional to the \ps (or vector) 
heavy-light decay constant as $L\to\infty$. 
At any fixed $L$, 
the large-$M$ behavior of these quantities 
is indeed described by HQET if it is the correct effective theory. 

Numerical tests of this equivalence were originally done in the quenched approximation \cite{Heitger:2004gb}, with $L \approx 0.2\,\fm$, 
profiting from the full improvement of the theory with mass-dependent 
improvement terms \cite{Heitger:2003ue}. Here we show more recent results 
with $\nf=2$ dynamical fermions. Based on the simulations 
\cite{Blossier:2012qu} the following results are due to 
Fritzsch, Garron and Heitger~\cite{test:nf2}.

The first two HQET predictions are 
\bes
    \Yrr(\theta_1,\theta_2) &=& Y_\mathrm{stat}(\theta_1,\theta_2) + \rmO(1/M)
    \\ 
    \Yvv(\theta_1,\theta_2)&=& Y_\mathrm{stat}(\theta_1,\theta_2) + \rmO(1/M)\,.
\ees
Due to spin symmetry, the static limit $Y_\mathrm{stat}(\theta_1,\theta_2)$ is the same for the vector channel and the \ps channel. It can be computed
by replacing the relativistic fields by the static ones. The comparison
is shown in \fig{f:RY}, extrapolating the relativistic results with
just a linear function in $1/z$ for $z\equiv ML \geq 5$. Results of
the extrapolation are in very good 
agreement with the numbers obtained directly in the static approximation, 
demonstrating at the same time the correctness of the effective theory and the usefulness of such observables for matching HQET to QCD.

\begin{figure}[ht!]
  \centering
  \includegraphics[width=0.47\textwidth]{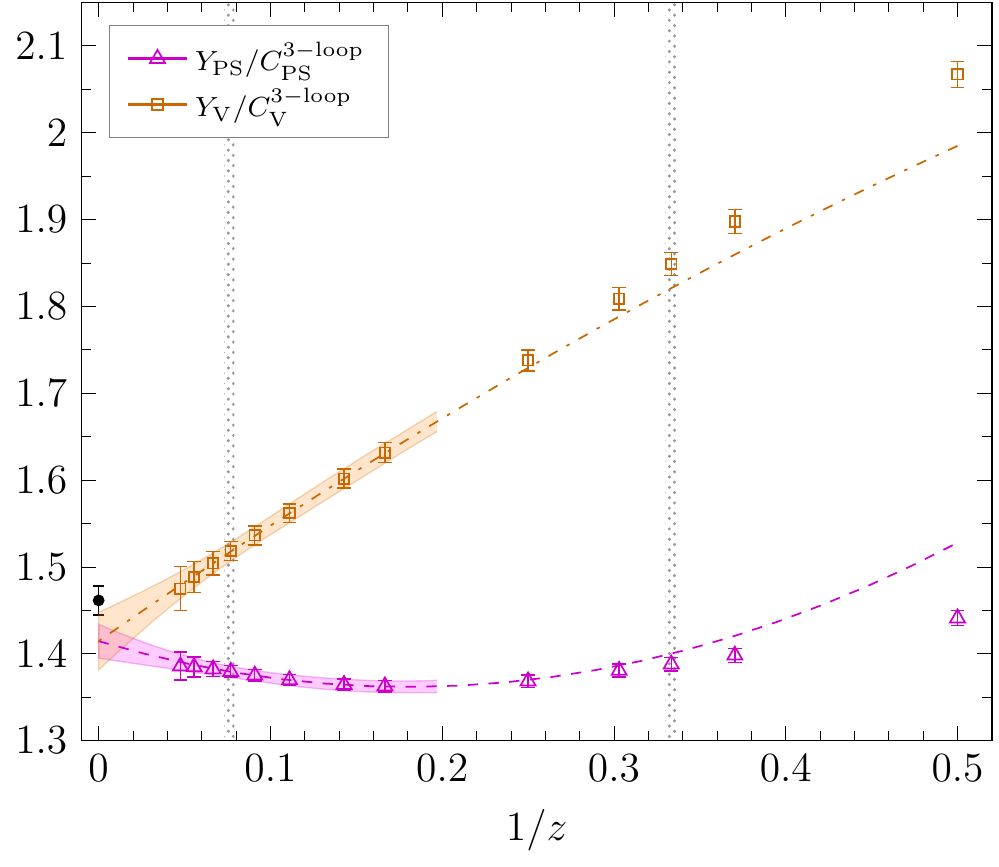}
  \caption{Comparison of static extrapolations of $\Yps(\theta)$ and $\Yv(\theta)$
           to the non-perturbative HQET results in the continuum (filled circle). Conversion functions $\Cps,\Cv$ evaluated at 3-loop  order of perturbation theory are used.
The extrapolation, quadratic in $1/z$, uses data with $1/z<0.2$
only. Graph provided by P.~Fritzsch, based on \cite{test:nf2}.
}
\label{f:YpsYv}
\end{figure}

Two more predictions read
\bes
    \frac{\Yps(\theta)}{\Cps(M/\Lambda)} &=& 
    X_\mathrm{stat}^\mathrm{RGI}(\theta) + \rmO(1/M) \,,
    \nonumber \\ \\ \nonumber
    \frac{\Yv(\theta)}{\Cv(M/\Lambda)} &=& 
    X_\mathrm{stat}^\mathrm{RGI}(\theta) + \rmO(1/M) \,.
\ees
The matrix element $X_\mathrm{stat}^\mathrm{RGI}$ is defined
as $\Yps$, but with the static quark field and
it is the renormalization group invariant one, 
see \eq{e:lammu}. For the time-component of the axial current,
the factor $\varphi_\mathrm{stat}(\gbar)$ in that equation 
and the renormalisation factor $\za^\mathrm{stat,SF}$ were determined 
non-perturbatively in a \textSF\ renormalization
scheme~\cite{DellaMorte:2006sv}. Thus $X_\mathrm{stat}^\mathrm{RGI}$
is known without perturbative uncertainties.

The comparison is shown in \fig{f:YpsYv}. 
The agreement between the extrapolation of 
relativistic results and the static effective theory 
is not as convincing as in \fig{f:RY}. In fact, for $\Yps$ there
is a disagreement, but only at the level of 
1-2 sigma. This may therefore be a statistical effect, but
it may also be due to the perturbative approximation of $\Cps$,
which is somewhat doubtful as discussed earlier. 
In comparison, in \fig{f:RY} the perturbative 
factors $\Cps,\Cv$ drop out. 

We have only shown two out of a number
of tests. Considering them all~\cite{Heitger:2004gb,test:nf2},
we conclude that the effective theory is very well tested,
but it is safer not to use conversion functions $\Cps,\Cv$
from perturbation theory, even if they are determined at 
three-loops, i.e. with relative errors of order $\gbar^6(\mstar)$.

\section{Numerical Simulations and results}
\label{s:res}
We now turn to a discussion of numerical results 
skipping most of the details of the Monte Carlo simulations.
These separate into two categories. One part concerns small
volume simulations $L\leq1\fm$ with \textSF\ boundary conditions.
These have been discussed in \cite{DellaMorte:2003jj,Hoffmann:2003mm,DellaMorte:2004hs,
DellaMorte:2005rd,DellaMorte:2005kg}. Many such
simulations had to be carried out in order to control the
renormalization and matching. They required tuning 
of the bare parameters such that continuum limits can be taken
at a fixed volume in physical units and at vanishing 
dynamical quark mass.

A second part is then necessary in 
large volume, with $L\geq2\fm, \; \mpi L\geq 4$. 
These simulations are rather universally useful and
also much more expensive in the numerical effort. 
Hence they have been carried out in coordination with
several European groups, by the CLS effort \cite{CLS},
see in particular~\cite{Fritzsch:2012wq}. 
The simulations were carried out down to
pion masses of 200MeV on lattices with up to $128\times 64^3$
points. They were possible due to a significant 
improvement of algorithms and their implementations 
\cite{Hasenbusch:2001ne, Hasenbusch:2002ai,Luscher:2007es,Luscher:2007se,Marinkovic:2010eg,Frommer:2013fsa,Omelyan2003272}
starting from the principle  
of the HMC algorithm \cite{Duane:1987de,Gottlieb:1987mq}.
For recent reviews covering lattice QCD algorithms we refer to 
\cite{Luscher:2010ae,Schaefer:2012tq}.
Some more details can be found in \cite{sfb:B2}.

All HQET computations and strategies were developed 
and tested in the quenched 
approximation, i.e. QCD with all valence quarks but a 
vanishing number
of sea quark flavors: $\nf=0$. We will mention
these results only for comparison, while we discuss 
the results with $\nf=2$ quark flavors
in more detail. A dynamical strange quark has  
not yet been included in the HQET simulations of the 
ALPHA collaboration.


\begin{figure*}
	\centering
	\includegraphics[width=0.69\textwidth]{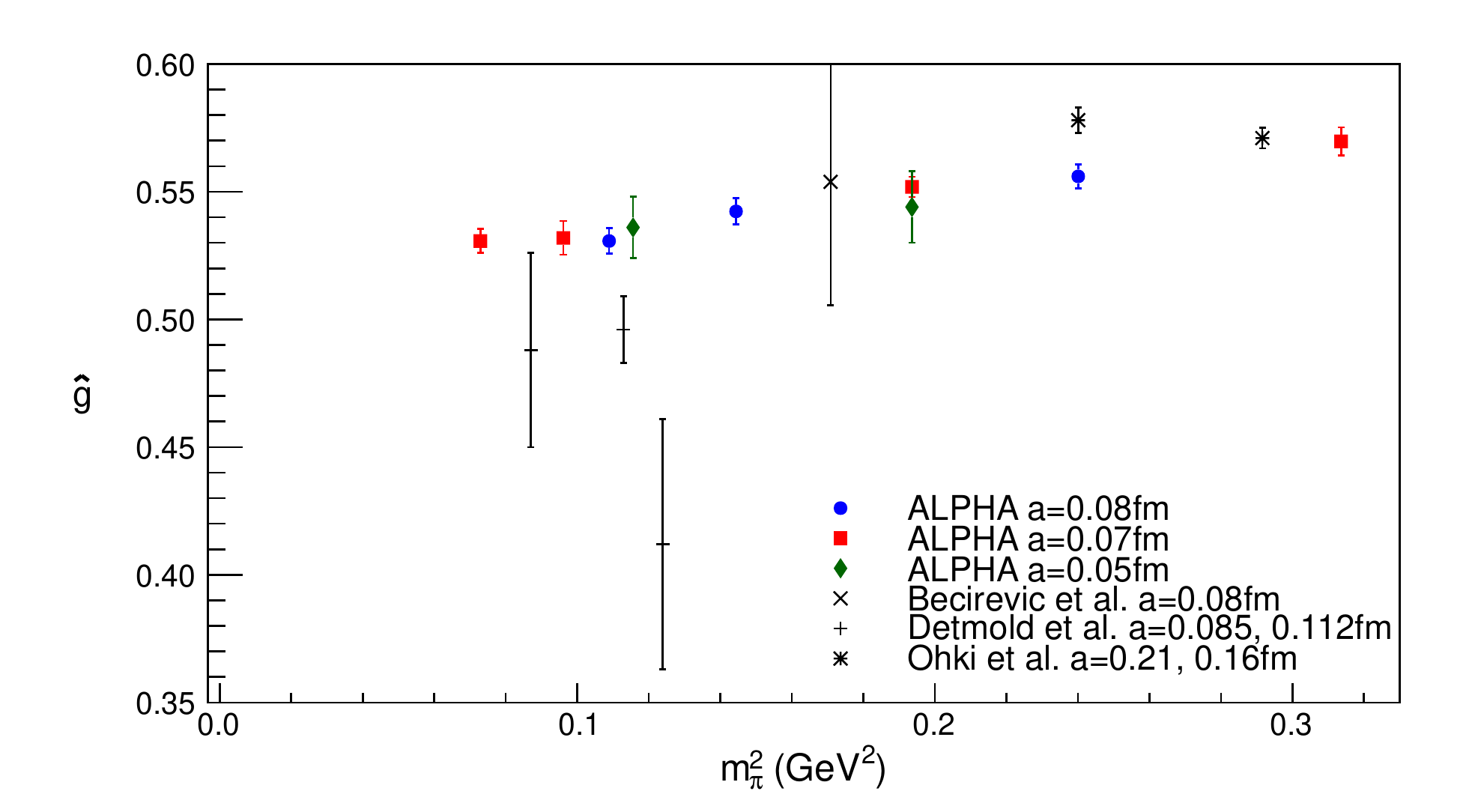} 
	\caption{\label{f:ghatsummary} A summary of unquenched lattice QCD results for $\hat{g}$. The results from the three lattice spacings used in \cite{Bernardoni:2014kla} are labeled `ALPHA'. Additionally
there are results of Ohki et al.~\protect\cite{Ohki:2008py},
Becirevic et al.~\protect\cite{Becirevic:2009yb} and 
Detmold et al.~\protect\cite{Detmold:2012ge}. 
For Ref.~\protect\cite{Detmold:2012ge}, which employs $\nf = 2+1$ dynamical flavors, we take the results for a single level of link smearing in the static action. Graph from \cite{Bernardoni:2014kla}.} 
\end{figure*}
\subsection{The B$^*$B $\pi$ coupling.}
\label{s:bbpi}

As explained above, the available large volume simulations still 
have unphysical quark masses and results need to be extrapolated
to the point where the light quark masses are the  physical ones. 
The natural way to carry out
such an extrapolation is with the help of a systematic expansion 
in the light quark mass. Again this means that one uses predictions
from an effective field theory which implements the expansion. 
It is heavy meson 
chiral perturbation theory (HMChPT). The fields in this effective theory 
are a triplet of pion fields as well as a (static) B-meson and a B$^*$ meson field. 
The expansion is a combined expansion in $\mhinv$, in 
the squared pion momenta and in the light quark mass, each counting as one 
power of the expansion variable, 
$\Lambda_\mathrm{QCD}/\mh = \rmO(\vecp^2/\Lambda_\mathrm{QCD}^2)=
\rmO(m_\mathrm{up}/\Lambda_\mathrm{QCD})$, where as before 
we work in the B-meson rest frame.\footnote{Usually,
the second term is written as $\rmO(\vecp^2/(8\pi^2 f_\pi^2) )$ 
and the expansion coefficients are assumed to be order one
in that variable. With a pion decay constant of
$f \sim \Lambda_\mathrm{QCD}/4$ for our previous 
estimate of $\Lambda_\mathrm{QCD}\sim 500\,\MeV$, 
there is a numerical difference. 
Similar ambiguities are present in a prefactor of  
$m_\mathrm{up}/\Lambda_\mathrm{QCD}$, but these numerical factors 
do not influence the structure of the expansion and are
not well known/defined anyway.} 
At lowest order, the effective theory contains five 
parameters, usually called low energy constants. They all
refer to the chiral limit and $\mh \to \infty$. There are
the pion decay constant, $f$,  the light-quark condensate, the mass of the B-meson and the three-point coupling 
coupling B$^*$ B  $\pi$ denoted by $\hat g$ \cite{Burdman:1992gh,Wise:1992hn,Yan:1992gz}.
Apart from  $\hat g$ the couplings can be taken
from experiment (the light quark condensate is removed
from the list by taking $\mpi^2$ instead of $m_\mathrm{up}$ as
mass-parameter).
Typical predictions of HMChPT are (with $\xi= \mpi^2/(8\pi^2 f_\pi^2)$)
\bes \label{e:mBHM}
   \nspace \mB = \mB^\mathrm{chir} + f_\pi\,\left(
        {3\sqrt{2}\pi^2 \hat{g}^2}  \xi^{3/2}\, 
        + \alpha_\mathrm{m}\, \xi \,+
        \,\rmO(\xi^2)
        \right)\, \nonumber \\ 
\ees
and
\bes \label{e:fBHM}
     \sqrt{\frac{\mB}{2}}  \fB &=&  
       \left. \sqrt{\frac{\mB}{2}}  \fB \right|^\mathrm{chir} 
       \, \times \\
       && \nspace\nspace \left[1-\frac{3}{4}\frac{1 + 3 \hat{g}^2}{2}  \xi\log( \xi ) + \alpha_\mathrm{f}\, \xi  + \rmO(\xi^2) \right]\,.
       \nonumber  
\ees

The coefficients of the leading non-analytic terms, 
$\xi^{3/2}$ and $\xi\log( \xi ) $, in \eq{e:mBHM} and 
in \eq{e:fBHM}, respectively, are given in terms of $\hat g$. 
This coupling is therefore better determined before using the
HMChPT formulae to extrapolate from unphysical quark masses 
to the physical one. 

A determination of $\hat g$ just means that HMChPT is 
matched to HQET at the lowest order in $\mhinv$; at this level HQET is regarded as the fundamental theory. 
The matching condition can be written to directly give $\hat g$ via
\bes
\hat{g} &=& \frac{1}{2}\langle B^0(\textbf{0}) | \opAk^\mathrm{du}(0) | B^{*+}_{k}(\textbf{0})\rangle, \qquad 
\\ {A}_{\mu}^\mathrm{du}(x) &=& {\psibar_{\rm d}}(x)\gamma_{\mu}\gamma_5\psi_{\rm u}(x),  
\label{e:ghatdeterm}
\ees
where $\psi_{\rm d}$($\psi_{\rm u}$) annihilates a down(up) quark and the index 
$k=1,\,2,\,3$ is not summed over. The 
non-relativistic normalization 
of states given earlier is used; 
in finite volume it is  
$\langle B^0(\textbf{p}) | B^{0}(\textbf{p}) \rangle = \langle B^*_k(\textbf{p}) | B^{*}_k(\textbf{p}) \rangle = 2L^{3}=2V$, 
where $L$ is the linear size of the torus. 

Profiting from a newly developed method for its computation \cite{Bulava:2011yz},
the matrix element \eq{e:ghatdeterm},  was 
determined in \cite{Bernardoni:2014kla} with a much better precision than it was possible before. A comparison to
other computations is shown in \fig{f:ghatsummary} for
various pion masses (i.e. dynamical quark masses) and 
lattice spacings.

\begin{table*}[t!]
\begin{center}
\renewcommand{\arraystretch}{1.25}
\begin{tabular}{@{\extracolsep{0.2cm}}lllll}
      i  & $\Phi^\mathrm{QCD}_i$ & c.f. &  large volume limit & 
      dominant sensitivity to\\ 
      \midrule \hline
      1 & $\left. -L \frac{\rmd}{\rmd x_0} \fa\right |_{x_0=T/2} $ &
          \eq{e_fa}
        & $L \mB$ & $m_\bare$ 
      \\[2mm] 
      2 & $\log(\Yr) $  & \eq{e_ratios}
        & $\log(L^{3/2}\fb\sqrt{\mB})$ &  $\ln Z_{A_0}^\hqet$  \\[2mm]
      3 & $\log\left(\fa(T/2,\theta_1)/\fa(T/2,\theta_2)\right)$ & \eq{e_ratios} 
        & 0 & $\ceff{A}{0}{1}$  \\[2mm]
      4 & $\frac14 \log \left[\frac{\fone(\theta_1)\kone(\theta_1)^3}{\fone(\theta_2)\kone(\theta_2)^3} \right]_{T=L/2}$ & \eq{e_fone},(\ref{e_kone}) 
        & 0 & $\omegakin$ \\[2mm]
      5 & $\frac34 \log[\fone/\kone]$ & \eq{e_fone},(\ref{e_kone}) 
        & $L\,(\mBstar-\mB)$ & $\omegaspin$ \\      
\bottomrule
\end{tabular}
\caption{Matching observables and some of their properties. 
\label{t:phii}
} 
\end{center}
\end{table*}

The low energy constant $\hat g$ is defined in the chiral
limit. Therefore,  the computations shown in \fig{f:ghatsummary}
are extrapolated to $\mpi^2=0$. 
The final uncertainties are dominated by the systematic
uncertainty in this step.
Ref.~\cite{Bernardoni:2014kla} obtained
\bes
  \hat{g} = 0.492(29)\,.
\ees
in the chiral limit.

\subsection{HQET parameters}
\label{s:param} 
For $\nf=0$ the HQET parameters $\omega_i$ have been determined in 
\cite{Blossier:2010jk}. Here we review the refined
strategy applied with two flavors of dynamical fermions \cite{Blossier:2012qu}.
It covered the three parameters in the action and the two parameters
of $A_0^\hqet$ needed for the computation of $\fb$.
All five used matching observables are similar to what we 
discussed in \sect{s:tests}. We list them in \tab{t:phii}. 
Note how three of them are small volume versions of physical
quantities which one would like to predict in large volume. 
If finite volume effects are truly small, these quantities are then 
predicted correctly with very small truncation errors of the $\mhinv$ 
expansion. Even with significant finite volume effects,
this property is expected to reduce truncation errors as 
compared to many other choices.

\begin{figure*}[t!]
  \includegraphics[width=0.333\textwidth]{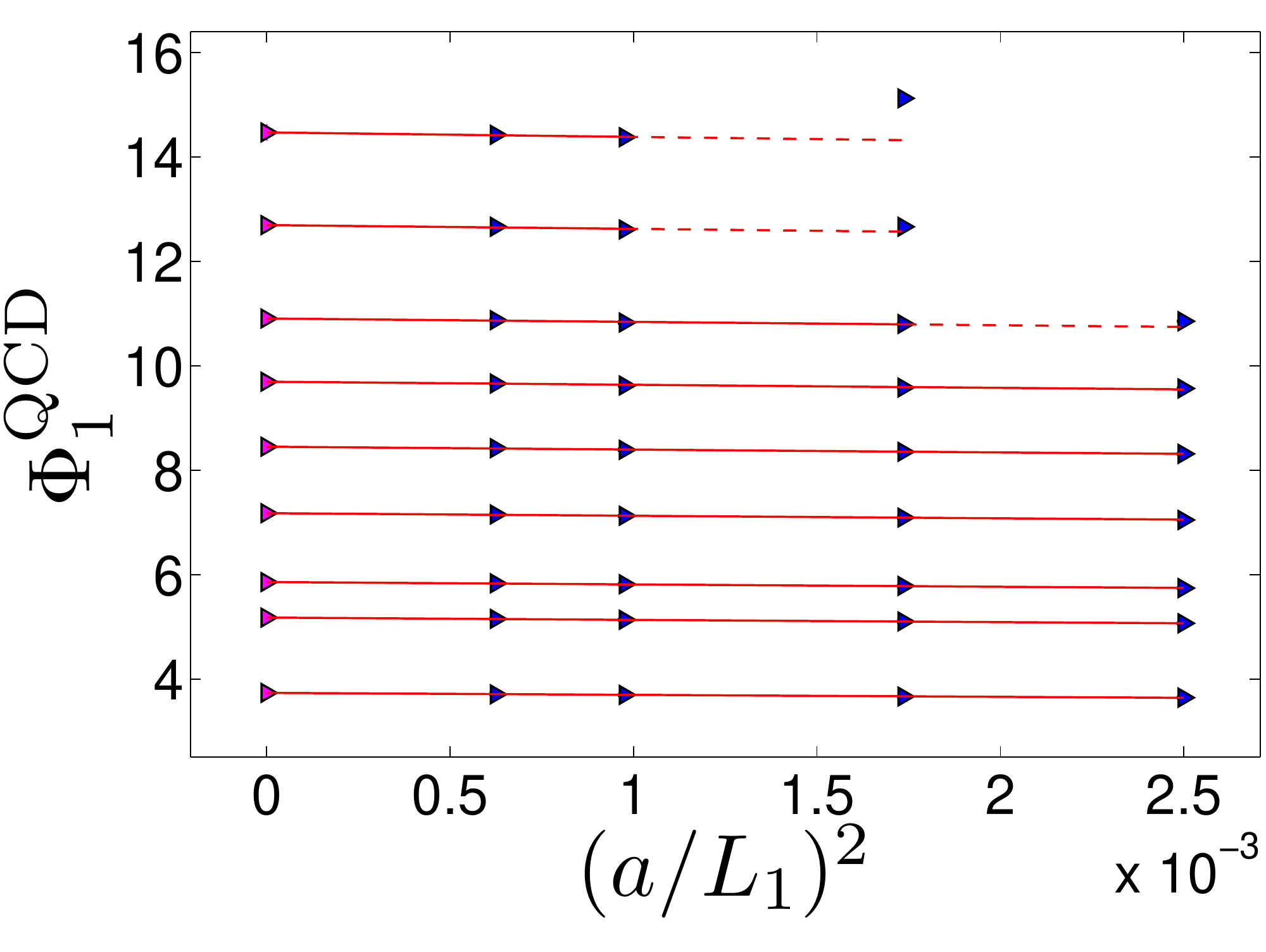}
  \includegraphics[width=0.333\textwidth]{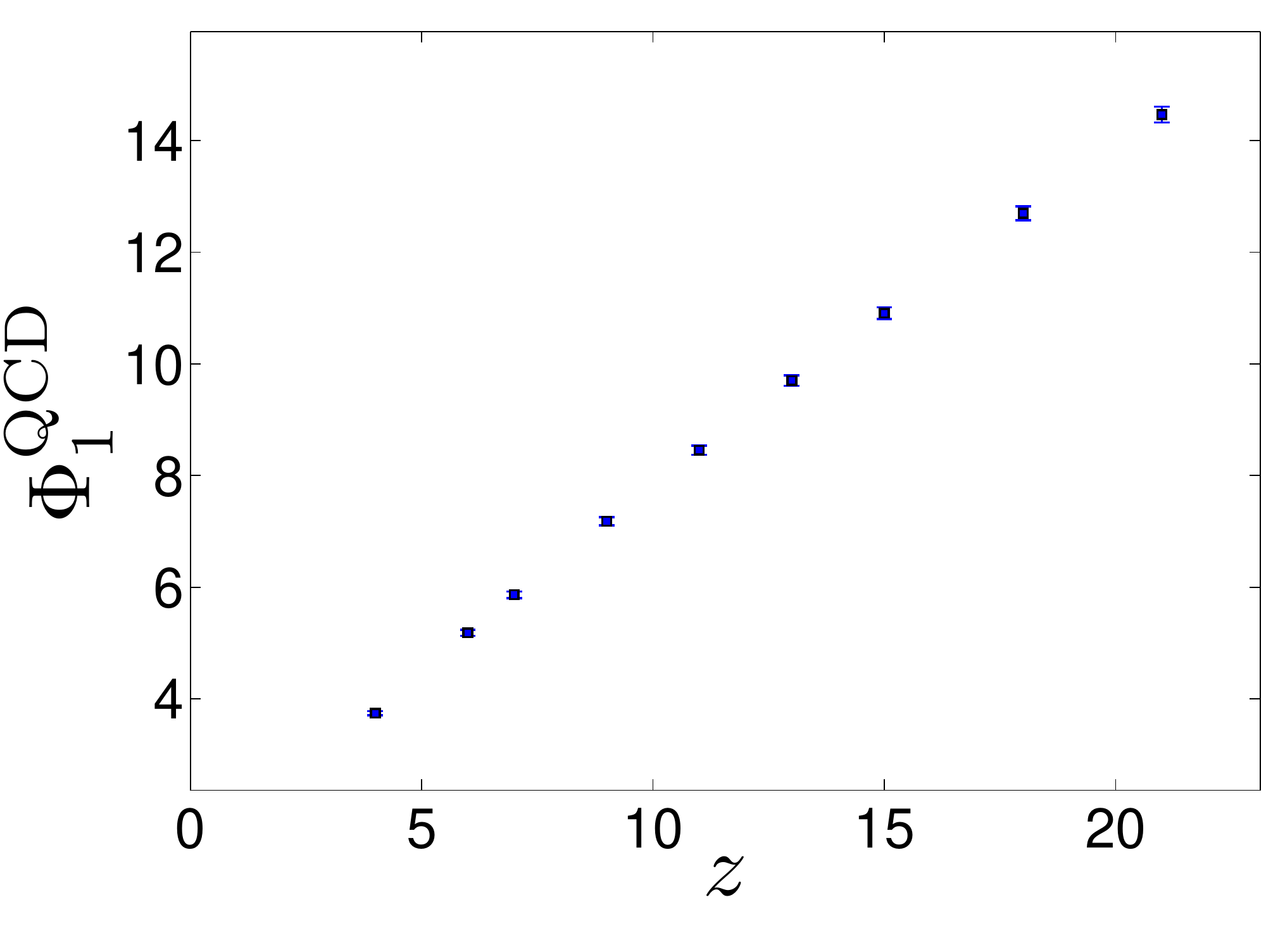}
  \includegraphics[width=0.333\textwidth]{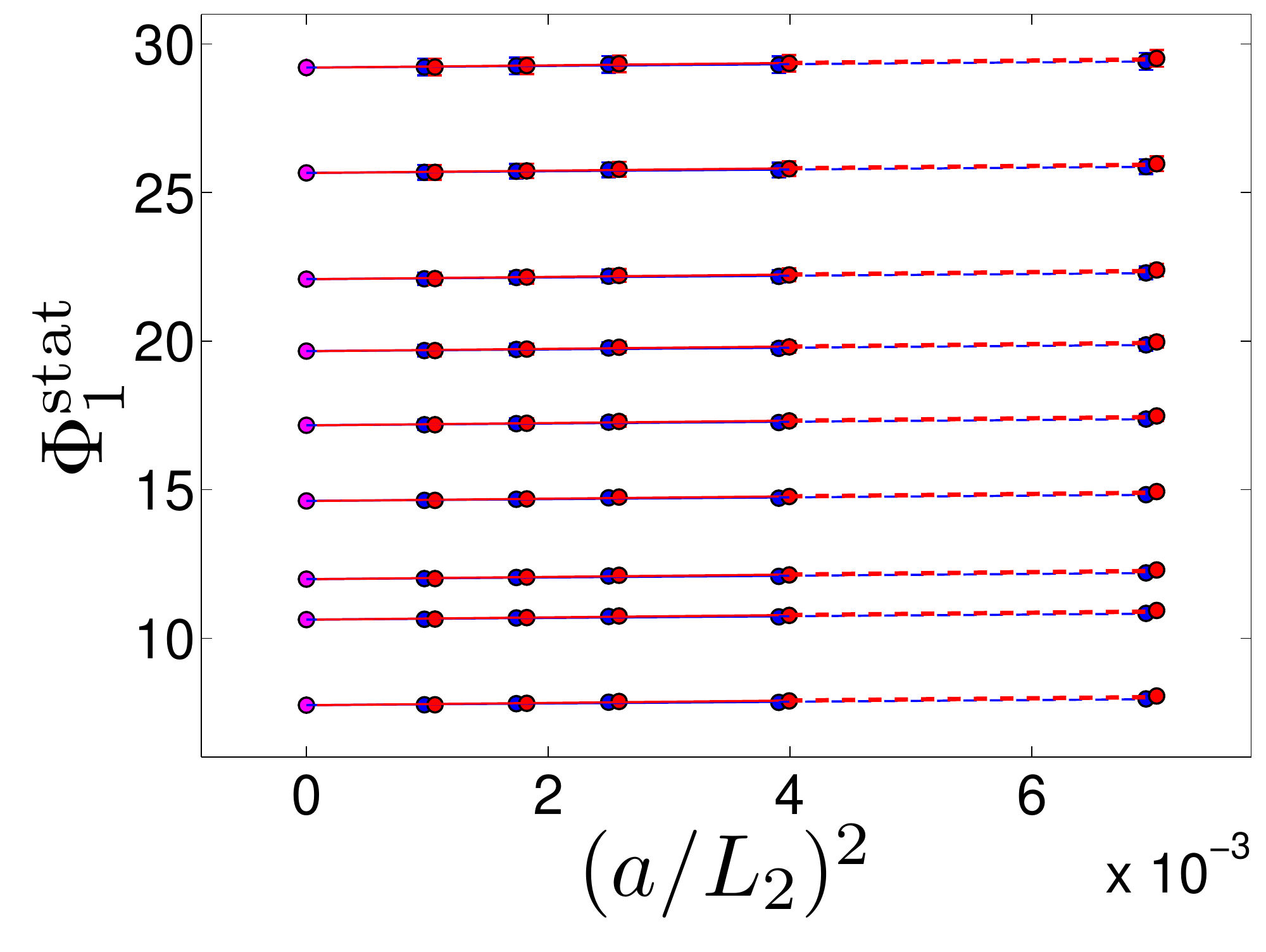}
  \\
  \includegraphics[width=0.333\textwidth]{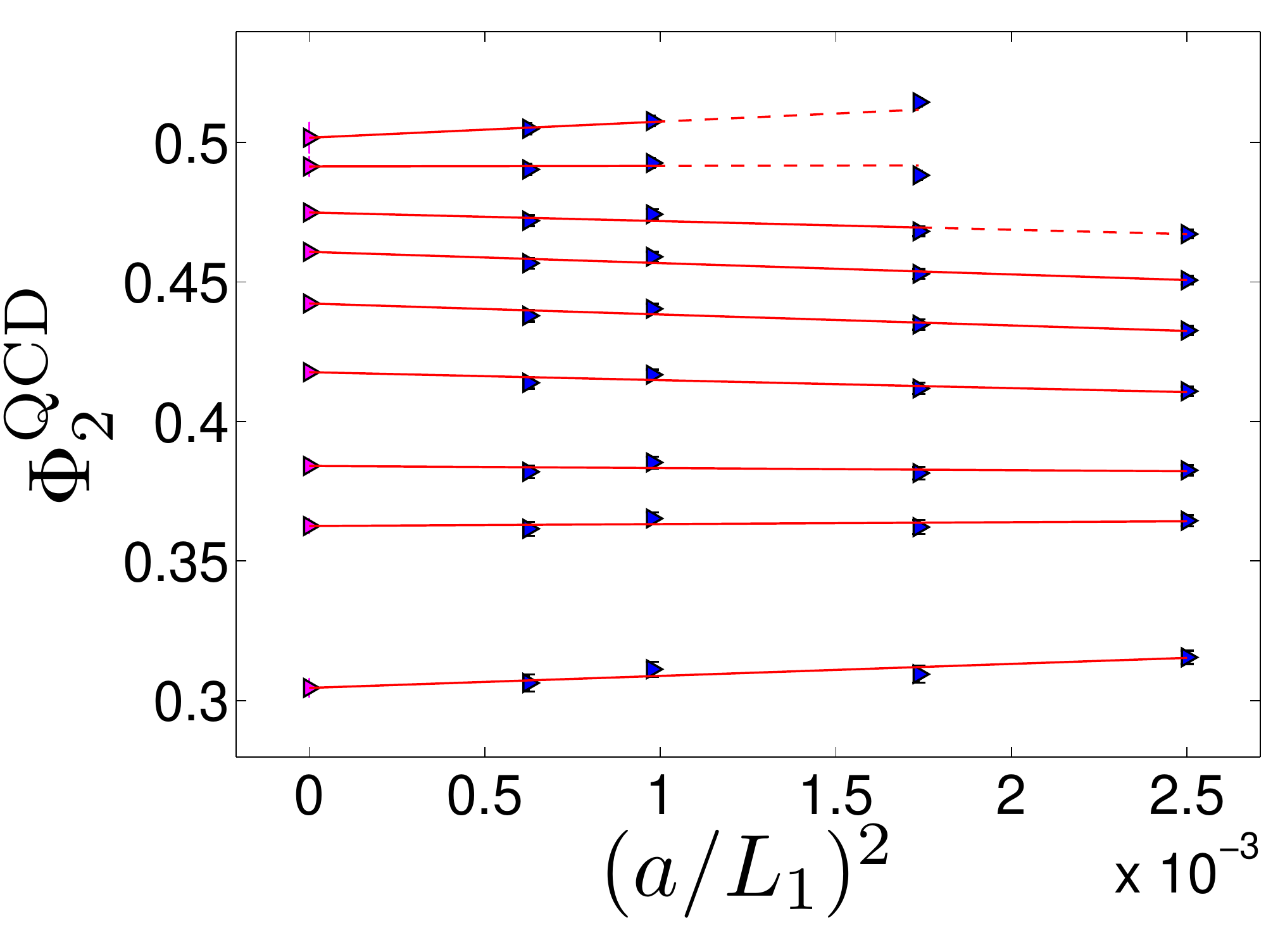}
  \includegraphics[width=0.333\textwidth]{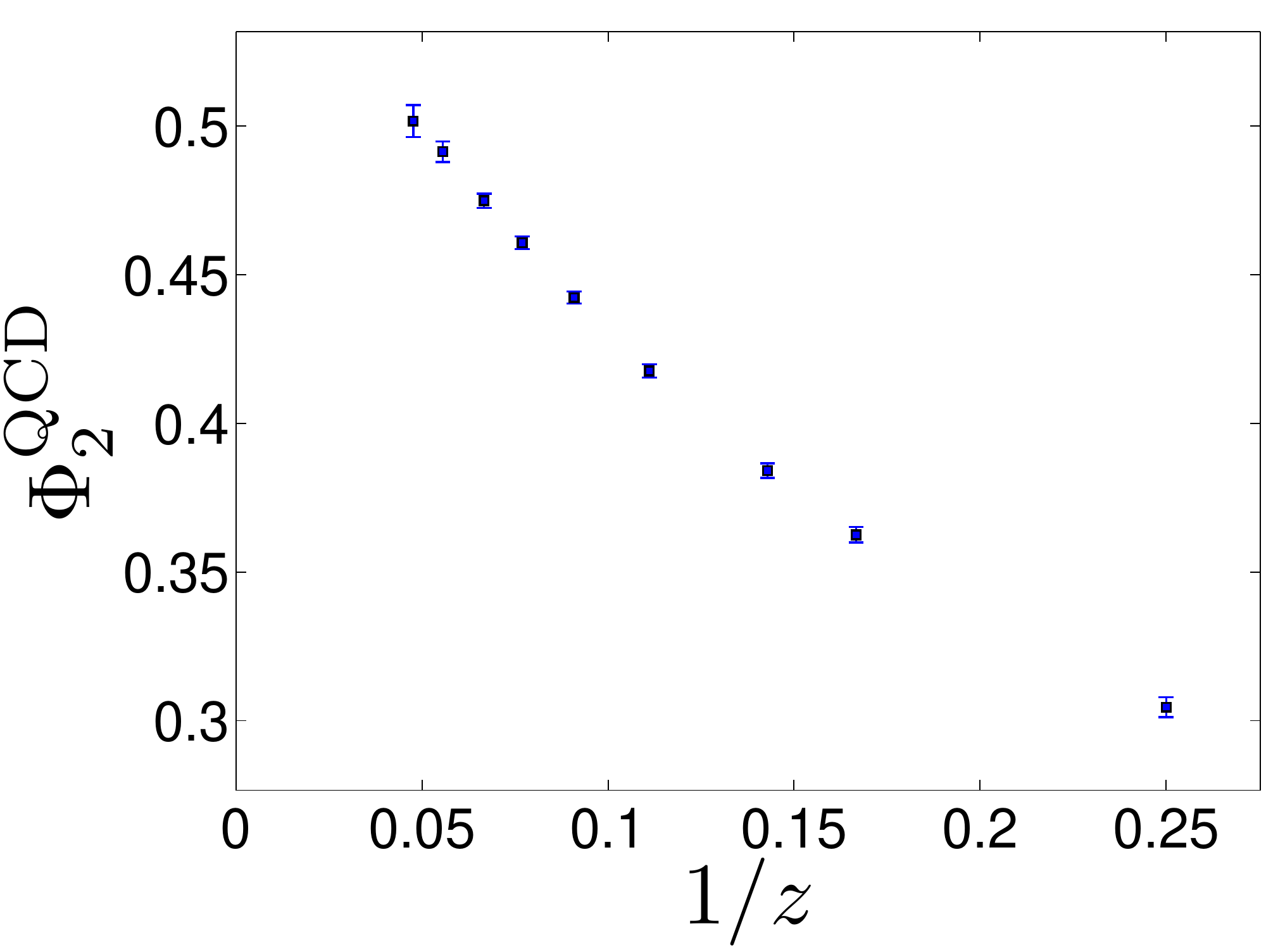}
  \includegraphics[width=0.333\textwidth]{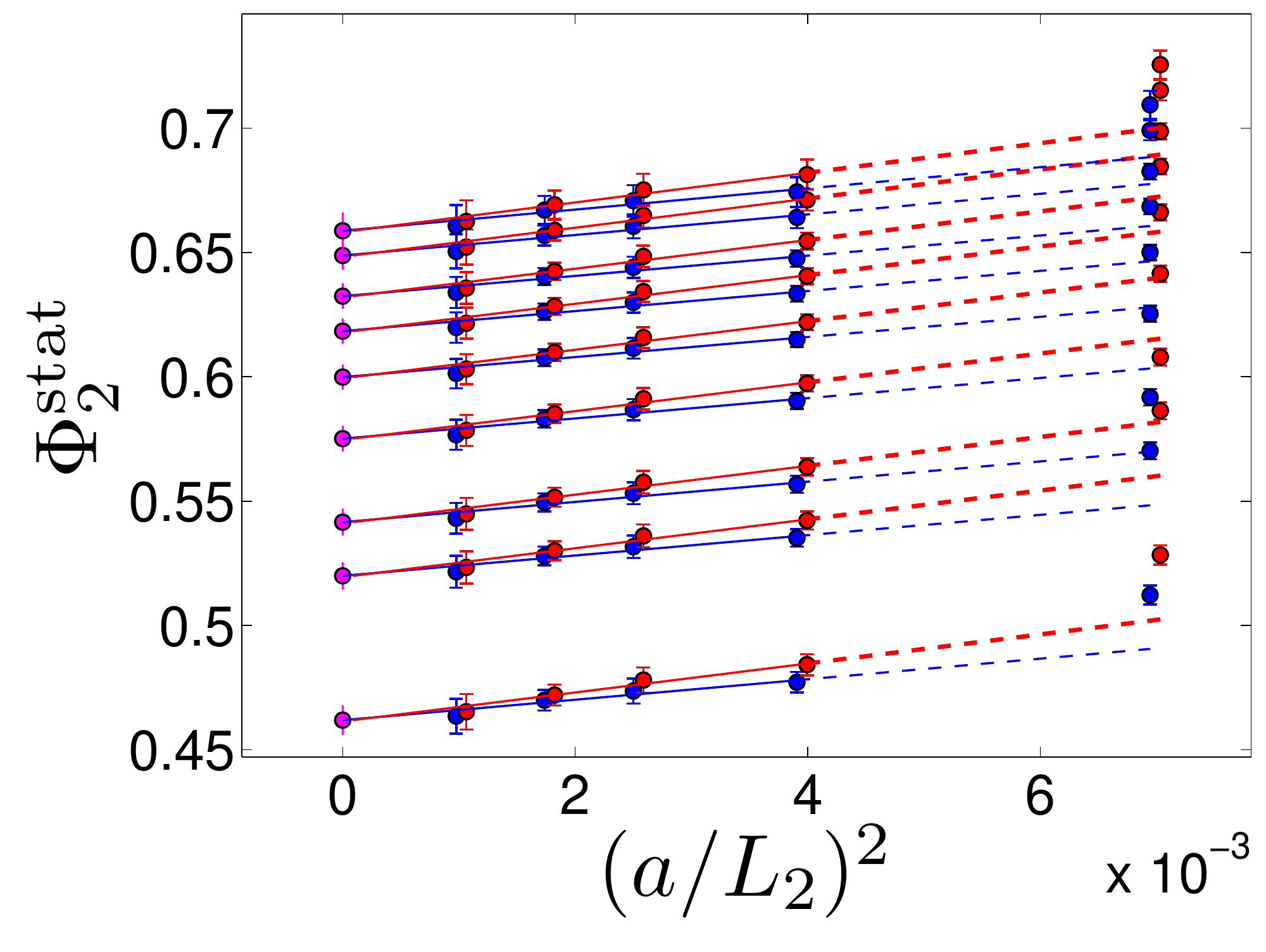}
 \\
  \caption{Continuum extrapolation of $\Phi_1$ and $\Phi_2$ in $L=L_1$ (left).
      Their resulting mass-dependence ($z=ML$, middle). 
      On the right we show
      the step scaling to obtain them in $L=L_2$ in LO HQET. 
      The b-quark in Nature has about $1/z\approx0.08$.
           \label{f:Phi12} }
\end{figure*}

In static order, only $\Phi_1,\,\Phi_2$  are needed and they are finite volume generalizations of mass and decay constant 
of the B-meson. In \fig{f:Phi12} we illustrate their determination. 
The observables are computed for various
fixed $z=M L_1$ and fixed $L_1$, but different resolutions 
$L_1/a=20\ldots40$. They are then extrapolated to the continuum limit
using the asymptotic dominance of $(a/L_1)^2$ corrections. 
In order to carry this step out, one needs to
control the non-perturbative determination of the RGI mass $M$ 
and one needs to know what fixed $L$ means in terms of the bare parameters. Indeed, fixed $L$ is only defined up to discretisation
errors. Here it was chosen to mean fixed $\bar g_\mathrm{SF}(L)$,
the running coupling in the \textSF\ renormalization scheme.
The required full control over the renormalization of QCD in the 
light sector had been gained in a series of earlier works. Some milestones are 
\cite{DellaMorte:2003jj,DellaMorte:2004bc,DellaMorte:2005kg,DellaMorte:2005rd,DellaMorte:2008xb,Fritzsch:2010aw},
reviewed in some detail in \cite{sfb:B2}.

The middle column of \fig{f:Phi12} shows the mass-dependence
of $\Phi_i$ for $L=L_1$ where we match. 
The expected $\Phi_1 = \rmO(M) = \rmO(z)$
and  $\Phi_2 = \rmO(1)$ is clearly visible, also the $\rmO(1/M) = \rmO(1/z)$ corrections are seen in $\Phi_2$. The right column
finally demonstrates the good control over the continuum
limit in the scaling step to $L_2$, \eq{eq:Phi_L2}. It is only in this last step where the restriction to the static approximation is relevant at all. Still, the graph does not look much different when 
the full system with five $\Phi_i$ is treated including the NLO
corrections \cite{Blossier:2012qu}. 

\begin{figure*}[p!]
\centering
   \includegraphics[width=0.400\textwidth]{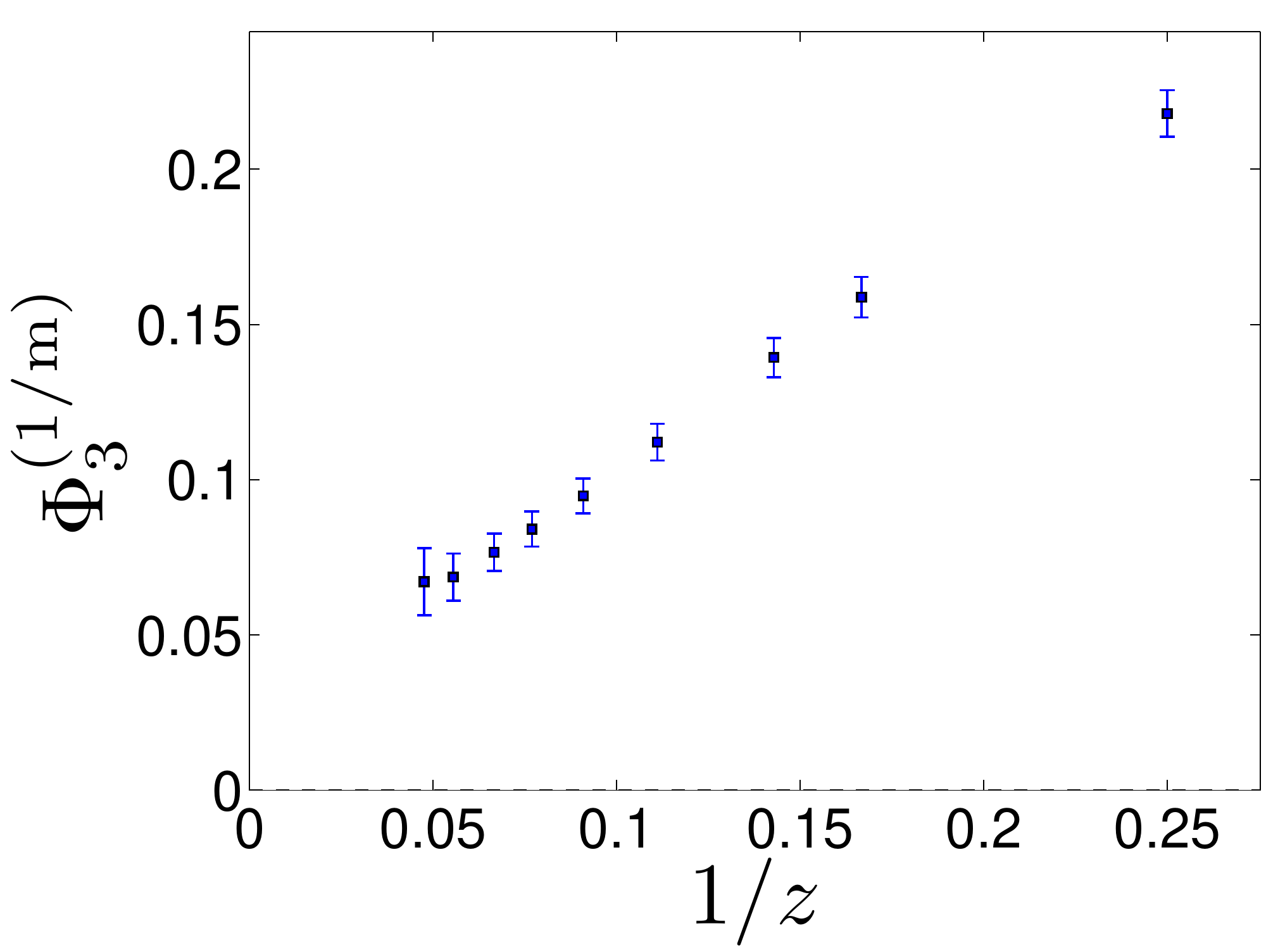}
  \includegraphics[width=0.400\textwidth]{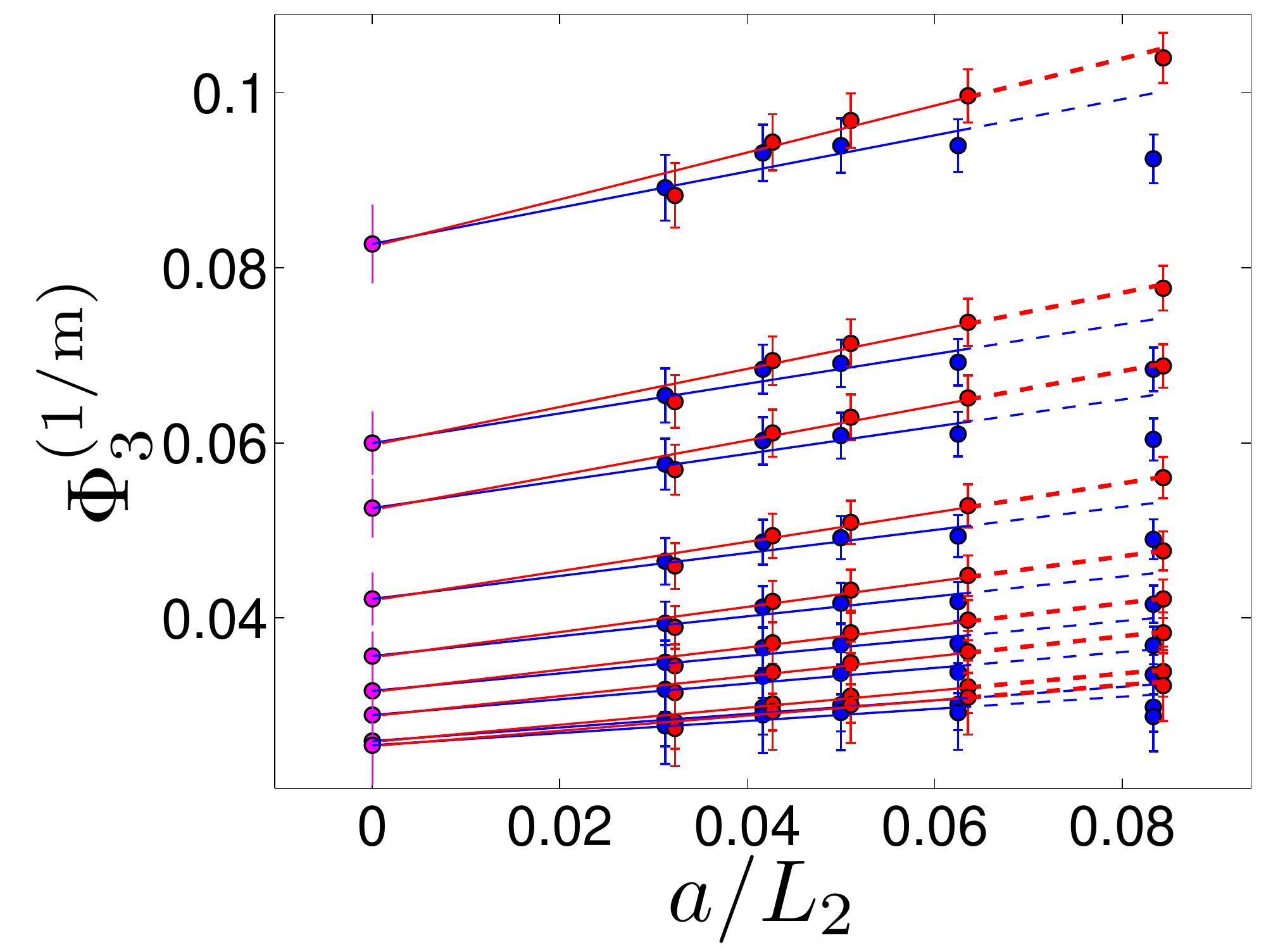}
  \\  
   \includegraphics[width=0.400\textwidth]{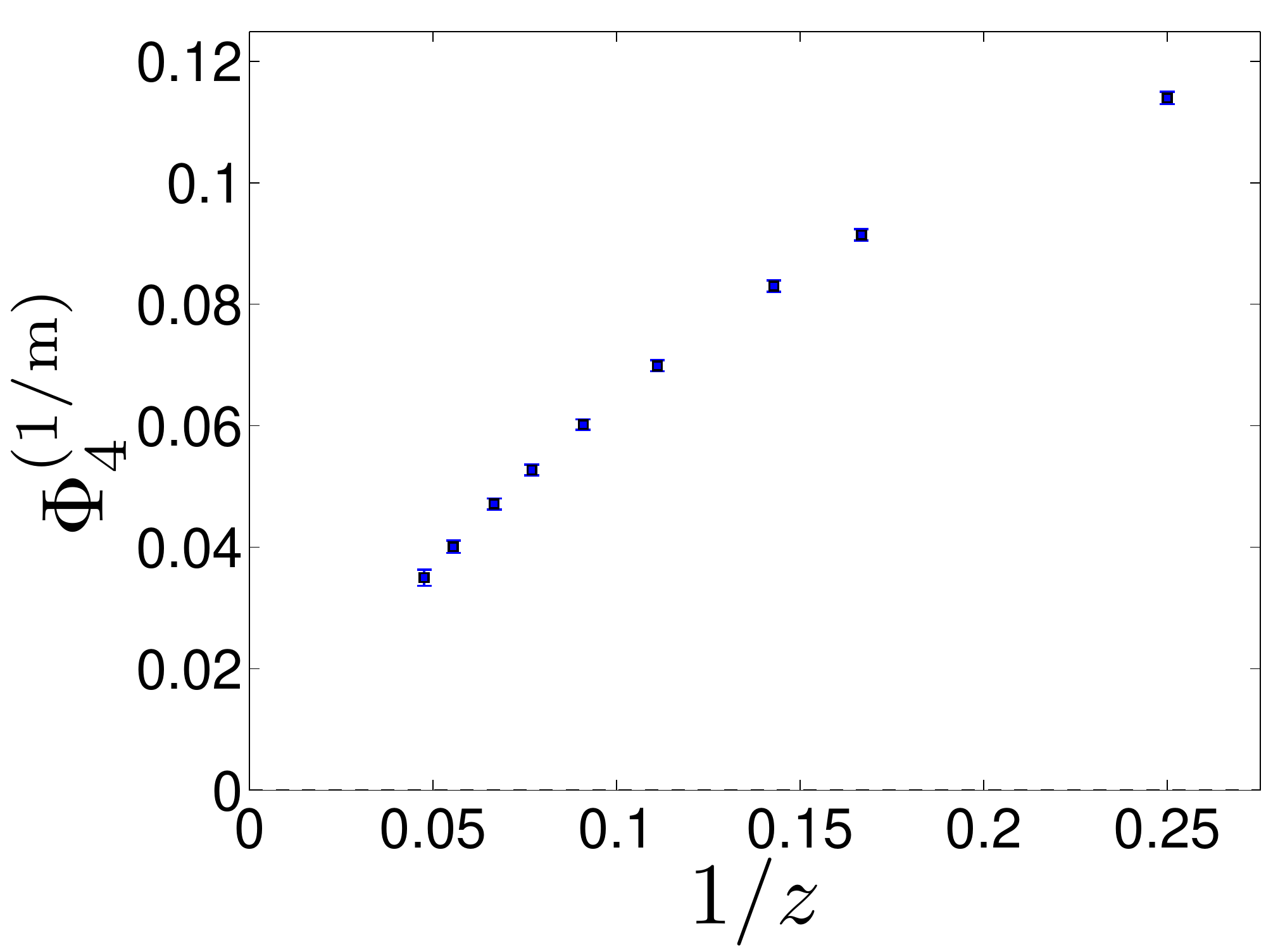}
  \includegraphics[width=0.400\textwidth]{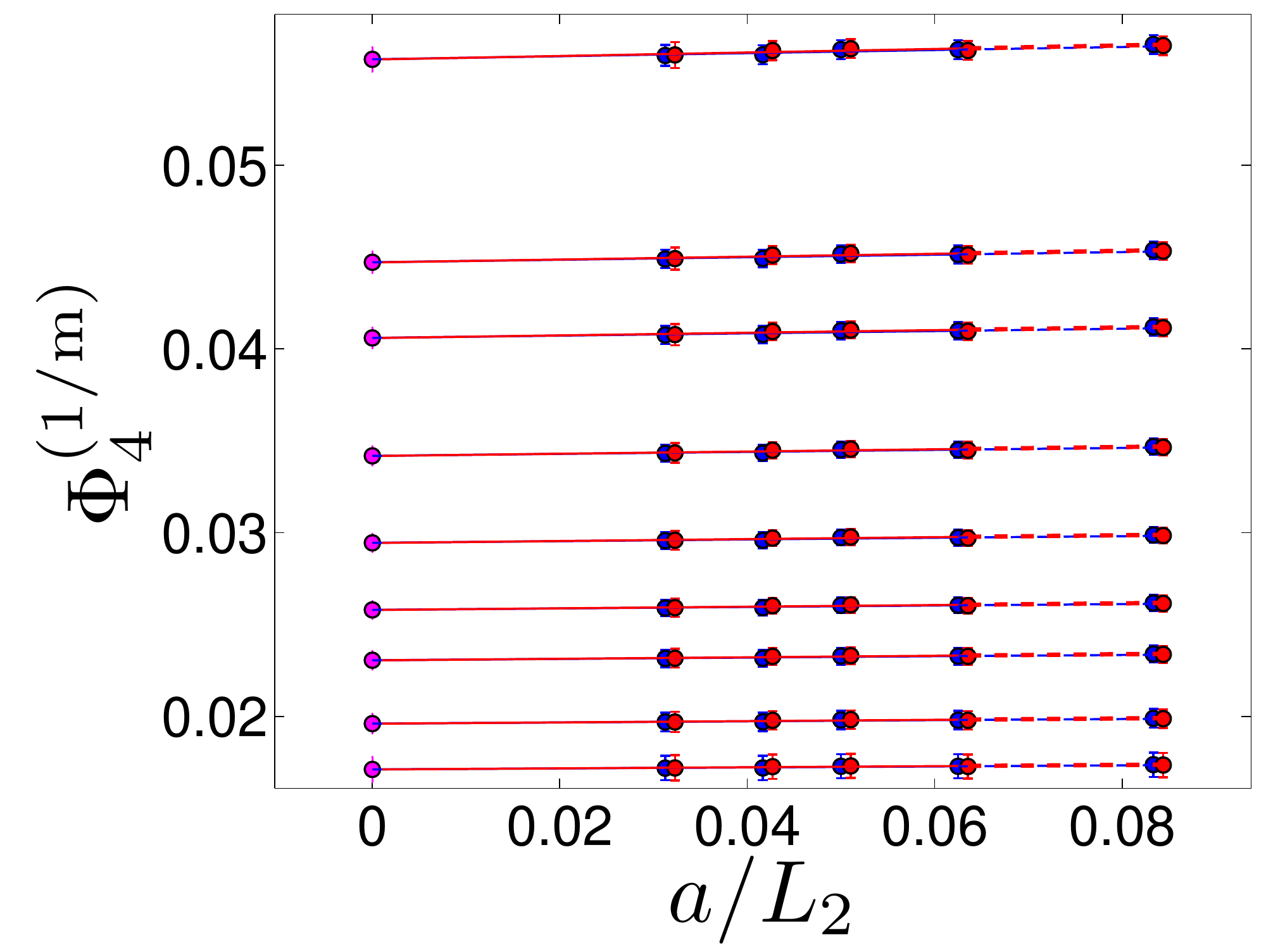}
  \\
  \includegraphics[width=0.400\textwidth]{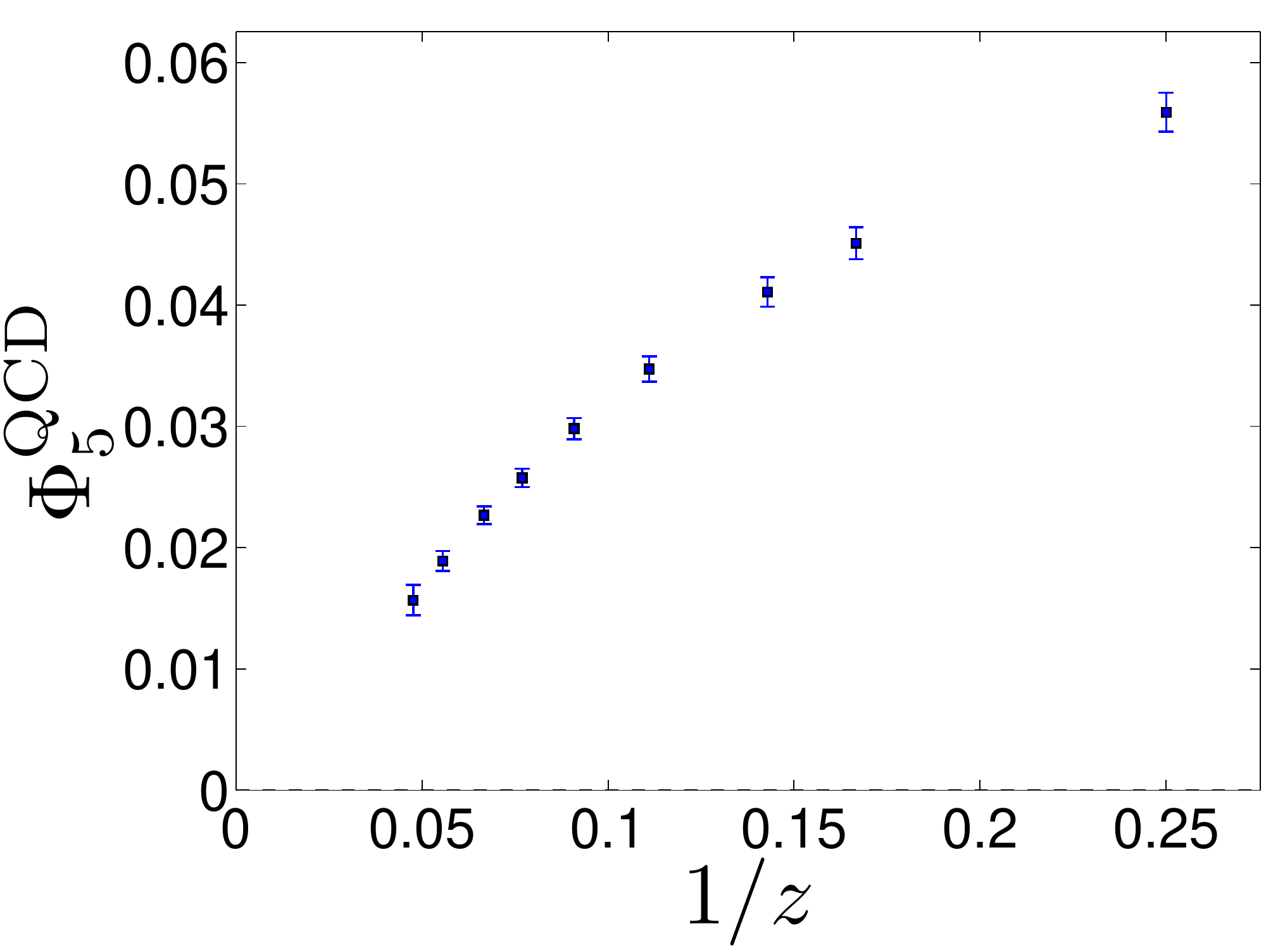}
  \includegraphics[width=0.400\textwidth]{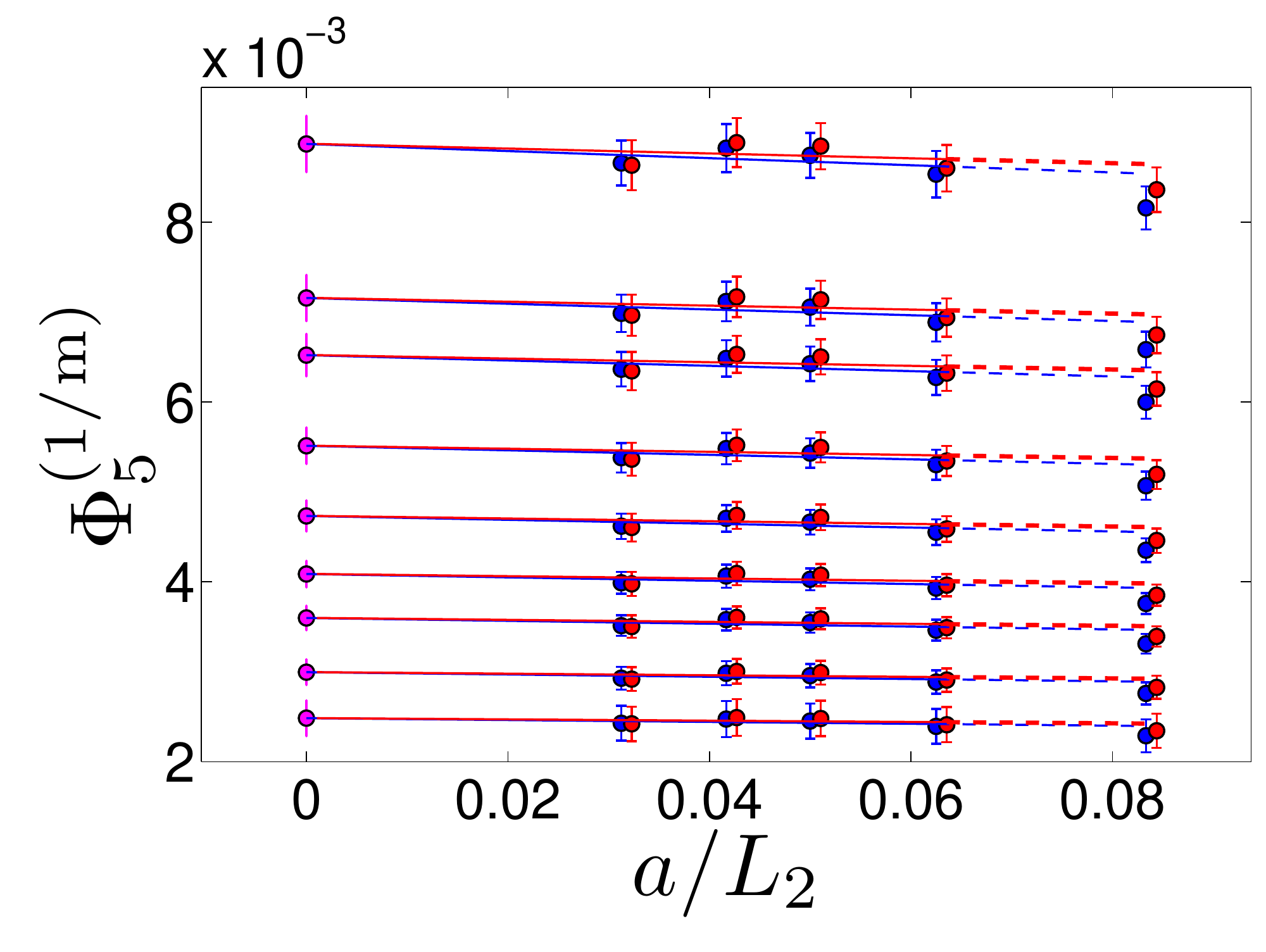}
  \\
  \caption{$\Phi^{(\mhinv)}_i(L_1) = \Phi_i^\qcd(L_1)-\eta_i(L_1),\; i=3,4$ in $L_1$ 
      on the left and their step scaling function to obtain
      $\Phi^{(\mhinv)}_i(L_2) = \Phi_i^\qcd(L_2)-\eta_i(L_2)$ 
      on the right. The bottom line shows $\Phi_5(L_1)$; spin symmetry
      means that its static part vanishes, $\eta_5=0$.
      The b-quark in Nature has about $1/z\approx0.08$.
           \label{f:Phi345} }
\end{figure*}

Let us now illustrate these NLO
corrections. \Fig{f:Phi345} shows the other three observables 
after subtracting their static part. They clearly exhibit
the $\rmO(1/M) = \rmO(1/z)$ behavior at large mass. Again
this represents a
confirmation of the correctness of HQET. 

With these steps carried out, it is only left to 
evaluate \eq{e:omegafin} at the desired lattice spacings
where large volume simulations are carried out. 
The HQET parameters are then known as a function of
the RGI mass $M$, more precisely $z=ML_1$ and
at a few values of the lattice spacing corresponding to 
integer $L_2/a$. By an interpolation the knowledge 
of the parameters is extended to all values of the 
lattice spacing within the range corresponding to  the 
accessible $L_2/a$. The correct
value $\Mbeauty$ for a b-quark in Nature still needs
to be determined by matching the meson mass as a 
function of $M$ to the experimental B-meson mass.
This step yields also the first prediction of the theory,
namely the physical value of the renormalized 
b-quark mass, which is one of the fundamental parameters
of QCD. We turn to it now.

\subsection{Mass of the b-quark}

\begin{table*}[t!]
\begin{center}
\renewcommand{\arraystretch}{1.25}
\begin{tabular}{@{\extracolsep{0.02cm}}lllllllllll}
\toprule
        & & & \multicolumn{2}{c}{$\mbeauty(\mbeauty)$[GeV], $\msbar$ } &  &
               \multicolumn{2}{c}{$\Mbeauty$[GeV]}   
      \\ 
     \cmidrule(lr){4-5}\cmidrule(lr){7-8}
    $\nf$  & ref. & remarks & {static} & {$\rmO(\mhinv)$} &
                            & {static} & {$\rmO(\mhinv)$} 
      \\ \midrule \hline
      0-2 &  \cite{Ryan:2001ej} & 2-loop subtracted; 2001 average & 4.30(10) & &
        & \\
      0 &  \cite{Heitger:2003nj} & NP & 4.12(7)(4) & &
        & \\
      0 &  \cite{DellaMorte:2006cb} & NP & 4.32(\phantom{1}5) & 4.35(\phantom{1}5) &
        & 6.81(\phantom{1}8) & 6.76(\phantom{1}9) \\
      2 & \cite{Bernardoni:2013xba} & NP & 4.21(11) & 4.21(11) &
        & 6.57(17) &  6.58(17)  \\
\bottomrule
\end{tabular}
\caption{HQET results for the RGI b-quark mass $\Mbeauty$ and 
$\mstar=\mbeauty(\mbeauty)$ in the  $\msbar$ scheme
computed using lattice HQET. Results with perturbative subtraction of 
$1/a$ divergences are labeled ``n-loop subtracted''. When renormalization 
and matching is treated non-perturbatively, we just have a label ``NP''.
\label{t:mbfb}
} 
\end{center}
\end{table*}

\subsubsection{Static order}
Lattice HQET computations of the b-quark mass have been carried out originally with a just perturbative subtraction of the 
$1/a$ divergence \cite{Martinelli:1998vt,Trottier:2001vj,DiRenzo:2000nd} in the static approximation. 
A continuum limit does not exist in this case, so a corresponding
extrapolation may not be performed. Instead one can check for stability under changes of $a$. When present, such a stability indicates that 
divergent terms as well as discretisation effects are small.
For comparison we include 
a world average of the year 2001 by S.~Ryan~\cite{Ryan:2001ej}
in our summary, \tab{t:mbfb}.

The first computations with non-perturbative renormalization
were still restricted to the quenched approximation.:
The initial static order result of~\cite{Heitger:2003nj}
was later extended to a full
NLO mass computation in \cite{DellaMorte:2006cb}. 

\subsubsection{HQET at NLO}
\begin{figure*}[p!]
   \centering
   \includegraphics[width=0.87\textwidth]{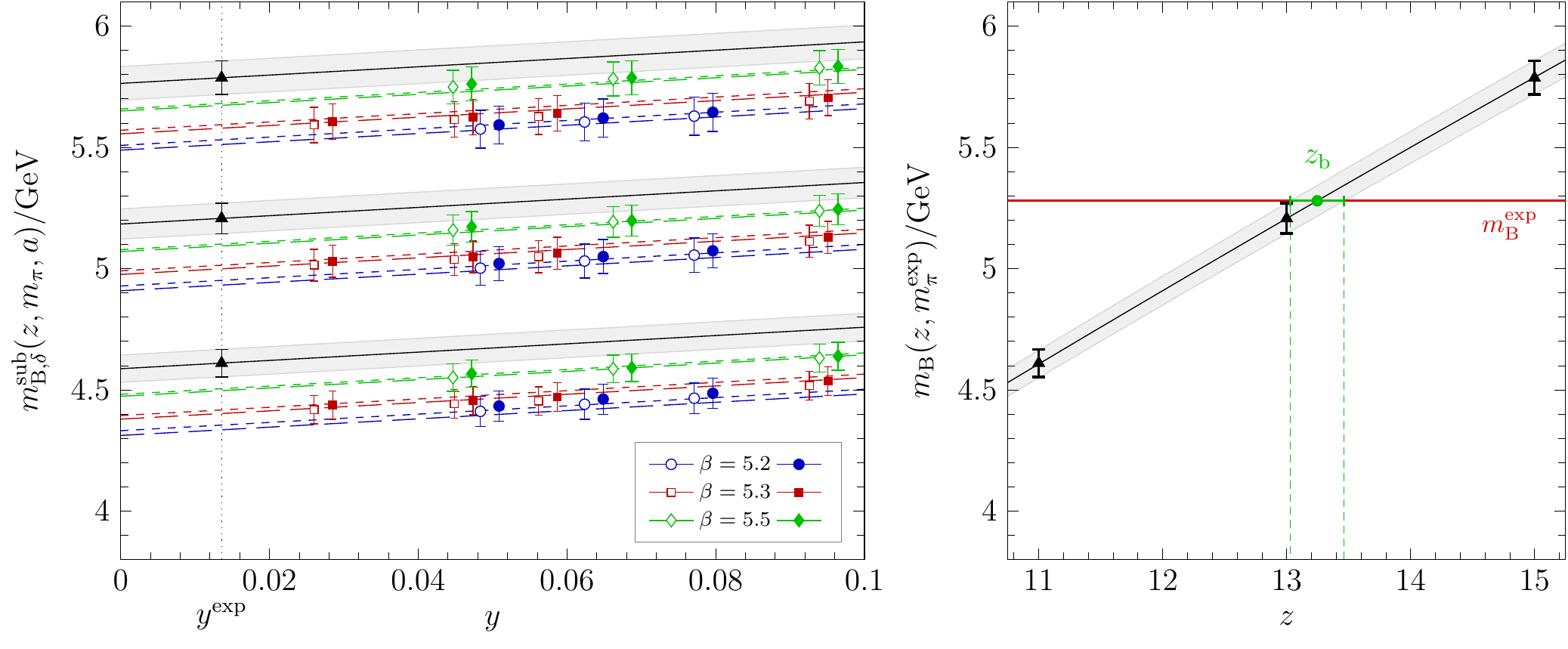}
   \caption{\label{f:mbextrap}
            Left: Chiral and continuum extrapolation of 
            $\mBd$ for the $z$-values used in the 
            determination of $\zb$. Open/filled symbols refer 
            to HYP1/HYP2 data 
            points as do long/short dashed curves, respectively. 
            The continuum pion mass dependence is given by the
            solid lines together with a shaded error band. 
            The triangle shows the estimated continuum limit
            at the physical pion mass.
            The figure uses a notation $y = \xi$ and $\delta$ is 
            an index for the different disretizations.
            \newline
            Right: Interpolation to $\zb$ by imposing \eq{e:zbcond}.
            Graphs from \cite{Bernardoni:2013xba}.
           }
\end{figure*}

Fairly recently a NLO HQET computation with $\nf=2$ flavors of 
dynamical fermions 
based on the parameter determination discussed in the previous section was completed  \cite{Bernardoni:2013xba}.
It used the GEVP method \cite{Blossier:2009kd} for better 
control of the computation of the B-meson mass (at a fixed
lattice spacing and other bare parameters such as $\mhbare$)
and exploited several CLS lattices. They cover 
three lattice spacings between $a=0.048$fm and $a=0.075$fm and several light-quark masses corresponding to
pion masses $190\,\MeV \leq \mpi \leq 440\,\MeV$.
At a fixed mass of the b-quark, parameterized by $z=ML$, 
these results needed to be extrapolated to the
physical pion mass and to the continuum limit (all volumes
are large enough to safely neglect finite volume corrections).
This extrapolation was performed in the form of one global fit: 
\bi 
\item[--]
The non-analytic term in the $\mpi^2$ expansion, \eq{e:mBHM}, discussed in and known from \sect{s:bbpi} is subtracted from the data.
\item[--]
The remaining mass-dependence is parameterized by a linear 
term in $\mpi^2$~($\propto \xi$). 
\item[--]
The lattice spacing dependence is parameterized by the 
leading term in the Symanzik expansion, $\propto a^2$.
\item[--]
The extrapolation is carried out simultaneously for two discretizations of HQET (HYP1/2). All fit-parameters except
for the coefficient of $a^2$ are common to both discretizations.
\item[--]
Physical units are taken from a previous extrapolation of the 
kaon decay constant to the physical point and continuum limit \cite{Fritzsch:2012wq}.
\ei 
This parameterization of lattice spacing and quark mass
dependence neglects higher order terms in 
$a^2$ and $\xi$, also the mixed term $a^2\times \xi$.
It fits the data very well. Data and fit, together with the 
continuum limit, are shown in  \fig{f:mbextrap} for 
three prescribed values of $z$.

The results at $a=0$ and the physical pion mass
were then interpolated in $z$ and finally $\zb = \Mbeauty L_1$
was determined by requiring
\bes \label{e:zbcond}
  \mB(\zb) = \mB^\mathrm{experimental} \,.
\ees 
for the interpolation function $\mB(z)$, see the right graph in 
\fig{f:mbextrap}.
The resulting number $\Mbeauty=6.58(17)$GeV is included in our 
summary, \tab{t:mbfb}. Repeating the analysis with all $\mhinv$ terms
dropped yields the static result. It is almost identical indicating
that NNLO terms are very small. Note that a naive estimate of these
terms, $\Lambda^3/\mbeauty^2 \approx 4$MeV, is indeed very small 
compared to the dominating statistical errors.  

The RGI mass $\Mbeauty$ can be changed easily to $m_*$ defined
earlier, using a knowledge of the $\Lambda$-parameter 
\cite{Fritzsch:2012wq} and inserting the $\msbar$ 
4-loop $\tau$ and $\beta$ 
functions \cite{MS:4loop1,MS:4loop2,MS:4loop3,Czakon:2004bu} into \eq{e:lammu} and \eq{e:gstarM}. 

These numbers (last row of \tab{t:mbfb}) are well controlled. 
Given the small difference to $\nf=0$ results, it is even 
very hard to imagine that the missing strange quark is 
a significant source of error. The numbers agree with other
determinations and the PDG average \cite{PDG2014} of 
$m_\mathrm{b*}=4.18(3)$. They thus 
boost our confidence in having systematic errors in
the determination of $\mbeauty$ under control. 

\begin{figure}[th!]
  \centering
  \includegraphics[width=0.47\textwidth]{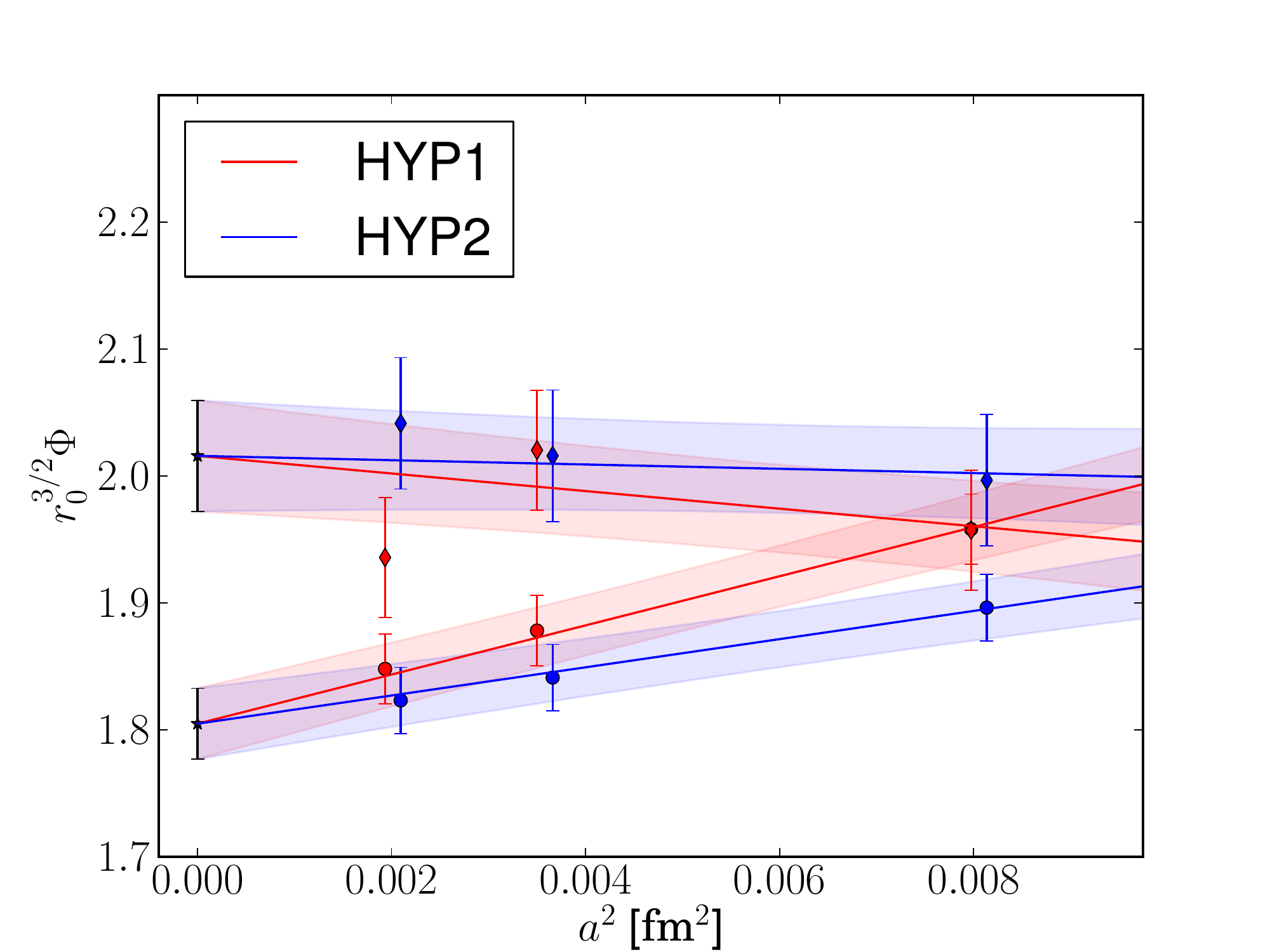}
  \caption{Continuum limit of the $\nf=0$ static matrix element 
          $\Phi = \langle 0| \opAstat |B \rangle $ in RGI     
          normalization. 
          The lower part shows the ground state
           matrix element and the upper part the first excited state
           with the quantum number of the B-meson.  Graph from     
           \cite{Blossier:2010mk}
           \label{f:contlimnf0} }
\end{figure}
\begin{figure}[th!]
  \centering
  \hspace*{-4mm}\includegraphics[width=0.52\textwidth]{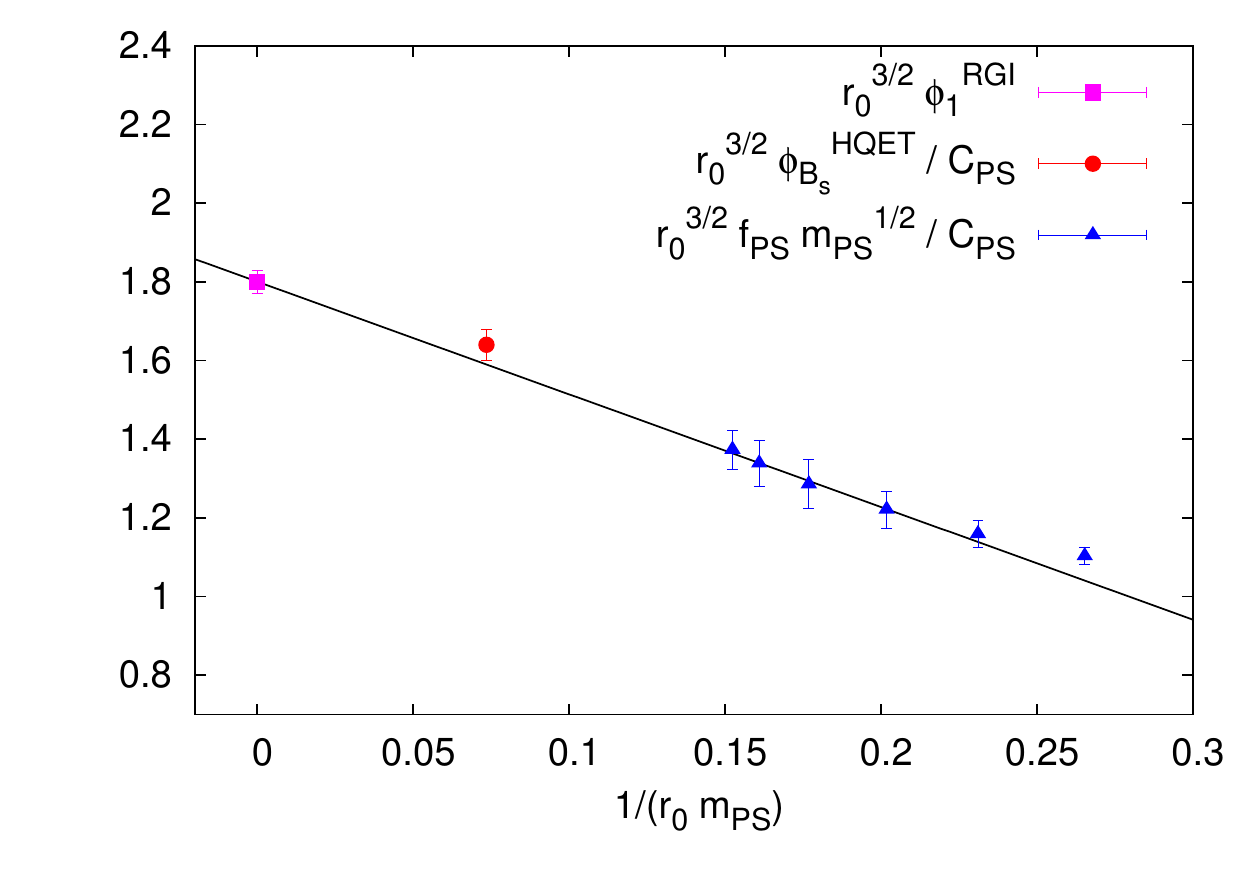}
  \caption{$\nf=0$ heavy quark mass dependence of the
  heavy-strange decay constant. All dimension-full quantities are expressed in units of 
$r_0 \approx 0.5\fm$~\cite{Sommer:1993ce}.
The charm quark is located 
  at about $m_\mathrm{PS} r_0=0.2$ and the b-quark at 
  $m_\mathrm{PS} r_0=0.07$. Graph from 
\cite{Blossier:2010mk}.
           \label{f:fbsnf0} }
\end{figure}

\begin{table*}[p!]
\begin{center}
\renewcommand{\arraystretch}{1.25}
\begin{tabular}{@{\extracolsep{0.02cm}}llp{0.08\textwidth}llllllllll}
\toprule    
        & & & \multicolumn{2}{c}{$\fb$[MeV] } & & 
              \multicolumn{2}{c}{$\fbs$[MeV] } & &
              \multicolumn{2}{c}{$\fbs/\fb$}   
      \\ 
     \cmidrule(lr){4-5}\cmidrule(lr){7-8}\cmidrule(lr){10-11}
    $\nf$  & ref. & remarks & {static} & {$\rmO(\mhinv)$} &
                            & {static} & {$\rmO(\mhinv)$}  &
                            & {static} & {$\rmO(\mhinv)$} 
      \\ \midrule \hline
    0 & \cite{reviews:beauty} & summary PT   
      &  \multicolumn{2}{l}{276(55)(19)} 
    & & & & & 1.22(4)(2) 
    \\  
    0 & \cite{Blossier:2010mk} & NP &  &  &  
        & 229(3) &  216(5) & &  &    
\\
    2 & \cite{Bernardoni:2014fva} & NP & 190(5)(2) & 186(13) &  
        & 226(6)(9) &  224(14) & & 1.189(24)(30) &  1.203(65)
\\
    2+1 & \cite{Aoki:2014nga} & PT, 1-loop & 219(17) &  &  
        & 264(19) &   & & 1.193(41) &  
\\        \midrule  
     2  & \cite{Aoki:2013ldr} & FLAG average$^1$  &
         \multicolumn{2}{c}{189(8)} & & 
         \multicolumn{2}{c}{228(8)} & &
         \multicolumn{2}{c}{1.206(24)} & &    
\\
     2+1  & \cite{Aoki:2013ldr} & FLAG  average$^1$ &
         \multicolumn{2}{c}{190.5(4.2)} & & 
         \multicolumn{2}{c}{227.7(4.5)} & &
         \multicolumn{2}{c}{1.202(22)} & &    
\\
\bottomrule
\end{tabular}
\caption{Results for the B-meson decay constant 
computed using lattice HQET in the upper part. 
Results with with completely perturbative
renormalization are labeled by ``PT''.
When renormalization 
and matching is treated non-perturbatively, we just label ``NP''.
\newline
$^1$ The FLAG averages in the lower part combine various methods, 
not just HQET.
\label{t:fb}
} 
\end{center}
\end{table*}

\begin{figure*}
  \small 
  \centering
  \includegraphics[width=0.87\textwidth]{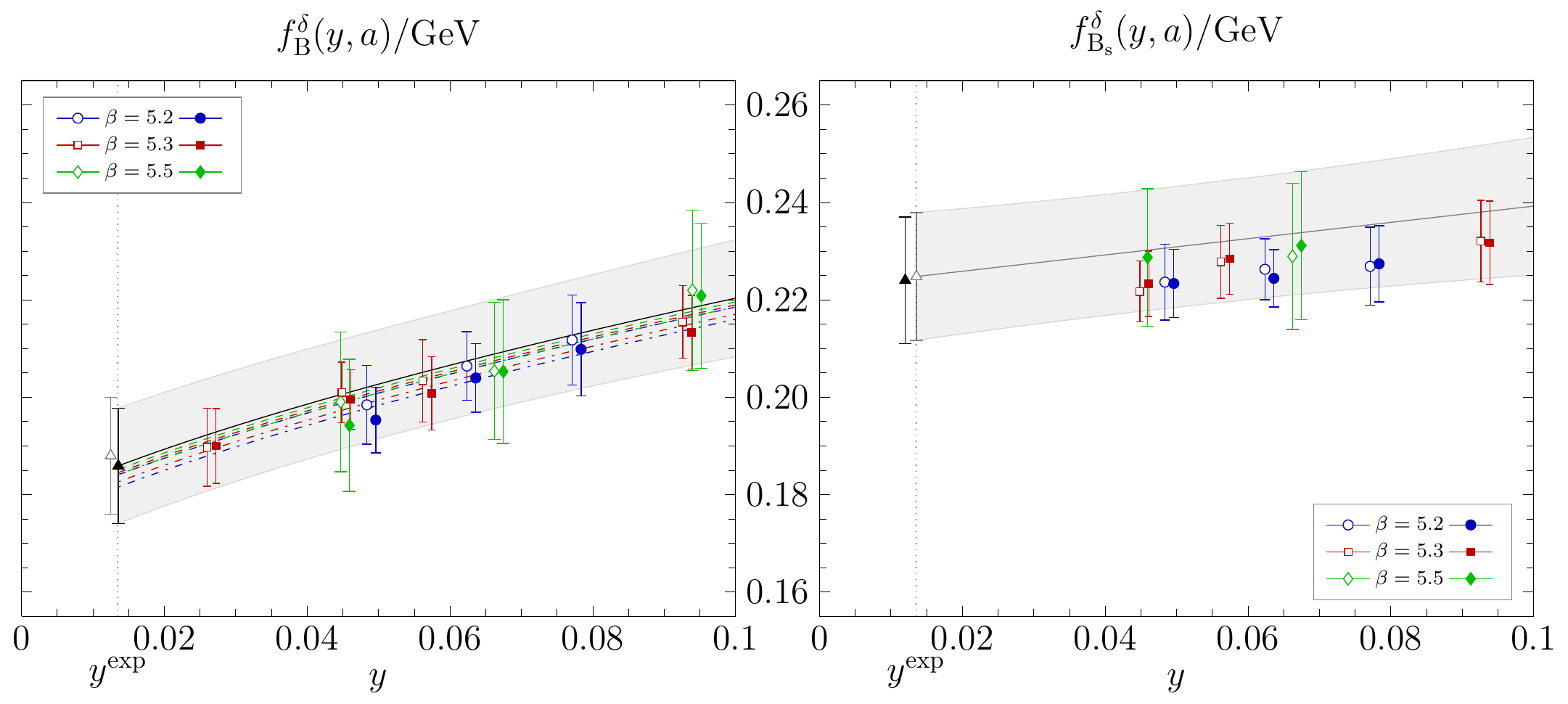}
  \vskip-0.8em
  \caption{Extrapolation of the $B$ (left panel) and $B_{\rm s}$ (right panel) 
           meson decay constant to the physical point. On the left, 
           the extrapolation using HM$\chi$PT at NLO (filled triangle) is 
           compared to a linear one (open triangle), in order to extract the systematic 
           error from truncating HM$\chi$PT at NLO. For $\fBs$ only a LO
           formula is known and shown. As a comparison also the final
           result, the continuum value of $\fBs=[\fBs/\fB]\fB$ is
           shown. Figure from \cite{Bernardoni:2014fva}.
           \label{f:extrfB} }
\end{figure*}

\subsection{B-meson decay constants}
\subsubsection{Mass-dependence}
Before discussing the most up-to-date $\nf=2$ 
results for the decay constants, 
we spend some time on an important 
theoretical aspect: what is the dependence of
the decay constants on the heavy quark mass?
Does the asymptotic mass-scaling \eq{e:f_asymptotoics} 
even reach down to $\mh=\mcharm$
and how far is the b-quark from the heavy quark limit?
With good statistical and systematic precision,
these questions have only been studied in the 
quenched approximation \cite{DellaMorte:2006cb,Blossier:2010mk}.
For qualitative earlier results we refer to \cite{reviews:beauty,Ryan:2001ej}. In \cite{Blossier:2010mk}
the static matrix elements were first determined 
at three different lattice spacings increasing the 
precision by a use of translation invariance as well as
the GEVP method. They were then
extrapolated to the continuum limit as shown in 
\fig{f:contlimnf0}. Only the B$_\mathrm{s}$ meson was investigated because the chiral limit is singular
in the quenched approximation, rendering computations
of the limit of very light quarks rather meaningless.

The static result for the ground state is compared to results 
at finite mass in \fig{f:fbsnf0}.  The graph shows 
$ f_\mathrm{PS} \sqrt{m_\mathrm{PS}} /\Cps
$,
where PS is a strange-heavy pseudo-scalar against 
the inverse heavy meson mass. At the lowest order,
the meson mass is proportional to the quark mass. 
Therefore $m_\mathrm{PS}$ can be taken as a proxy for the 
quark mass.
The conversion function $\Cps$ is computed with the full 
perturbative knowledge, i.e. it has relative
errors of order $\alpha(\mh)^4$. However, as discussed
before, for b- or c- quark masses it is not at all
clear that the perturbative errors are negligible.

The graph shows a very consistent picture,
with the static limit computation on the left being in line
with the relativistic data points on the right {\em and}
an HQET computation with non-perturbative matching and 
including the $\mhinv$ terms at $m_\mathrm{PS} = \mB$.
Given the somewhat uncertain status of the perturbative
$\Cps$, the good agreement with a simple linear behavior
in $1/m_\mathrm{PS} \sim \mhinv$ over
a large range is in fact somewhat surprising.

In the above discussion of the mass-dependence it is hard to get
around a perturbatively extracted $\Cps$.\footnote{Non-perturbative
definitions of $\Cps$ do not separate the logarithmic corrections 
from power corrections.}
Without it taking care of the 
logarithmic modifications of the naive scaling law
the curve in \fig{f:fbsnf0} would not
even have a finite limit at $\mhinv=0$. 
However, for predictions at the b-quark mass 
(or any other large mass), the non-perturbative 
HQET parameters $\omega_i$ 
from \sect{s:param} can be used. From now on we 
only discuss ALPHA collaboration results obtained in this 
entirely non-perturbative way. Only in the comparison to results 
in the literature perturbatively renormalized results by other groups
are shown. 

\subsubsection{HQET results with non-perturbative parameters}

The physical value of the b-quark mass (parameterized by $z_\mathrm{b}$) is known from the previous section. All HQET parameters
are then interpolated to that quark mass.  
Once they are known, the computation of the 
decay constants is in principle straight forward. However, again
care has to be taken about several limits.

\begin{figure}[th]
  \centering
  \includegraphics[width=0.45\textwidth]{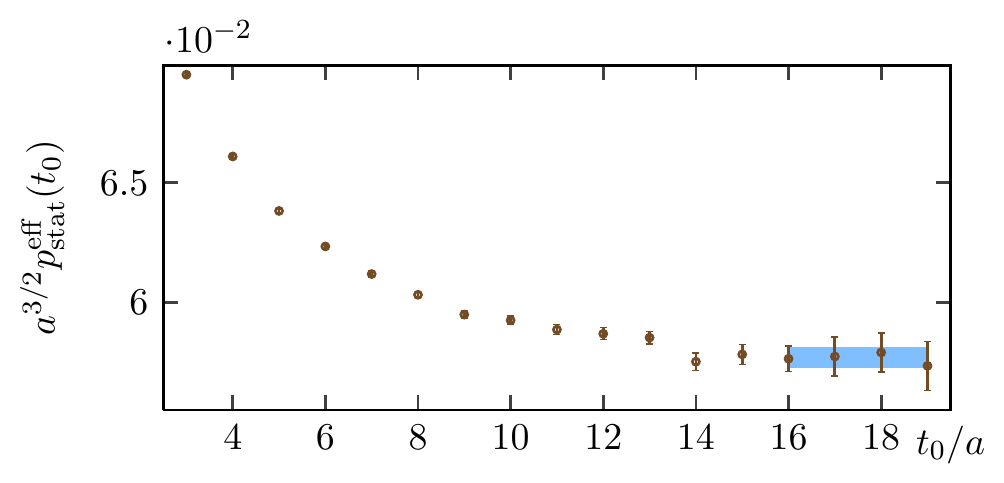}\\[0.2em]
  \includegraphics[width=0.45\textwidth]{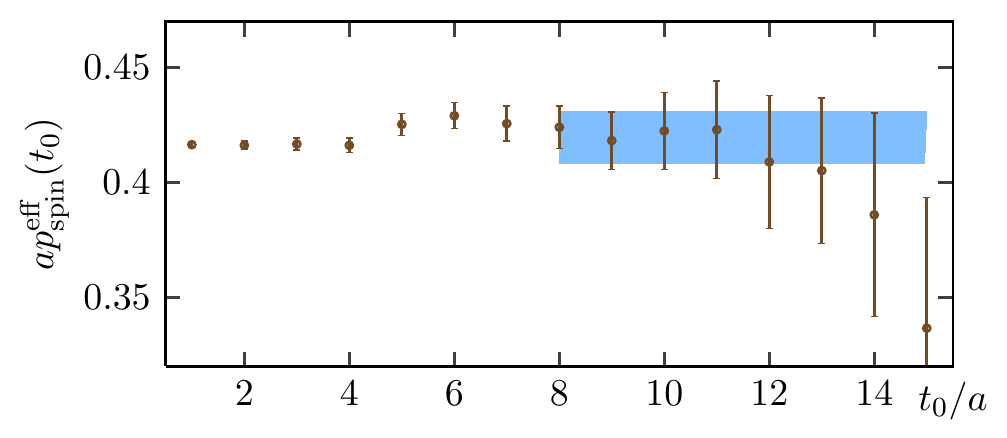}\\[-0.8em]
  \caption{Plateau averages after applying the GEVP analysis to data
           obtained on the $\Nf=2$ CLS ensemble N6 ($a=0.048\,\fm$, 
           $\mpi=340\,\MeV$). The upper plot shows 
           the B$_\mathrm{s}$-meson static matrix element 
           $p_{\rm stat}$. The lower plot shows 
           the chromo-magnetic matrix element $p_{\rm spin}$. 
           Note \eq{e:fbdecomposed} for how these terms enter
           in the expression for the decay constant. Figure from 
           \cite{Bernardoni:2014fva}.
          }
  \label{f:plat_N6}
\end{figure}

1) for each lattice spacing, volume and light quark mass, the decay constant
has to be extracted from the large-time limit of correlation functions. One usually forms ratios of correlation functions
which have the decay constant as their large time (=$t$) limit and
asymptotic corrections to it of order\footnote{The
pre-factor $\Delta E\, t$ is present since we are dealing
with an effective theory computation, 
see \cite{Bernardoni:2014fva}.}
$\Delta E_{2,1} \,t \,\exp(-\Delta E_{2,1}\, t)$. 
Here, $t$ is the time extent of the two-point functions
entering and  $\Delta E_{n,1} = E_n-E_1$ is the difference
of the indicated energy levels;  $\Delta E_{2,1}$ is around 600 MeV. 
Alternatively, with the GEVP method, 
the dominant corrections can be changed to 
$\rmO(\Delta E_{n,1} \,t_0 \,\exp(-\Delta E_{n,1}\, t_0))$
with a larger $n$.
At the same time,  
a second time separation $t_0$ is present in the GEVP which in
practice is $t_0=t/2$.
For a proper explanation we
have to refer to \cite{Blossier:2009kd,Bernardoni:2014fva}.
The analysis in the latter reference estimates the correction
term from the results at smaller $t_0$ and then uses
$t_0$ large enough such that the estimated correction is
a factor three below the statistical error of the final
result. We show two such plateau analysis for the 
B$_\mathrm{s}$ meson in \fig{f:plat_N6}. The plotted
quantities $p_\stat,p_\spin$ 
enter the decay constant through the HQET expansion
\bes
   \fBs  &=& \exp( \chi)\big/\sqrt{a^3\,\mBs/2}   \,, \nonumber
   \\ \label{e:fbdecomposed} 
   \chi &=&  \ln(Z_{A_0}^\hqet) +
   \ln( a^{3/2}p_\stat) \\ && \;\;
              + \omegakin  \, p_{\rm kin}  
              + \omegaspin \, p_{\rm spin}
              + \ceff{A}{0}{1}    \, p_{\rm A^{(1)}_0} \,.
              \nonumber
\ees
The chosen start of the plateau averages 
according to the explained criterion look overly 
conservative in the plot, but 
on the other hand, excited state contaminations are clearly visible
in the static piece, $p_\stat$, and they are 
also hard to rule out for the 
$\mhinv$ term shown. 

In summary, obtaining the ground state 
matrix element is far from trivial, but within the rather conservative errors of \cite{Bernardoni:2014fva} it appears 
to be under control. We do not want to hide that the statistical
errors for the B-meson (rather than B$_\mathrm{s}$) are somewhat larger, see Figure 1 of \cite{Bernardoni:2014fva}.

2) 
A continuum extrapolation has to be carried out and

3) the results have to be extrapolated 
to the physical quark mass. 

Just like in the case of the meson mass, 2) and 3) are carried out in one global fit. 
As seen in
\fig{f:extrfB}, the lattice spacing dependence is not
significant at all and one could just average the numbers
at the different lattice spacings. However, this would
not account for the uncertainty in the statement that
the a-dependence is insignificant. Therefore, as above, an 
$a^2$ term is fitted. The light-quark-mass dependence is parameterized
as a linear dependence and alternatively 
as the one predicted by heavy meson chiral perturbation theory. 
The difference in the extrapolated value is very small
and accounted for in the errors.

4) In principle we should also worry about effects of the 
finite volume, but these are very small for the considered 
volumes.

The resulting numbers are listed in \tab{t:fb} together 
with the previously computed $\nf=0$ decay constants.~\footnote{Earlier 
$\nf=0$ static results are summarised in \cite{reviews:beauty}. A later
static decay constant at a single lattice spacing is found in \cite{Bowler:2000xw}. The results of \cite{Blossier:2010mk} are much more precise and do have a continuum limit extrapolation.
}
The latter were quoted as~\cite{Blossier:2010mk}
\bea\label{fBsstat}
r_0^{3/2}\,f^{\rm stat}_{\rm B_{\rm s}}\sqrt{m_{\rm B_s}} &=&2.14(4)\,,\\
r_0^{3/2}\,f^{\rm HQET}_{\rm B_{\rm s}}\sqrt{m_{\rm B_s}} &=&2.02(5)\,.
\eea
We combine them with the present estimate 
$r_0=0.50(2)$~fm (see \cite{Sommer:2014mea}) to
$f^{\rm stat}_{\rm B_{\rm s}}=229(3)(10)$ MeV and
$f^{\rm HQET}_{\rm B_{\rm s}}=216(5)(9)$ MeV.
In the table we also show older numbers as well as
the recent computation by the RBC/UKQCD 
collaboration~\cite{Albertus:2010nm,Aoki:2014nga}.
All of these use a perturbative renormalization. 
The apparent difference between 2+1 and 2 flavor results
in the static approximation are likely due to the
different renormalization and matching. We remind the reader 
that static results have a dependence on the matching condition
which is of order $\mhinv$. It would therefore be premature
to conclude that the difference between the static numbers in 
the table is due to perturbative renormalization. 

As a separate result, it was also observed, in the continuum limit of the $\nf=0$ theory,  that the decay constant of the excited state is larger than the ground state decay constant 
by the following amount~\cite{Blossier:2010mk}:
\be
{f^{\rm stat}_{\rm B'_{\rm s}}\sqrt{m_{\rm B'_s}}\over f^{\rm stat}_{\rm B_{\rm s}}\sqrt{m_{\rm B_s}}}= 1.24(7) \quad \mbox{for }\;\nf=0 \,.
\ee
A careful continuum limit was needed to observe this splitting of
decay constants, see \fig{f:contlimnf0}. Still it is in agreement
with an older investigation where $a$-effects
were not yet controlled~\cite{Burch:2008qx}.

\section{Conclusions, opportunities and challenges}

So far, numerical results of non-perturbative HQET have been obtained
only in the theory with at most two dynamical 
quark flavors. The dominant impact of the project described here 
concerns the concept, the methodology
and the qualitative features of the HQET expansion.
It has been shown that this works out, in principle and in practice.
Having said that, we would like to emphasize, however,
that the numbers in \tab{t:mbfb} and \tab{t:fb}  provide a very 
valuable crosscheck on other
results for flavor physics, even if they are obtained with 
just up and down quarks as dynamical quarks. The reason 
is that there is overwhelming numerical evidence that strange
quark loop effects are strongly suppressed when one looks at the 
low energy properties of QCD \cite{Aoki:2013ldr,Bruno:2014ufa}. 
Therefore, it is very reasonable to expect present uncertainties
to be dominated by the errors in renormalization and 
matching or in the assumptions made on
the size of lattice spacing effects or the associated extrapolations.
In these respects, HQET is on very solid theoretical grounds.
There are no doubts about its renormalizability and the existence
of the continuum limit, even if these properties
are not proven to all orders of perturbations theory (see \sect{s:status}). Renormalization and 
matching is carried out non-perturbatively without any
compromise.

\begin{figure}[t!]
  \centering
  \vspace*{-4mm}
  \includegraphics[width=0.52\textwidth]{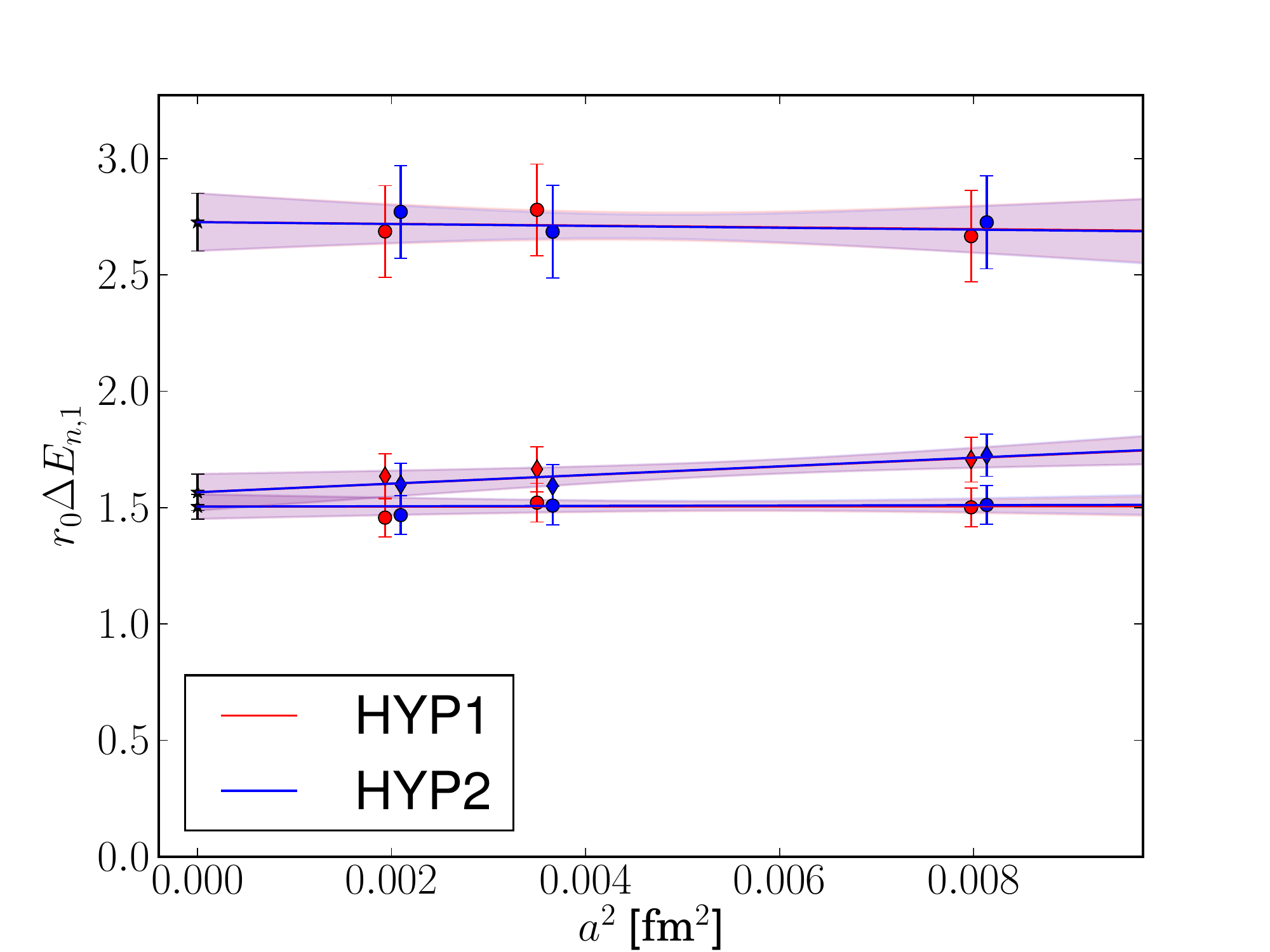}
  \caption{Continuum limits (stars) of 
  $\Delta E_{n,1}^{\rm stat}$ (circles) and $\Delta E_{2,1}^{\rm HQET}$ 
  (diamonds). Shown are the results for both HYP1 (red, shifted to the 
  left) and HYP2 (blue, shifted to the right). Graph from 
  \cite{Blossier:2010vz}. 
           \label{f:contlimDeltaE} }
\end{figure}

A crucial question concerning any expansion is of course 
how large truncation errors are. First of all, we do not expect 
the HQET expansion to be a convergent one. This does not matter,
good asymptotic expansions are just fine, especially if
one is able to compute only a few terms anyway. Looking at the 
size of the computed corrections, it is important to remember 
that they do depend on the matching conditions (see \sect{s:strat}). 
The way we (mostly) perform the matching, HQET is used 
to compute the finite size effects between e.g. a finite volume
definition of a decay constant and the true large volume one.
Apart from the fact that for B-physics we are expanding in 
a truly small parameter $\Lambda/\mbeauty \approx 1/10$,
this is one explanation for the very small NLO corrections
that were found. However, there are cases where a priory we do
not have a reason to expect such additional suppressions 
due to the matching condition. One such case is the splitting between
ground state and first (``radially'') excited state of a B-meson~\cite{Blossier:2010vz}. Its continuum limit in the quenched 
approximation is shown in the lower part of \fig{f:contlimDeltaE}.
It is remarkable that there is no
significant difference between the pure static result and
the one including $\mhinv$ corrections after the continuum limit
has been taken.
In numbers, the static splitting is 
$r_0\Delta E_{2,1}^{\rm stat}=1.50(5)$
while $r_0\Delta E_{2,1}^{(1/m)} = 0.03(6)$ which is a bit 
smaller than expected from a simple
$1.50 \times \Lambda/\mbeauty \approx 0.15 $ estimate. 

There are good opportunities for improving the present numerical 
results in the near future. They arise because the uncertainties are 
(apart from the truncation error which we discussed is small)
by far dominated by statistical errors apart from, maybe, the one
coming from the omission of the strange quark vacuum polarization.
The latter will be removed by the new set of large volume 
CLS simulations \cite{Bruno:2014jqa} (and the associated matching
program) and there is also plenty
of room to enhance the precision in the finite volume matching and step scaling. 

There is also a very good opportunity to help in the 
understanding of the so-called $V_\mathrm{ub}$ puzzle, which
says that this CKM matrix element differs between different 
determinations. Form factors of semi-leptonic B-decays are a 
key to understanding the puzzle. Work on them requires an
extended program to match the full set of currents 
\cite{DellaMorte:2013ega,Hesse:2012hb,Korcyl:2013ara,Korcyl:2013ega}
as well as the computation of form factors in large volume 
HQET \cite{Bahr:2012qs,Bahr:2012vt,Bahr:2014iqa}. We refer to \cite{sfb:C1b} for a review.

Let us finish with a challenge for the future. 
It is the exponential deterioration
of the signal-to-noise ratio, \eq{e:SN}. There are two issues. 
The first is that $E_\mathrm{stat}$ has a linear divergence 
which means the problem becomes worse as we approach the continuum 
limit further. As we have explained 
in \sect{s:hqetlat} this issue is not so severe since the coefficient
of the $1/a$ divergence is very small for the HYP2 static action.
However, the finite part is not small and leads to 
rather short plateaus, see \fig{f:plat_N6}. Formula (\ref{e:SN})
holds for our standard estimators of the correlation functions
and the standard importance sampling. We 
should not give up to think about whether there are ways around it
and see long plateaus with small errors in the future.
\\[3ex]
{\bf Acknowledgements}
\\[1ex]
We would like to acknowledge our collaborators in this project,
F. Bahr, F. Bernardoni, B. Blossier, J. Bulava, M. Della Morte,  M. Donnellan, S. D\"urr, P. Fritzsch, N. Garron, A. G\'erardin, D. Hesse, J. Heitger, G. von Hippel,  A. Joseph, A. J\"uttner, H. B. Meyer, M. Papinutto, A. Ramos, J. Rolf, S. Schaefer,  A. Shindler and H. Simma as well as 
many many fruitful discussions with other colleagues in
the ALPHA Collaboration. We would further like to thank 
Martin L\"uscher for advise, encouragement and for 
founding the ALPHA collaboration. We are grateful to 
Martin Beneke and Patrick Fritzsch for very useful comments 
on earlier versions of this text.

We are grateful for the support of the 
Deutsche Forschungsgemeinschaft (DFG)
in the SFB/TR~09 ``Computational Particle Physics'' and 
we have profited from the scientific exchange 
in the SFB. 

Lastly, the results described here are also
due to a lot of support in the form of computational ressources.
We
gratefully acknowledge the Gauss Centre for Supercomputing (GCS)
for providing computing time through the John von Neumann Institute for
Computing (NIC) on the GCS share of the supercomputer JUQUEEN at J\"ulich
Supercomputing Centre (JSC). GCS is the alliance of the three national
supercomputing centres HLRS (Universit\"at Stuttgart), JSC (Forschungszentrum
J\"ulich), and LRZ (Bayerische Akademie der Wissenschaften), funded by the
German Federal Ministry of Education and Research (BMBF) and the German State
Ministries for Research of Baden-W\"urttemberg (MWK), Bayern (StMWFK) and
Nordrhein-Westfalen (MIWF).
We acknowledge PRACE for awarding us access to resource JUQUEEN in Germany at J\"ulich  and to resource SuperMUC based in Germany at LRZ, Munich. We thank DESY for access to the PAX cluster in Zeuthen.





\end{document}